\documentclass[traditabstract,longauth]{aa}  

\usepackage{graphicx}
\usepackage{txfonts}
\begin{document}
\title{GOODS--{\it Herschel}: an infrared main sequence for star-forming galaxies\thanks{{\it Herschel} is an ESA space observatory with science instruments provided by European-led Principal Investigator consortia and with important participation from NASA.}}

\author{D.~Elbaz\inst{1}
\and M.~Dickinson\inst{2}
\and H.S.~Hwang\inst{1}
\and T.~D\'iaz-Santos\inst{3}
\and G.~Magdis\inst{1}
\and B.~Magnelli\inst{4}
\and D.~Le Borgne\inst{5}
\and F.~Galliano\inst{1}
\and M.~Pannella\inst{1}
\and P.~Chanial\inst{1}
\and L.~Armus\inst{6}
\and V.~Charmandaris\inst{3,7}
\and E.~Daddi\inst{1}
\and H.~Aussel\inst{1}
\and P.~Popesso\inst{4}
\and J.~Kartaltepe\inst{2}
\and B.~Altieri\inst{8}
\and I.~Valtchanov\inst{8}
\and D.~Coia\inst{8}
\and H.~Dannerbauer\inst{1}
\and K.~Dasyra\inst{1}
\and R.~Leiton\inst{1,9}
\and J.~Mazzarella\inst{10}
\and D.M.~Alexander\inst{11}
\and V.~Buat\inst{12}
\and D.~Burgarella\inst{12}
\and R.-R.~Chary\inst{6}
\and R.~Gilli\inst{13}
\and R.J.~Ivison\inst{14,15}
\and S.~Juneau\inst{16}
\and E.~Le Floc'h\inst{1}
\and D.~Lutz\inst{4}
\and G.E.~Morrison\inst{17,18}
\and J.R.~Mullaney\inst{1}
\and E.~Murphy\inst{6}
\and A.~Pope\inst{19}
\and D.~Scott\inst{20}
\and M.~Brodwin\inst{2}
\and D.~Calzetti\inst{19}
\and C.~Cesarsky\inst{1}
\and S.~Charlot\inst{5}
\and H.~Dole\inst{21}
\and P.~Eisenhardt\inst{22}
\and H.C.~Ferguson\inst{23}
\and N.~F{\"o}rster Schreiber\inst{4}
\and D.~Frayer\inst{24}
\and M.~Giavalisco\inst{19}
\and M.~Huynh\inst{6}
\and A.M.~Koekemoer\inst{23}
\and C.~Papovich\inst{25,26}
\and N.~Reddy\inst{2}
\and C.~Surace\inst{12}
\and H.~Teplitz\inst{6}
\and M.S.~Yun\inst{19}
\and G.~Wilson\inst{19}
}
\institute{Laboratoire AIM-Paris-Saclay, CEA/DSM/Irfu - CNRS - Universit\'e Paris Diderot, CE-Saclay, pt courrier 131, F-91191 Gif-sur-Yvette, France\\
 \email{delbaz@cea.fr}
      \and National Optical Astronomy Observatory, 950 North Cherry Avenue, Tucson, AZ 85719, USA
	\and Department of Physics and Institute of Theoretical \& Computational Physics, University of Crete, GR-71003 Heraklion, Greece
    \and Max-Planck-Institut f\"ur Extraterrestrische Physik (MPE), Postfach 1312, 85741, Garching, Germany
	\and Institut d'Astrophysique de Paris, UMR 7095, CNRS, UPMC Univ. Paris 06, 98bis boulevard Arago, F-75014 Paris, France
	\and Spitzer Science Center, California Institute of Technology, Pasadena, CA 91125, USA
	\and IESL/Foundation for Research and Technology - Hellas,  GR-71110, Heraklion, Greece and Chercheur Associ\'e, Observatoire de  Paris, F-75014, Paris, France
	\and Herschel Science Centre, European Space Astronomy Centre, Villanueva de la Ca\~nada, 28691 Madrid, Spain
   \and Astronomy Department, Universidad de Concepci\'{o}n, Casilla 160-C, Concepci\'{o}n, Chile
   \and IPAC, California Institute of Technology, Pasadena, CA, 91125, USA
   	\and Department of Physics and Astronomy, Durham University, South Road, Durham, DH1 3LE, U.K.
	\and Laboratoire d'Astrophysique de Marseille, OAMP, Universit\'e Aix-marseille, CNRS, 38 rue Fr\'ed\'eric Joliot-Curie, 13388 Marseille cedex 13, France
	\and INAF-Osservatorio Astronomico di Bologna, via Ranzani 1, I-40127 Bologna, Italy
	\and UK Astronomy Technology Centre, Royal Observatory, Blackford Hill, Edinburgh EH9 3HJ, UK
	\and Institute for Astronomy, University of Edinburgh, Royal Observatory, Blackford Hill, Edinburgh EH9 3HJ, UK
	\and Steward Observatory, University of Arizona, 933 North Cherry Avenue, Tucson, AZ 85721, USA
	\and Institute for Astronomy, University of Hawaii, Honolulu, HI 96822, USA
	\and Canada-France-Hawaii Telescope, Kamuela, HI 96743, USA
	\and Department of Astronomy, University of Massachusetts, Amherst, MA 01003, USA
	\and Department of Physics \& Astronomy, University of British Columbia, 6224 Agricultural Road, Vancouver, BC V6T~1Z1, Canada
	\and Institut d'Astrophysique Spatiale (IAS), b\^atiment 121, Universit\'e Paris-Sud 11 and CNRS (UMR 8617), 91405 Orsay, France
	\and Jet Propulsion Laboratory, California Institute of Technology, 4800 Oak Grove Drive, Pasadena, CA 91109, USA
	\and Space Telescope Science Institute, 3700 San Martin Drive, Baltimore, MD 21228, USA
	\and National Radio Astronomy Observatory, P.O. Box 2, Green Bank, WV 24944, USA 
	\and Department of Physics and Astronomy, Texas A\&M University, College Station, TX 77845-4242, USA
	\and George P. and Cynthia Woods Mitchell Institute for Fundamental Physics and Astronomy, Texas A\&M University, College Station, TX 77845-4242, USA
}
\date{Received 11 May 2011; accepted 3 August 2011}
 
\abstract
{We present the deepest 100 to 500\,$\mu$m far-infrared observations obtained with the {\it Herschel} Space Observatory as part of the GOODS--{\it Herschel} key program, and examine the infrared (IR) 3--500\,$\mu$m spectral energy distributions (SEDs) of galaxies at 0 $<$ $z$ $<$ 2.5, supplemented by a local reference sample from {\it IRAS}, {\it ISO}, {\it Spitzer} and {\it AKARI} data.  We determine the projected star formation densities of local galaxies from their radio and mid-IR continuum sizes.

We find that the ratio of total IR luminosity to rest-frame 8\,$\mu$m luminosity, $IR8$ ($\equiv$$L_{\rm IR}^{\rm tot}$/$L_8$), follows a Gaussian distribution centered on $IR8$=4 ($\sigma$=1.6) and defines an IR main sequence for star-forming galaxies independent of redshift and luminosity. Outliers from this main sequence produce a tail skewed toward higher values of $IR8$. This minority population ($<$20\,\%) is shown to consist of starbursts with compact projected star formation densities. $IR8$ can be used to separate galaxies with normal and extended modes of star formation from compact starbursts with high--$IR8$, high projected IR surface brightness ($\Sigma_{\rm IR}$$>$3$\times$10$^{10}$ L$_{\odot}$kpc$^{-2}$) and a high specific star formation rate (i.e., starbursts). The rest-frame, UV-2700\,\AA\ size of these distant starbursts is typically half that of main sequence galaxies, supporting the correlation between star formation density and starburst activity that is measured for the local sample.

Locally, luminous and ultraluminous IR galaxies, (U)LIRGs ($L_{\rm IR}^{\rm tot}$$\geq$10$^{11}$L$_{\odot}$), are systematically in the starburst mode, whereas most distant (U)LIRGs form stars in the ``normal'' main sequence mode. This confusion between two modes of star formation is the cause of the so-called ``mid-IR excess'' population of galaxies found at $z$$>$1.5 by previous studies. Main sequence galaxies have strong polycyclic aromatic hydrocarbon (PAH) emission line features, a broad far-IR bump resulting from a combination of dust temperatures ($T_{\rm dust}$$\sim$15 -- 50 K), and an effective $T_{\rm dust}$$\sim$31 K, as derived from the peak wavelength of their infrared SED. Galaxies in the starburst regime instead exhibit weak PAH equivalent widths and a sharper far-IR bump with an effective $T_{\rm dust}$$\sim$40 K. Finally, we present evidence that the mid-to-far IR emission of X-ray active galactic nuclei (AGN) is predominantly produced by star formation and that candidate dusty AGNs with a power-law emission in the mid-IR systematically occur in compact, dusty starbursts. After correcting for the effect of starbursts on $IR8$, we identify new candidates for extremely obscured AGNs.
}
\keywords{Galaxies: evolution -- Galaxies: active -- Galaxies: starburst -- Infrared: galaxies}

\maketitle

\section{Introduction}
\label{SEC:intro}
It is now well established that $\sim$85\,\% of the baryon mass contained in present-day stars formed at 0$<$$z$$<$2.5 (see, e.g., Marchesini et al. 2009 and references therein) and that most energy radiated during this epoch by newly formed stars was heavily obscured by dust. To understand how present-day galaxies were made, it is therefore imperative to accurately determine the bolometric output of dust, hence the total IR luminosity, $L_{\rm IR}^{\rm tot}$, integrated from 8 to 1000\,$\mu$m. In the past, this key information on the actual star formation rate (SFR) experienced by distant galaxies was determined by extrapolating observations in the mid-IR and sub-millimeter (sub-mm) or by correcting their UV luminosities for extinction. These extrapolations implied that the number density per unit comoving volume of luminous IR galaxies (LIRGs, 10$^{11}$$\leq$$L_{\rm IR}$/L$_{\sun}$$<$10$^{12}$) was 70 times larger at $z$$\sim$1, i.e., $\sim$ 8 Gyr ago, when LIRGs were responsible for most of the cosmic SFR density per unit co-moving volume (see e.g., Chary \& Elbaz 2001 -- hereafter CE01, Le Floch et al. 2005, Magnelli et al. 2009). Earlier in the past, at $z$$\sim$2, sub-mm and {\it Spitzer} observations revealed that the contribution to the cosmic SFR density of even more active objects, the ultraluminous IR galaxies (ULIRGs, $L_{\rm IR}$$\geq$10$^{12}$ L$_{\sun}$), was as important as for LIRGs (Chapman et al. 2005, Papovich et al. 2007, Caputi et al. 2007, Daddi et al. 2007a, Magnelli et al. 2009, 2011). However, none of these studies used rest-frame far-IR measurements of individual galaxies at wavelengths where the IR spectral energy distribution (SED) of star-forming galaxies is known to peak. At best, they relied on stacking of far-IR data from individually undetected sources.

With the launch of the {\it Herschel} Space Observatory (Pilbratt et al. 2010), it has now become possible to measure the total IR luminosity of distant galaxies directly. Using shallower {\it Herschel} data than the present study, Elbaz et al. (2010) showed that extrapolations of $L_{\rm IR}^{\rm tot}$ from the mid-IR (24\,$\mu$m passband), which was done under the assumption that the IR SEDs of star-forming galaxies remained the same at all epochs, were correct below $z$$\lesssim$1.3, with an uncertainty of only 0.15 dex. However, the extension of this assumption to (U)LIRGs at $z$$\gtrsim$1.3, in large part relying on stacking, failed by a factor 3-5 typically (Elbaz et al. 2010, Nordon et al. 2010). This finding confirmed the past discovery of a so-called ``mid-IR excess'' population of galaxies (Daddi et al. 2007a, Papovich et al. 2007, Magnelli et al. 2011): the 8\,$\mu$m rest-frame emission of $z$$\sim$2 (U)LIRGs was excessively strong compared to the IR SED of local galaxies with equivalent luminosities when deriving $L_{\rm IR}^{\rm tot}$ from the radio continuum at 1.4 GHz, from stacked measurements from {\it Spitzer}-MIPS 70\,$\mu$m, or from the UV luminosity corrected for extinction.

Various causes have been invoked to explain this ``mid-IR excess'' population: \textit{(i)} an evolution of the IR SEDs of galaxies; \textit{(ii)} the presence of an active galactic nucleus (AGN) heating dust to temperatures of a few 100 K; or \textit{(iii)} limitations in local libraries of template SEDs, i.e., the $k$-correction effect on distant galaxies probing regimes where the SEDs were not accurately calibrated. Evidence pointing toward an important role played by obscured AGN to explain these discrepancies (point \textit{(ii)}) came from the stacking of {\it Chandra} X-ray images at the positions of the most luminous $z$$\sim$2 BzK galaxies (Daddi et al. 2007a). The most luminous of these  distant galaxies were detected in both the soft (0.5--2 keV) and hard (2--7 keV) X-ray channels of {\it Chandra} and exhibited a flux ratio typical of heavily obscured ($N_H$$\geq$10$^{23}$ cm$^{-2}$) or even Compton thick AGN ($N_H$$\geq$10$^{24}$ cm$^{-2}$). Surprisingly, however, a high fraction of the same objects, when observed in mid-IR spectroscopy with the {\it Spitzer} IR spectrograph (IRS), were found to possess intense polycyclic aromatic hydrocarbon (PAH, L\'eger $\&$ Puget 1984, Puget $\&$ L\'eger 1989, Allamandola et al. 1989) broad lines with equivalent widths strongly dominating over the hot to warm dust continuum (Rigby et al. 2008, Farrah et al. 2008, Murphy et al. 2009, Fadda et al. 2010, Takagi et al. 2010). Deeper Chandra observations have since showed that only $\sim$25\,\% of the $z$$\sim$2 BzK-selected mid-IR galaxies hosted heavily obscured AGN, the rest being otherwise composed of relatively unobscured AGNs and star-forming galaxies (Alexander et al. 2011). This would instead favor points \textit{(i)} or \textit{(iii)} above. 

In this paper, we present the deepest 100 to 500\,$\mu$m far-IR observations obtained with the {\it Herschel} Space Observatory as part of the GOODS--{\it Herschel} Open Time Key Program with the PACS (Poglitsch et al. 2010) and SPIRE (Griffin et al. 2010) instruments. Thanks to the unique power of {\it Herschel} to determine the bolometric output of star-forming galaxies, we demonstrate that incorrect extrapolations of $L_{\rm IR}^{\rm tot}$ from 24\,$\mu$m observations at $z$$\gtrsim$1.5, and the associated claim for a ``mid-IR excess''  population, do not indicate a drastic evolution of infrared SEDs, nor the ubiquity of warm AGN-heated dust dominating the mid-IR emission. Instead, we show that the 8\,$\mu$m bolometric correction factor ($IR8$$\equiv$$L_{\rm IR}^{\rm tot}$/$L_8$) is universal  in the range 0$<$$z$$\leq$2.5, hence defining an IR ``main sequence'' (MS). We show that past incorrect extrapolations resulted from the confusion between galaxies with extended star formation and those with compact starbursts, which exhibit notably different infrared SEDs.

We present evidence that this IR main sequence is directly related to the redshift dependent SFR -- M* relation (Noeske et al. 2007, Elbaz et al. 2007, Daddi et al. 2007a, 2009, Pannella et al. 2009, Magdis et al. 2010a, Gonzalez et al. 2011) and is able to separate galaxies between those experiencing a ``normal'' mode of extended star formation and starbursts with compact projected star formation densities. This distinction between a majority of ``main sequence'' (MS) galaxies and a minority of compact ``starbursts'' (SB) is analogous to the recent finding of two regimes of star formation in the Schmidt-Kennicutt (SK) law, with MS galaxies following the classical SK relation while the SFR of SB galaxies is an order of magnitude greater than expected from their projected gas surface density (Daddi et al. 2010, Genzel et al. 2010). To separate these two star-formation modes, the GOODS--{\it Herschel} observations of  distant galaxies are supplemented by a reference sample of local galaxies using a compilation of data from {\it IRAS}, {\it AKARI}, {\it Spitzer}, SDSS and radio observations. 

The GOODS--{\it Herschel} observations and catalogs are presented in Section~\ref{SEC:data}. The main limitation of the {\it Herschel} catalogs, the confusion limit, and a ``clean index'' identifying sources with robust photometry are discussed in Sect.~\ref{SEC:clean}. The high- and low-redshift galaxy samples are introduced in Section~\ref{SEC:highlowz} together with a description of the method used to compute total IR luminosities, stellar masses and photometric redshifts. The IR main sequence is presented in Section~\ref{SEC:IR8} where the so-called ``mid-IR excess problem'' is addressed and a solution proposed using the $IR8$ bolometric correction factor. This parameter, which relies on the same rest-frame wavelengths independent of galaxy redshift, is used to separate star-forming galaxies in two modes: a main sequence and a starburst mode. In the following sections, $IR8$ is shown to correlate closely with the IR surface brightness, hence with the projected star formation density, and with the starburst intensity, that we quantify here with a parameter named ``starburstiness'', for local (Section~\ref{SEC:compactness}) and distant (Section~\ref{SEC:MSSB}) galaxies. It is shown that galaxies exhibiting enhanced $IR8$ values are undergoing a compact starburst phase. The universality of $IR8$ among main sequence star-forming galaxies is used to produce a prototypical IR SED for galaxies in the main sequence mode of star formation in Section~\ref{SEC:sed}. We combine {\it Spitzer} and {\it Herschel} photometry in many passbands for galaxies at $0 < z < 2.5$ to derive composite SEDs for both main sequence and starburst galaxies. Finally, galaxies exhibiting an AGN signature are discussed in Section~\ref{SEC:AGN}, where we present a technique to identify obscured AGN candidates that would be unrecognized by previous methods.

We use below a cosmology with $H_0$=70~$\mathrm{km s^{-1} Mpc^{-1}}$, $\Omega_M=0.3$, $\Omega_\Lambda=0.7$ and we assume a Salpeter initial mass function (IMF, Salpeter 1955) when deriving SFRs and stellar masses.
   \begin{figure*}
   \centering
	\includegraphics*[width=3.2cm]{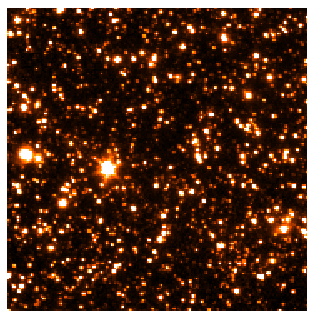}
	\includegraphics*[width=3.2cm]{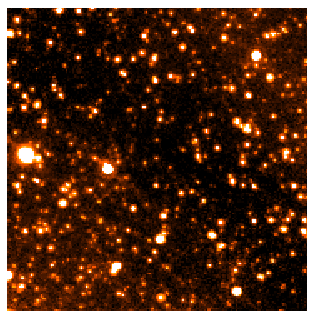}
	\includegraphics*[width=3.2cm]{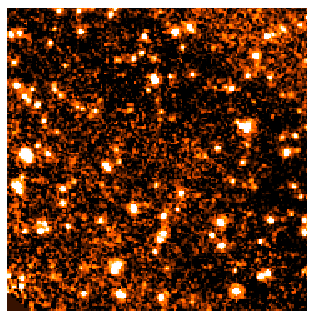}
	\includegraphics*[width=3.2cm]{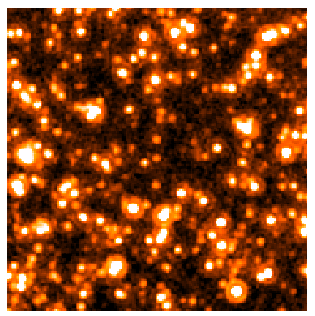}
	\includegraphics*[width=3.2cm]{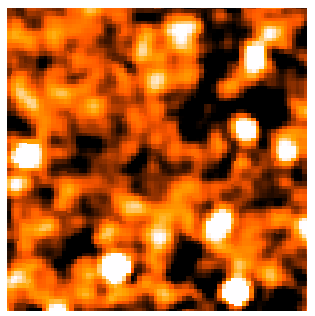}
	\includegraphics*[width=3.2cm]{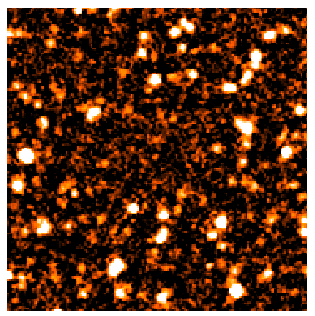}
	\includegraphics*[width=3.2cm]{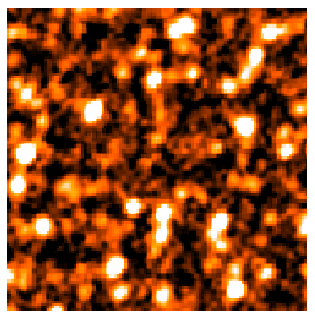}
	\includegraphics*[width=3.2cm]{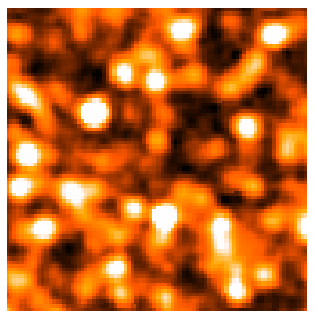}
	\includegraphics*[width=3.2cm]{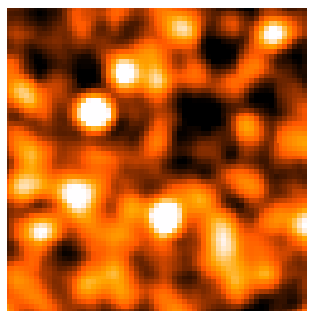}
	\includegraphics*[width=3.2cm]{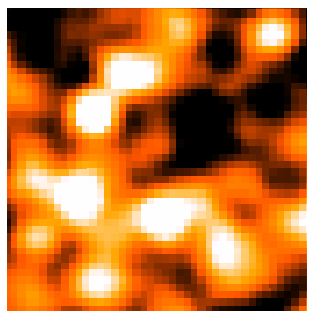}
      \caption{Postage stamp images of the same 5\arcmin$\times$5\arcmin\,region of the GOODS--north field ranging from 3.6\,$\mu$m (upper-left) to 500\,$\mu$m (bottom-right). {\it Upper panel:} five {\it Spitzer} images (85cm telescope diameter) obtained with IRAC at 3.6 and 8\,$\mu$m, the IRS peak-up array at 16\,$\mu$m and MIPS at 24 and 70\,$\mu$m (from upper-left to upper-right). {\it Bottom panel:} five {\it Herschel} images (3.5m telescope diameter) obtained with PACS at 100 and 160\,$\mu$m and SPIRE at 250, 350 and 500\,$\mu$m (from bottom-left to bottom-right).}
         \label{FIG:stamps}
   \end{figure*}
   \begin{figure}[h!]
   \centering
   	\includegraphics*[width=9.0cm]{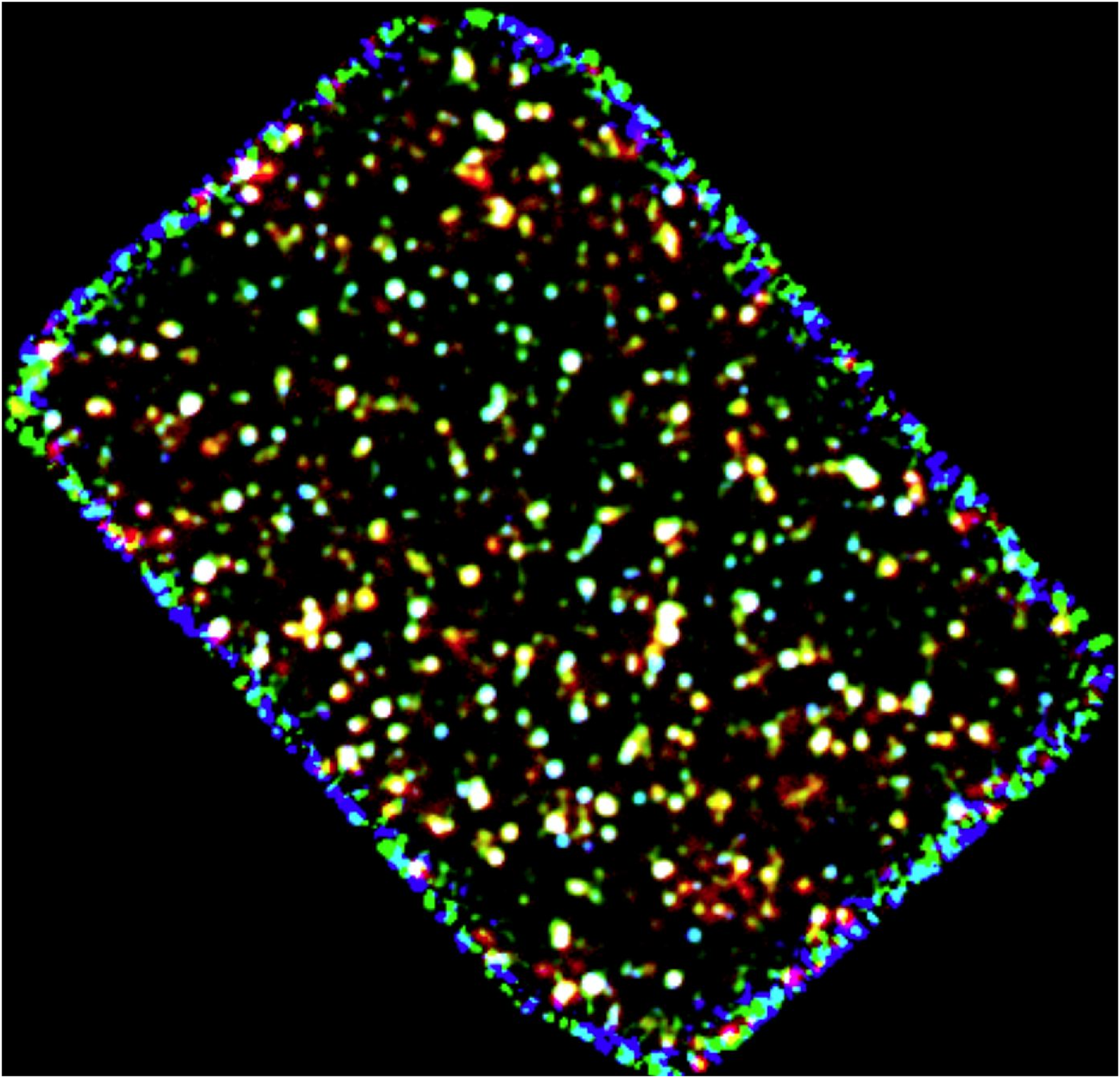}
   \caption{Composite three color image of the GOODS--north field (10$\arcmin$$\times$15$\arcmin$) at 100\,$\mu$m (blue), 160\,$\mu$m (green) and 250\,$\mu$m (red). North is up and east is left.}
         \label{FIG:GN}
   \end{figure}

   \begin{figure}[h!]
   \centering
	\includegraphics*[width=8.5cm]{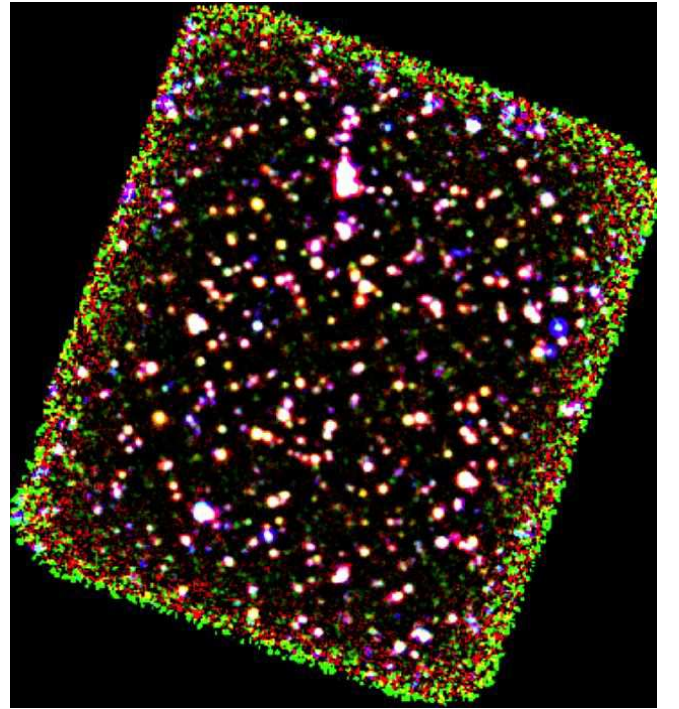}
      \caption{Composite three color image of the GOODS--south field (10$\arcmin$$\times$10$\arcmin$) at 24\,$\mu$m (blue), 100\,$\mu$m (green) and 160\,$\mu$m (red). North is up and east is left.}
         \label{FIG:GS}
   \end{figure}

\section{GOODS--{\it Herschel} data and catalogs}
\label{SEC:data}
\subsection{Observations}
The sample of high-redshift galaxies analyzed here consists of galaxies observed in the two Great Observatories Origins Deep Survey (GOODS) fields in the Northern and Southern hemispheres. Observations with the {\it Herschel} Space Observatory were obtained as part of the open time key program GOODS--{\it Herschel} (PI D.Elbaz), for a total time of 361.3 hours. PACS observations at 100 and 160\,$\mu$m cover the whole GOODS--north field of 10$\arcmin$$\times$16$\arcmin$ and part of GOODS--south, i.e., 10$\arcmin$$\times$10$\arcmin$ (but reaching the largest depths over $\sim$64 arcmin$^2$). When considering the total observing times of 124 hours in GOODS--N and 206.3 hours in GOODS--S (including 2.6 and 5 hours of overheads), the PACS GOODS--{\it Herschel} observations reach a total integration time per sky position of 2.4 hours in GOODS--N and of 15.1 hours in GOODS--S, i.e., 6.3 times longer. Due to the larger beam size and observing configuration, the SPIRE observations of GOODS--N cover a field of 900 arcmin$^2$, hence largely encompassing the central 10$\arcmin$$\times$16$\arcmin$, for a total observing time of 31.1 hours and an integration time per sky position of 16.8 hours.

Fig.~\ref{FIG:stamps} shows a montage of images (each 5$\arcmin$$\times$5$\arcmin$) from {\it Spitzer}--IRAC at 3.6\,$\mu$m to SPIRE at 500\,$\mu$m. This illustrates the impact of the increasing beam size as a function of wavelength:  the number of sources that are clearly visible at each wavelength increases when going from the longest to the shortest wavelengths (with the exception of the 70\,$\mu$m image, which comes from {\it Spitzer} and not {\it Herschel}). Composite three color images of GOODS--N at 100--160--250\,$\mu$m and GOODS--S at 24--100--160\,$\mu$m are shown in Figs.~\ref{FIG:GN} and~\ref{FIG:GS}.

\subsection{Catalogs}
\subsubsection{Source extraction}
Flux densities and their associated uncertainties were obtained from point source fitting using 24\,$\mu$m prior positions. For the largest passbands of SPIRE (i.e., 350 and 500\,$\mu$m), the 24\,$\mu$m priors are much too numerous and would lead to an over-deblending of the actual sources. Hence, we defined priors with the following procedure. For PACS-100\,$\mu$m, we used MIPS-24\,$\mu$m priors down to the 3$\sigma$ limit and imposing a minimum flux density of 20\,$\mu$Jy. For PACS-160\,$\mu$m and SPIRE-250\,$\mu$m, we restricted the 24\,$\mu$m priors to the 5$\sigma$ (30\,$\mu$Jy) limit (reducing the number of priors by about 35\,\%). For SPIRE-350 and 500\,$\mu$m, we kept only the 24\,$\mu$m priors for sources with a S/N ratio greater than 2 at 250\,$\mu$m. These criteria were chosen from Monte Carlo simulations (see Sect.~\ref{SEC:simulations}) to avoid using too many priors that would result in subdividing flux densities artificially, while producing residual maps (after PSF subtracting the sources brighter than the detection limit) with no obvious sources remaining.
   \begin{figure}
   \centering
	\includegraphics*[width=9cm]{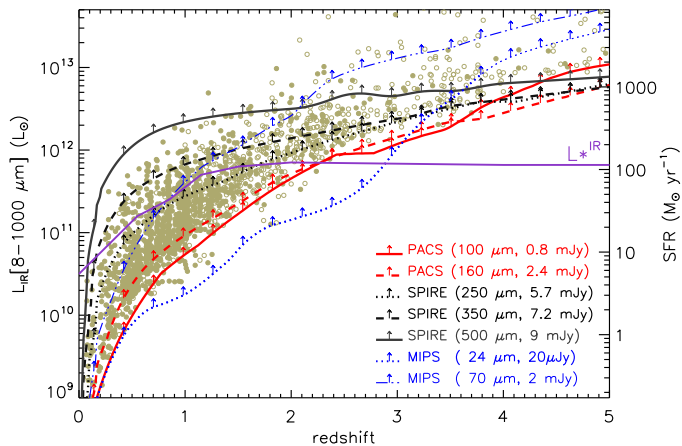}
      \caption{GOODS--{\it Herschel} detection limits (from Table~\ref{TAB:depths}) and total IR luminosities of the {\it Herschel} sources as a function of redshift (filled dots: spectroscopic, open dots: photometric). The right axis is the SFR derived from SFR[M$_{\sun}$yr$^{-1}$]=1.72$\times$10$^{-10}$$\times$$L_{\rm IR}^{\rm tot}$[L$_{\sun}$] (Kennicutt 1998a). These limits were computed assuming the local library of template SEDs of CE01. For comparison, the {\it Spitzer} MIPS-24\,$\mu$m detection limit is represented as well as the knee of the total IR luminosity function, as derived by Magnelli et al. (2009, 2010).}
         \label{FIG:limits}
   \end{figure}

Fig.~\ref{FIG:limits} can be used to infer the reliability of the {\it Spitzer}-MIPS 24\,$\mu$m images of the two GOODS fields for identifying potential blending issues with {\it Herschel}. It shows that the 20\,$\mu$Jy depth at 24\,$\mu$m (3$\sigma$) reaches fainter sources than any of the {\it Herschel} bands, down to the confusion level and up to a redshift of $z$$\sim$3. The technique used to estimate total IR luminosities for the {\it Herschel} sources is discussed in Sect.~\ref{SEC:method}. Hence the positions of 24\,$\mu$m sources can be used to perform robust PSF fitting source detection and flux measurements on the {\it Herschel} maps. We validated the efficiency of this technique by checking that no sources remain in the residual images after subtracting the detected sources or by independently extracting sources using a blind source extraction technique (Starfinder: Diolaiti et al. 2000). Although it is the case that most {\it Herschel} sources have a 24\,$\mu$m counterpart, a few 24\,$\mu$m-dropout galaxies were found, i.e., galaxies detected by {\it Herschel} but not at 24\,$\mu$m. This will be the subject of a companion paper (Magdis et al. 2011). But these objects represent less than 1\,\% of the {\it Herschel} sources.

\begin{table*}[htdp]
\caption{Depths of the GOODS--{\it Herschel} catalogs$^{\mathrm{(a)}}$ in the GOODS--north and south fields. }
\begin{center}
\begin{tabular}{crrrrrrrrrrrrrrr}
\hline 
\hline 
                  &             & \multicolumn{7}{c}{GOODS--north}                                                       & \multicolumn{7}{c}{GOODS--south} \\ 
$\lambda$ & FWHM & Depth & \multicolumn{4}{c}{All sources}        & \multicolumn{2}{c}{Clean $+$ $z$}& Depth &\multicolumn{4}{c}{All sources}& \multicolumn{2}{c}{Clean $+$ $z$}  \\
 ($\mu$m)&(\arcsec)&   &    Nb & N$_{z}^{spec}$ & \% $z_{spec}$ & \% $z_{all}$  & Nb   &  \% $z_{sp}$  &  & Nb & N$_{z}^{spec}$ & \% $z_{spec}$ & \% $z_{all}$ & Nb & \% $z_{sp}$  \\ 
(1)&(2)& (3)  &  (4) & (5) & (6) & (7)  & (8)   &  (9)  & (10) & (11) & (12) & (13) & (14) & (15) & (16) \\ \hline
  3.6 &  1.6$^{\mathrm{(a)}}$ & 17\,$\mu$Jy &  24476 &   3489 &   14 &   55 &   --   &  --  & 20\,$\mu$Jy &  13791 &   2933 &   21 &   99 &   --   &   --   \\
  4.5 &  1.6$^{\mathrm{(a)}}$ & 24\,$\mu$Jy &  19957 &   3445 &   17 &   63 &   --   &  --  & 35\,$\mu$Jy  &  10203 &   2808 &   28 &   99 &   --   &   --   \\
  5.8 &  1.7$^{\mathrm{(a)}}$ & 129\,$\mu$Jy &   8182 &   2648 &   32 &   76 &   --   &  --  & 137\,$\mu$Jy  &   4005 &   1943 &   49 &  100 &   --   &   --   \\
    8 &  2.0 & 150\,$\mu$Jy &   6020 &   2324 &   39 &   84 &   --   &  --  & 134\,$\mu$Jy &   3519 &   1760 &   50 &  100 &   --   &   --   \\
   16 &  4.0 & 32\,$\mu$Jy  &   1297 &    870 &   67 &   90 &   --   &  --  & 52\,$\mu$Jy   &    883 &    571 &   65 &   83 &   --   &   --   \\
   24 &  5.7 & 21\,$\mu$Jy  &   2575 &   1284 &   50 &   91 &   --   &  --  & 20\,$\mu$Jy   &   2063 &   1054 &   51 &   81 &   --   &   --   \\
   70 & 18.0 & 2.4 mJy     &    150 &    118 &   79 &   83 &     94 &   98 & 3.1 mJy     &    456 &     77 &   17 &  100 &     50 &   17 \\ 
  100 &  6.7 & 1.1 mJy     &   1095 &    693 &   63 &   93 &    959 &   72 & 0.8 mJy     &    531 &    375 &   71 &   91 &    485 &   77 \\
  160 & 11.0 & 2.7 mJy     &    781 &    517 &   66 &   94 &    355 &   77 & 2.4 mJy     &    296 &    216 &   73 &   90 &    170 &   84 \\
  250 & 18.1 & 5.7 mJy$^{\mathrm{(b)}}$   &    374 &    251 &   67 &   94 &    194 &   80 &   --    &   --   &   --   & --   &   -- &   --   & -- \\
  350 & 24.9 & 7.2 mJy$^{\mathrm{(b)}}$   &    173 &    114 &   66 &   94 &     91 &   78 &   --    &   --   &   --   & --   &   -- &   --   & -- \\
  500 & 36.6 & 9 mJy$^{\mathrm{(b)}}$    &     24 &     11 &   46 &   96 &     11 &   73 &   --    &   --   &   --   & --   &   -- &   --   & -- \\
 All  & --   &   --    &   1263 &    776 &   61 &   89 &    990 &   72 &   --    &    555 &    385 &   69 &   91 &    498 &   41 \\
 \hline
 \end{tabular}
\end{center}
\textbf{Notes.} \textit{Column definitions:} Col.(1) central wavelength of the passband; Col.(2) full width half maximum (FWHM) of the point spread function (PSF) in the passband. In the shortest bands, the FWHM is limited by the under-sampling of the PSF; Col.(3) depth of the image at that wavelength, i.e., flux density of the faintest sources of the catalog. The depths listed in the table correspond to the 3$\sigma$ limit. Due to local noise variations in the maps, some sources with slightly fainter flux densities may lie above this signal-to-noise threshold; Col.(4) total number of point sources above the 3-$\sigma$ limit. For the {\it Herschel} PACS and SPIRE passbands, we list the number of sources identified using a PSF-fitting based on {\it Spitzer}-MIPS 24\,$\mu$m prior positions, themselves resulting from {\it Spitzer}-IRAC 3.6\,$\mu$m priors; Col.(5) number of sources identified with an optical counterpart having a spectroscopic redshift; Col.(6) fraction of sources with an optical counterpart having a spectroscopic redshift; Col.(7) fraction of sources with an optical counterpart having either a spectroscopic or a photometric redshift; Col.(8) number of sources used in the present study, i.e., sources which are both ``clean'' (not polluted by bright neighbors as discussed in Sect.~\ref{SEC:clean}) and for which a redshift either photometric or spectroscopic was measured; Col.(9) fraction of the sources listed in column (8) for which a spectroscopic redshift was determined. Cols.(10) to (16) for GOODS--south are defined as Cols.(3) to (9) for GOODS--north.\\
$^{\mathrm{(a)}}$ The characteristics of the GOODS--{\it Herschel} PACS 100, 160\,$\mu$m and SPIRE 250, 350 and 500\,$\mu$m images and catalogs are given in the bottom part of the table. The SPIRE images extend over 30\arcmin$\times$30\arcmin\ but we restricted the present analysis, hence also the number of sources given in the table, to the same field as the one covered with the PACS bands and to the sources with a 24\,$\mu$m counterpart. The upper part of the table lists the characteristics of the {\it Spitzer} IRAC 3.6, 4.5, 5.8, 8\,$\mu$m, IRS peakup array 16\,$\mu$m and MIPS 24, 70\,$\mu$m images and catalogs. \\
$^{\mathrm{(b)}}$ The depth of the SPIRE catalogs given here applies only to the sub-sample of ``clean'' galaxies, located in isolated areas as probed by the density maps obtained from the shorter wavelengths starting at 24\,$\mu$m (see Sect.~\ref{SEC:clean}). This explains why they are much lower than the statistical confusion limits as listed by Nguyen et al. (2010) of 29, 31 and 34 mJy (5$\sigma_{\rm conf}$ confusion limits).
\label{TAB:depths}
\end{table*}

\subsubsection{Limiting depths of the catalogs and flux uncertainties}
\label{SEC:simulations}
The noise in the {\it Herschel} catalogs results from the combined effects of (1) instrumental effects $+$ photon noise, (2) background fluctuations due to the presence of sources below the detection threshold (photometric confusion noise, see Dole et al. 2004), (3) blending due to neighboring sources, above the detection threshold (source density contribution to the confusion noise). In both PACS and SPIRE images (except at 100\,$\mu$m in GOODS--N), the depths of the GOODS--{\it Herschel} observations are always limited by confusion, i.e., (2) and (3) are always stronger than (1). Global confusion limits have been determined for PACS (Berta et al. 2011) and SPIRE (Nguyen et al. 2010). However these global definitions assume no a priori knowledge on the local projected densities of sources, as if e.g., 500\,$\mu$m sources were distributed in an independent manner with respect to shorter wavelengths such as the 250 and 350\,$\mu$m ones, or even down to 24\,$\mu$m. Moreover, the flux limit associated to source blending, (3), is often artificially set to be the flux density above which 10\,\% of the sources are blended, even though statistical studies, such as the present one, could afford higher fractions as long as the photometric uncertainty is well controlled.
Actual observations instead demonstrate that shorter wavelengths do provide a good proxy for the density field of longer wavelengths (see Fig.~\ref{FIG:stamps}). Hence we define the 3$\sigma$ (or 5$\sigma$) sensitivity limits of the GOODS--{\it Herschel} catalogs as the flux densities above which at least 68\,\% of the sources can be extracted with a photometric accuracy better than 33\,\% on the basis of Monte Carlo simulations and we use the positions of 24\,$\mu$m sources as priors to extract sources from PSF fitting. Individual sources are attributed a ``clean'' flag depending on the underlying density field as defined in Sect.~\ref{SEC:clean}.

Flux uncertainties were derived in two independent ways. First \textit{(i)}, we added artificial sources into the real {\it Herschel} images and applied the source extraction procedure. This process was repeated a large number of times (Monte Carlo -- MC -- simulations). Second \textit{(ii)}, we measured the local noise level at the position of each source on the residual images produced after subtracting sources detected above the detection threshold. The first technique gives a noise level for a given flux density averaged over the whole map, while the second one provides a local noise estimate. In the MC simulations, we define the 3$\sigma$ (or 5$\sigma$) sensitivity limits in all bands as the flux densities above which a photometric accuracy better than 33\,\% (or 20\,\%) is achieved for at least 68\,\% of the sources in the faintest flux density bin (as in Magnelli et al. 2009, 2011). 

Technique \textit{(i)} provides a statistical noise level attributed for a given flux density which accounts for all three noise components but is independent of local variations of the noise. The histogram of the output - input flux densities of the MC simulations follows a Gaussian shape whose $rms$ was used to define the typical limiting depths of the {\it Herschel} catalogs listed in Col.(3) of Table~\ref{TAB:depths}. All GOODS--{\it Herschel} images (except the PACS--100\,$\mu$m image in GOODS--N) reach the 3$\sigma$ confusion level, i.e., the flux density for which the photometric accuracy is better than 33\,\% for at least 68\,\% of the sources is more than three times higher than the instrumental noise level. 

In technique \textit{(ii)}, only the noise components (1) and (2) are taken into account, since the objects participating in the third component (source blending) have been subtracted to produce the residual images. However, imperfect subtraction of sources, due to local blending, may inflate the local residuals in the maps after source subtraction. In the PACS images and catalogs, both techniques result in very similar noise levels. A statistical limiting depth was computed by convolving the residual images with the PACS beam at each wavelength and measuring the $rms$ of the distribution of individual pixels. This method resulted in the same depths as in technique \textit{(i)} and listed in Col.(3) of Table~\ref{TAB:depths}. Instead, for the SPIRE data, local noise estimates in the residual maps were found to be systematically lower than those measured with technique \textit{(i)}. On average, sources with a SPIRE flux density corresponding to the detection threshold of 3$\sigma$ in the MC simulations are found to present a local signal-to-noise ratio of 5 in the residual maps. For SPIRE sources, this implies that we consider only sources above the 5$\sigma$ limit in the residual maps, to be consistent with the 3$\sigma$ limit resulting from the MC simulations.

\begin{figure*}
   \centering
	\includegraphics*[width=18cm]{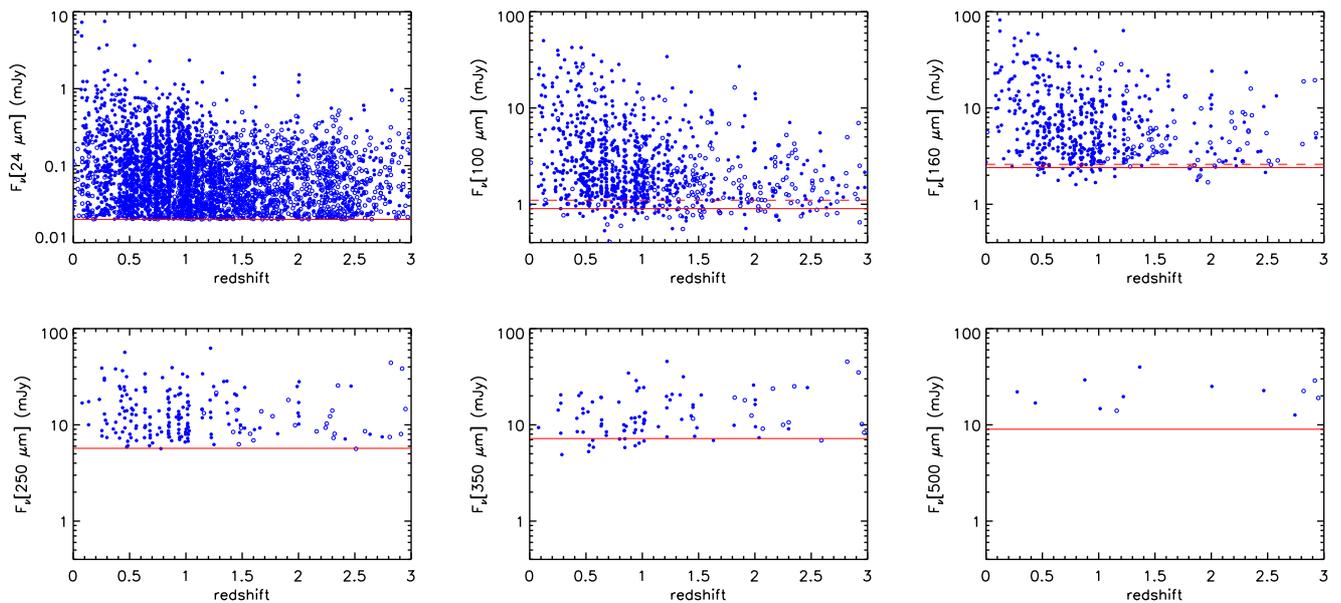}
      \caption{Distribution of the 'clean' GOODS--{\it Herschel} and {\it Spitzer}-MIPS 24\,$\mu$m flux densities as a function of redshift (spectroscopic--filled and photometric--open symbols). Non ``clean'' galaxies (see Sect.~\ref{SEC:clean}) are not represented. The average 3--$\sigma$ depths of the surveys listed in Table~\ref{TAB:depths} are shown with horizontal lines (the dashed lines at 100 and 160\,$\mu$m represent the shallower GOODS--N depths). Due to local noise variations, some 3-$\sigma$ sources can be detected below these typical limiting flux densities. Sources above $z$$\sim$3 are not represented due to redshift uncertainties.
}
         \label{FIG:fluxes}
   \end{figure*}

Due to local noise variations in the maps, there can be small numbers of sources with flux densities slightly fainter than the nominal detection limits, which explains the presence of sources below the horizontal lines in Fig.~\ref{FIG:fluxes}.

\subsection{Local confusion limit and ``clean index''}
\label{SEC:clean}
The main source of uncertainty, in the SPIRE images in particular, comes from the high source density relative to the beam size, i.e., the so-called confusion limit (see Condon 1974). Assuming that this limit applies equally at all positions of the sky, Nguyen et al. (2010) estimated that the floor below which SPIRE sources may not be extracted is $\sim$ 30 mJy, corresponding to 5$\sigma_{\rm conf}$ confusion limits of 29, 31 and 34 mJy/beam for beams of 18.1\arcsec, 24.9\arcsec and 36.6\arcsec\ FWHM at 250, 350 and 500\,$\mu$m respectively.

However, this ``global confusion limit'' is defined assuming no a priori knowledge on the projected density map of the underlying galaxy population. If one instead assumes that shorter wavelengths, at a higher spatial resolution, can be used to define the local galaxy density at a given galaxy position, then a ``local confusion limit'' can be defined. In practice, this means that not all SPIRE sources are located at a place where several bright PACS or MIPS--24\,$\mu$m fall in the SPIRE beam. Following this recipe, Hwang et al. (2010a) defined a ``clean index'' that was attributed to all individual {\it Herschel} detections under the following conditions: a 500\,$\mu$m source is flagged as ``clean'' if its 24\,$\mu$m prior has at most one bright neighbor in the {\it Spitzer}-MIPS 24\,$\mu$m band (where ``bright'' means an F$_{24}$$>$50\% of the central 24\,$\mu$m source) within 20$\arcsec$ (1.1$\times$FWHM of {\it Herschel} at 250\,$\mu$m) and no bright neighbor in each one of the shorter {\it Herschel} passbands, i.e., at 100, 160, 250, 350 and 500\,$\mu$m within 1.1$\times$FWHM of {\it Herschel} in these passbands (see Table~\ref{TAB:depths}). As a result, we only kept 11 clean sources at 500\,$\mu$m for which we consider that the photometry is reliable. The criterion becomes less critical for the shorter bands, since we only consider the presence of bright neighbors at shorter wavelengths. As a result, the number of 350\,$\mu$m detections is an order of magnitude larger than at 500\,$\mu$m. This ``local confusion limit'' was empirically defined after visually inspecting the data for all individual sources but a more detailed investigation of this quality flag using simulations of the actual GOODS sources both spatially and in redshift confirms its robustness (Leiton et al. 2011, in prep.). For galaxies for which this ``clean index'' condition is not met in some bands, unphysical jumps in the IR SED are observed. This may lead to wrong estimates of the dust temperature for example, systematically shifting it to colder values, since source blending affects preferentially the longest wavelengths.

With the sensitivity limits of GOODS--{\it Herschel}, Fig.~\ref{FIG:limits} shows that below a redshift of $z$$\sim$3 the shortest wavelengths are always deeper than the longest ones, hence one can take advantage of these higher resolution images to better constrain the confusion limit at local, instead of global, scales. Moreover, we note that the fluxes in e.g., the SPIRE bands are not independent of those measured in the 24\,$\mu$m and 100\,$\mu$m passbands. They even follow a tight correlation (see Elbaz et al. 2010 and the present analysis), again in the redshift range of interest here, i.e., $z$$\sim$0--2.5. Hence it is possible to map the density of IR sources and to flag sources in relatively isolated areas, with respect to similarly bright or brighter IR sources. If the ``clean index'' did not reject efficiently problematic measurements, this would result in an increase of the dispersion in the figures presented in this paper. Since we will show that these dispersions are quite small already, if this effect was corrected, it would only reinforce our results. Typically, half of the {\it Herschel} sources detected at $\lambda$$>$160\,$\mu$m survive this criterion (see Table~\ref{TAB:depths}). 

\section{High- and low-redshift galaxy samples}
\label{SEC:highlowz}
\subsection{The GOODS--{\it Herschel} galaxy sample}
\label{SEC:GH}

Both GOODS fields have been subject to intensive follow-up campaigns, resulting in a spectroscopic redshift completeness greater than 70\,\% for the {\it Herschel} sources (Table~\ref{TAB:depths}). We use a compilation of 3630 and 3018 spectroscopic redshifts for GOODS--N (Cohen et al. 2000, Wirth et al. 2004, Barger, Cowie \& Wang 2008, and Stern et al. in prep.) and GOODS--S (Le F\`evre et al. 2004, Mignoli et al. 2005, Vanzella et al. 2008, Popesso et al. 2009, Balestra et al. 2010, Silverman et al. 2010, and Xia et al. 2010) respectively. Photometric redshifts and stellar masses are computed in both fields from U-band to IRAC 4.5\,$\mu$m photometric data using Z-PEG (Le Borgne  \& Rocca-Volmerange 2002). The templates used for both photometric redshifts and stellar mass estimates are determined from PEGASE.2 (Fioc \& Rocca-Volmerange 1999) are were produced using nine scenarios for the star formation history (see Le Borgne  \& Rocca-Volmerange 2002) with various star-formation efficiencies and infall timescales, ranging from a pure starburst to an almost continuous star-formation rate, aged between 1 Myr and 13 Gyr (200 ages). There is no constraint on the formation redshift. The templates are required to be younger than the age of the Universe at any redshift.

The redshift distributions of the sources individually detected in each of the {\it Herschel} bands, as well as at 24\,$\mu$m with {\it Spitzer}, are presented in Fig.~\ref{FIG:fluxes} for both fields. This illustrates the relative power of these bands to detect sources as a function of redshift. While the 500\,$\mu$m band samples sources at all redshifts from $z$=0 to 4, it only provides a handful of objects: 24 galaxies in total within the 10\arcmin$\times$15\arcmin\ size of the GOODS--N field, with only 11 flagged as clean, 73\,\% of which have a spectroscopic redshift determination. In comparison, more than a thousand sources are detected in the 100\,$\mu$m band, the vast majority being flagged as clean and 72\,\% having a spectroscopic redshift.

Table~\ref{TAB:depths} also lists the characteristics of the other IR catalogs that we use in the present study.  The GOODS {\it Spitzer} IRAC catalogs were created using SExtractor (Bertin \& Arnouts 1996), detecting sources in a weighted combination of the 3.6 and 4.5\,$\mu$m images, with matched-aperture photometry in the four IRAC bands, using appropriate aperture corrections to total flux. The {\it Spitzer} 24\,$\mu$m and 70\,$\mu$m catalogs (Magnelli et al. 2011) use data from the {\it Spitzer} GOODS and FIDEL programs (PI: M.\ Dickinson).  Sources detected in the IRAC images are used as priors to extract the 24\,$\mu$m fluxes, and then in turn a subset of those 24\,$\mu$m sources are used as priors to extract fluxes at 70\,$\mu$m.  The 16\,$\mu$m data comes from {\it Spitzer} IRS peak-up array imaging (Teplitz et al. 2011);  here again, 16\,$\mu$m catalog fluxes are extracted using IRAC priors.  In this study, we make particular use of the {\it Spitzer} data to quantify the redshift dependence of the IR SEDs while minimizing mid-infrared k-corrections by measuring the rest-frame 8\,$\mu$m emission of galaxies at $z$$\sim$0, 1 and 2 from their observed fluxes in the IRAC-8\,$\mu$m, IRS-16\,$\mu$m and MIPS-24\,$\mu$m passbands.  Table~\ref{TAB:depths} also gives the spectroscopic (\%$z_{spec}$), and photometric $+$ spectroscopic (\%$z_{all}$) completeness of the IR catalogs from 3.6 to 500\,$\mu$m within the fiducial GOODS area.   As noted previously, the SPIRE images of GOODS--N cover a wider field, but here we do not count the sources detected outside the regular GOODS area.

Known AGN were excluded from the sample and will be discussed separately in Section~\ref{SEC:AGN}. X-ray/optical AGN were identified from one of the following criteria: $L_{\rm X}$[0.5-8.0 keV] $>$ 3$\times$10$^{42}$ ergs s$^{-1}$, a hardness ratio (ratio of the counts in the 2-8 keV to 0.5-2 keV passbands) higher than 0.8, N$_{\rm H}\geq$10$^{22}$ cm$^{-2}$, or broad/high-ionization AGN emission lines (Bauer et al. 2004). We also excluded power-law AGN, i.e., galaxies showing a rising continuum emission in the IRAC bands due to hot dust radiation (see definition in Sect.~\ref{SEC:AGN}).

\subsection{Total infrared luminosities}
\label{SEC:method}
Total IR luminosities, $L_{\rm IR}^{\rm Herschel}$, for GOODS--{\it Herschel} galaxies were determined by allowing the normalization of the CE01 template SEDs to vary and choosing the one that minimizes the $\chi^2$ fit to the {\it Herschel} measured flux densities. At the highest redshifts considered in the present analysis ($z \approx 2.5$), the {\it Herschel} 100\,$\mu$m passband samples rest-frame mid-IR wavelengths. Hence, to avoid mixing galaxies with and without direct far-IR detections, we require at least one photometric measurement at wavelengths longer than 30\,$\mu$m in the rest-frame.  This excludes a few high redshift galaxies detected only at 100\,$\mu$m. Total IR luminosities, $L_{\rm IR}^{\rm Herschel}$, were integrated from 8 to 1000\,$\mu$m on the best-fitting normalized CE01 SED. When only one or two {\it Herschel} measurements are available above 30\,$\mu$m, the degeneracy of the fit being large, we use the standard CE01 technique, i.e., we use the SED with the closest luminosity from the CE01 library without allowing any renormalization.

In order to quantify the impact of the choice of a given set of SEDs to fit the {\it Herschel} measurements and determine $L_{\rm IR}^{\rm Herschel}$, we have repeated the same exercise with another SED library from Dale \& Helou (2002, DH02). The ratio of the $L_{\rm IR}^{\rm Herschel}$ values derived with one or the other family of SEDs has a median of 1 and a dispersion of 12\,\%--rms. The uncertainty in the determination of $L_{\rm IR}^{\rm Herschel}$ is therefore dominated by the actual error bars on the {\it Herschel} flux measurements rather than by the choice of the SED library. In order to account for the latter source of uncertainty, we have generated a series of 100 realizations of the {\it Herschel} flux measurements  assuming a Gaussian distribution within their error bars and determined 100 values of $L_{\rm IR}^{\rm Herschel}$ by fitting those realizations independently. The final $L_{\rm IR}^{\rm Herschel}$ associated to a given galaxy is the median of the 100 Monte Carlo estimates and its error bar is the rms around the median. This procedure was repeated for each individual galaxy.

Since we will compare the distant GOODS--{\it Herschel} galaxies to a reference sample of local galaxies for which $L_{\rm IR}^{\rm tot}$ is estimated from {\it IRAS} measurements alone, as a consistency check we computed the total IR luminosity that we would obtain for the GOODS--{\it Herschel} galaxies if we had used Eq.~\ref{EQ:sanders} (taken from Sanders \& Mirabel 1996),
\begin{equation}
\begin{array}{l}
L_{\rm IR}/L_{\odot} = 4 \pi D_{lum}^2[{\rm m}] \left[1.8\times10^{-14} ( FIR [{\rm W m}^{-2}] ) \right] / 3.826\times10^{26} \\
{\rm where~} FIR = 13.48 F_{12\mu m} + 5.16 F_{25\mu m} + 2.58 F_{60\mu m} + F_{100\mu m}~,
\end{array}
\label{EQ:sanders}
\end{equation}
as a proxy for the derivation of the 8 -- 1000\,$\mu$m luminosity, instead of the actual integral over the IR SED. The {\it IRAS} flux densities $F_{12\mu m}$, $F_{25\mu m}$, $F_{60\mu m}$ and $F_{100\mu m}$ in Eq.~\ref{EQ:sanders} are in Jy. Both techniques give equivalent total IR luminosities within 5\,\%, hence again the dominant cause of discrepancy in the comparison is related to flux uncertainties.

\begin{table}[t!]
\caption{Number of galaxies and total IR luminosity range of the local galaxy samples.}
\begin{center}
\begin{tabular}{ccccrrrr}
\hline
\hline
Local             & (2)    &  (3)                        &  (4)      &  \multicolumn{4}{c}{$log_{10}$($L_{\rm IR}$/L$_{\odot}$)} \\ 
samples & Nb & R$_{\rm radio}$ & FEE & $<$10 & 10--11 & 11 -- 12 & $\geq$12 \\ 
\hline
{\it ISO}     & 150 & 11 & 0 & 36 & 56 & 14 & 45 \\
{\it AKARI} & 287 & 47 & 0  & 63 & 164 & 55 & 9 \\ 
{\it Spitzer} & 211 & 58 & 211 & 0 & 44 & 154 & 13 \\ 
\hline
Total   &  648  &  116  & 211 &  99  & 264  & 223 & 67 \\
\hline
\end{tabular}
\end{center}
\textbf{Notes}. Column (2) total number of objects for each local sample; Col.(3) number of galaxies with a radio continuum (1.4 GHz) size estimate; Col.(4) number of galaxies for which a fraction of extended emission (FEE) was measured (D\'iaz-Santos et al. 2010), i.e., fraction of the mid-IR continuum at 13.2\,$\mu$m more extended than the \textit{Spitzer}/IRS resolution of $3.6\,\arcsec$ (see Sect.~\ref{SEC:compactness}).
\label{TAB:local}
\end{table}%

\subsection{Local galaxy reference sample}
\label{SEC:local}
The local galaxy reference sample that we use in this paper consists of galaxies detected with the {\it Infrared Space Observatory} ({\it ISO}), {\it AKARI}, and {\it Spitzer}. Their rest-frame 8\,$\mu$m luminosities and total IR luminosities are compared to those of the GOODS--{\it Herschel} galaxies. Galaxies with direct IRAC--8\,$\mu$m measurements from {\it Spitzer} are supplemented with galaxies with {\it ISO} 6.75$\mu$m and {\it AKARI} 9$\mu$m photometry, for which pseudo-IRAC 8\,$\mu$m luminosities, $L_8$, were computed using the IR SED of M82 (F\"orster Schreiber et al. 2001, Elbaz et al. 2002). The {\it ISO} and {\it AKARI} samples span a wide range of relatively low luminosity galaxies, together with a sample of ULIRGs, while the {\it Spitzer} sample contains a quite complete sample of local LIRGs (see Table~\ref{TAB:local}). 

\subsubsection{Local {\it ISO} galaxy sample}
\label{SEC:ISOlocal}
The mid-IR luminosities of this sample of 150 galaxies described in CE01 and Elbaz et al. (2002) were obtained from measurements taken with {\it ISO}. The sample includes 110 galaxies closer than 300 Mpc and spanning a wide range of mid-IR luminosities estimated from ISOCAM-LW2 (5--8.5\,$\mu$m, centered at 6.75\,$\mu$m) and 41 ULIRGs, at distances 80 to 900 Mpc, with mid-IR luminosities determined with the PHOT-S spectrograph of ISOPHOT (Rigopoulou et al. 1999). We refer to CE01 for a discussion of the conversion of the PHOT-S spectra into broadband luminosities equivalent to the LW2 filter. Pseudo-IRAC 8\,$\mu$m luminosities, $L_8$, were estimated by first convolving the ISOCAM CVF spectrum of M82 (F\"orster Schreiber et al. 2001, Elbaz et al. 2002) to the ISOCAM-LW2 and IRAC-8\,$\mu$m bandpasses and then normalizing the resulting luminosities to the observed luminosity for each of the 150 galaxies, in order to derive their $L_8$. Since both filters are wide and largely overlapping, the conversion depends very little on the exact shape of the spectrum used for the conversion and we checked that indeed using the CE01 SEDs (for example) instead of that for M82 would make negligible differences with respect to the actual dispersion of galaxies in the $L_{\rm IR}^{\rm tot}$ -- $L_8$ diagram. Total IR luminosities, $L_{\rm IR}^{\rm tot}$, were derived from the four {\it IRAS} band measurements using Eq.~\ref{EQ:sanders}.

\subsubsection{Local {\it AKARI} galaxy sample}
\label{SEC:AKARIlocal}
Galaxies with mid-infrared (9\,$\mu$m) measurements from {\it AKARI} were cross-matched with the {\it IRAS} Faint Sources Catalog ver. 2 (FSC-2; Moshir, Kopman \& Conrow 1992) and with spectroscopic redshifts from the Sloan Digital Sky Survey Data Release 7 (SDSS DR7; Abazajian et al. 2009) supplemented by a photometric sample of galaxies with redshifts available in the literature (Hwang et al. 2010b). For both {\it IRAS} and {\it AKARI}, we consider only the sources with reliable flux densities\footnote[1]{Flux quality flags are either ``high'' or ``moderate'' for {\it IRAS} sources and ``high'' for {\it AKARI} sources}. A total of 287 galaxies have 9\,$\mu$m flux densities from the {\it AKARI}/Infrared Camera (IRC, Onaka et al. 2007) Point Source Catalog (PSC  ver. 1.0, Ishihara et al. 2010) reaching a detection limit of 50 mJy (5$\sigma$) with a uniform distribution over the whole sky and closer than $\sim$450 Mpc ($z$$<$0.1). As in Sect.~\ref{SEC:ISOlocal}, pseudo-IRAC 8\,$\mu$m luminosities, $L_8$, were computed by convolving the ISOCAM CVF spectrum of M82 with the {\it AKARI}-IRC 9\,$\mu$m bandpass to estimate the conversion factor between the IRC--9\,$\mu$m and IRAC--8\,$\mu$m luminosities assuming the same IR SED for all galaxies.  The effective wavelength of the {\it AKARI} 9\,$\mu$m passband is 8.6\,$\mu$m (Ishihara et al. 2010), not far from that of the IRAC-8\,$\mu$m filter (7.9\,$\mu$m, Fazio et al. 2004). 

Total IR luminosities were computed from the four {\it IRAS} bands using Eq.~\ref{EQ:sanders}. The IRC--9\,$\mu$m measurements were not used in the computation of $L_{\rm IR}^{\rm tot}$. Far-IR measurements were supplemented with the {\it AKARI}/Far-Infrared Surveyor (FIS; Kawada et al. 2007) all-sky survey Bright Source Catalogue (BSC ver. 1.0\footnote[2]{http://www.ir.isas.jaxa.jp/AKARI/Observation/PSC/Public/RN/AKARI-FIS\_BSC\_V1\_RN.pdf})  that contains 427 071 sources, with measured flux densities at 65, 90, 140 and 160\,$\mu$m. We used the supplementary far-IR measurements for 16\,\% of the sample for which there is no 12\,$\mu$m nor 25\,$\mu$m reliable measurement from IRAS. We checked the consistency of these IR estimates from {\it AKARI} with those obtained from {\it IRAS} alone and found that {\it AKARI} luminosities were systematically lower by 10\,\%. We corrected those 16\,\% galaxies by this factor.

\subsubsection{Local {\it Spitzer} galaxy sample}
\label{SEC:SPITZERlocal}
A sample of 202 {\it IRAS} sources, consisting of 291 individual galaxies (some blended at {\it IRAS} resolution), were observed with the IR spectrograph (IRS) on-board \textit{Spitzer} as part of the Great Observatories All-sky LIRG Survey project (GOALS; Armus et al. 2009). The sources were drawn from the \textit{IRAS} Revised Bright Galaxy Sample (RBGS; Sanders et al. 2003) and represent a complete sub-sample of systems ($z\,<\,0.088$) with IR luminosities originally defined to be in the range of $10^{11}$\,L$_{\odot}$\,$\leq$\,$L_{\rm IR}$\,$\leq$\,10$^{13}$\,L$_{\odot}$. The GOALS sample includes 200 LIRGs and 22 ULIRGs. The total IR luminosities of the systems were derived using their \textit{IRAS} measurements and Eq. 1 (see Armus et al. 2009 for further details on this calculation).

Using the spectral images obtained with the short-low module of IRS, D\'iaz-Santos et al. (2010) measured the spatial extent of the light radiated in the mid-IR continuum at 13.2\,$\mu$m of a sub-sample of 211 individual galaxies (closer than 350 Mpc) for which data were available at the time of publication and sources could be detected. We use these size estimates in our analysis regarding the link between star formation compactness and the $IR8$ ratio. This fraction of extended emission (FEE) is directly related to the spatial distribution of the star formation regions and presents the advantage of being measured in a wavelength range not affected by the presence/absence of emission lines such as PAHs. For the multiple systems unresolved by IRAS, D\'iaz-Santos et al. (2010) distributed the total IR luminosity between galaxies proportionally to their \textit{Spitzer}/MIPS--24\,$\mu$m fluxes. Due to this redistribution of the luminosity, there are now 44 galaxies with IR luminosities less than 10$^{11}$ L$_{\odot}$ in our sample. Added to these normal star-forming galaxies, the present sample finally includes 154 LIRGs and 13 ULIRGs (with 10$^{12}$$\leq$$L_{\rm IR}^{\rm tot}$/L$_{\odot}$$<$4$\times$10$^{12}$). IRAC-8\,$\mu$m luminosities for these galaxies are from Mazzarella et al. (in prep). Stellar masses were derived by cross-matching the GOALS sample with 2MASS and converting the Ks luminosities into stellar masses (excluding remnants) using a using a mass-to-light ratio $M_*$/$L_{Ks}$=0.7 M$_{\odot}$/L$_{K,\odot}$ computed from PEGASE 2 (Fioc \& Rocca-Volmerange 1997, 1999) assuming a Salpeter IMF and an age of 12 Gyr.
\section{Universality of $IR8$ (=$L_{\rm IR}$/$L_{8}$): an IR main sequence}
\label{SEC:IR8}
   \begin{figure*}[ht!]
   \centering
\includegraphics*[width=9cm]{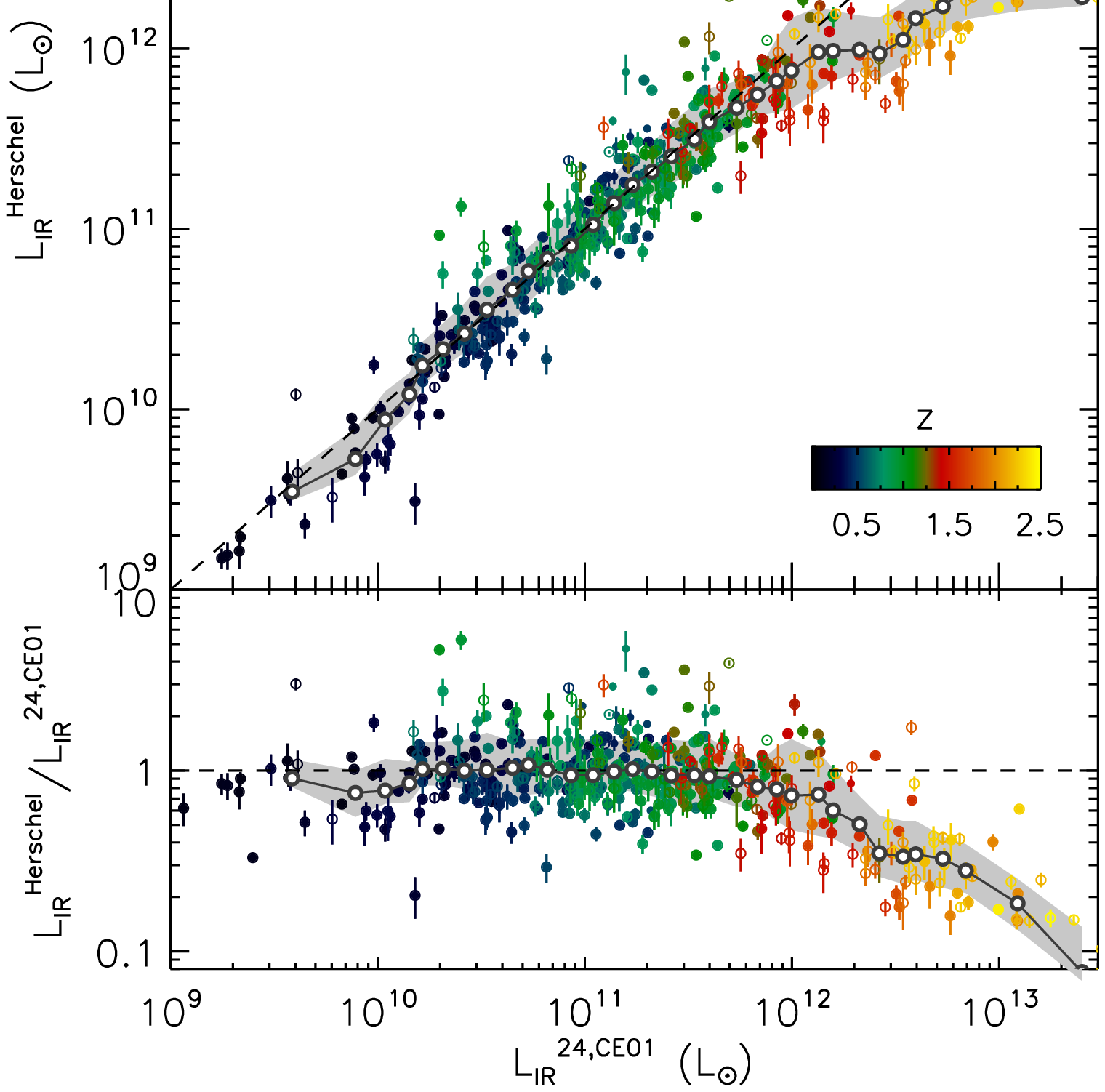}
\includegraphics*[width=9cm]{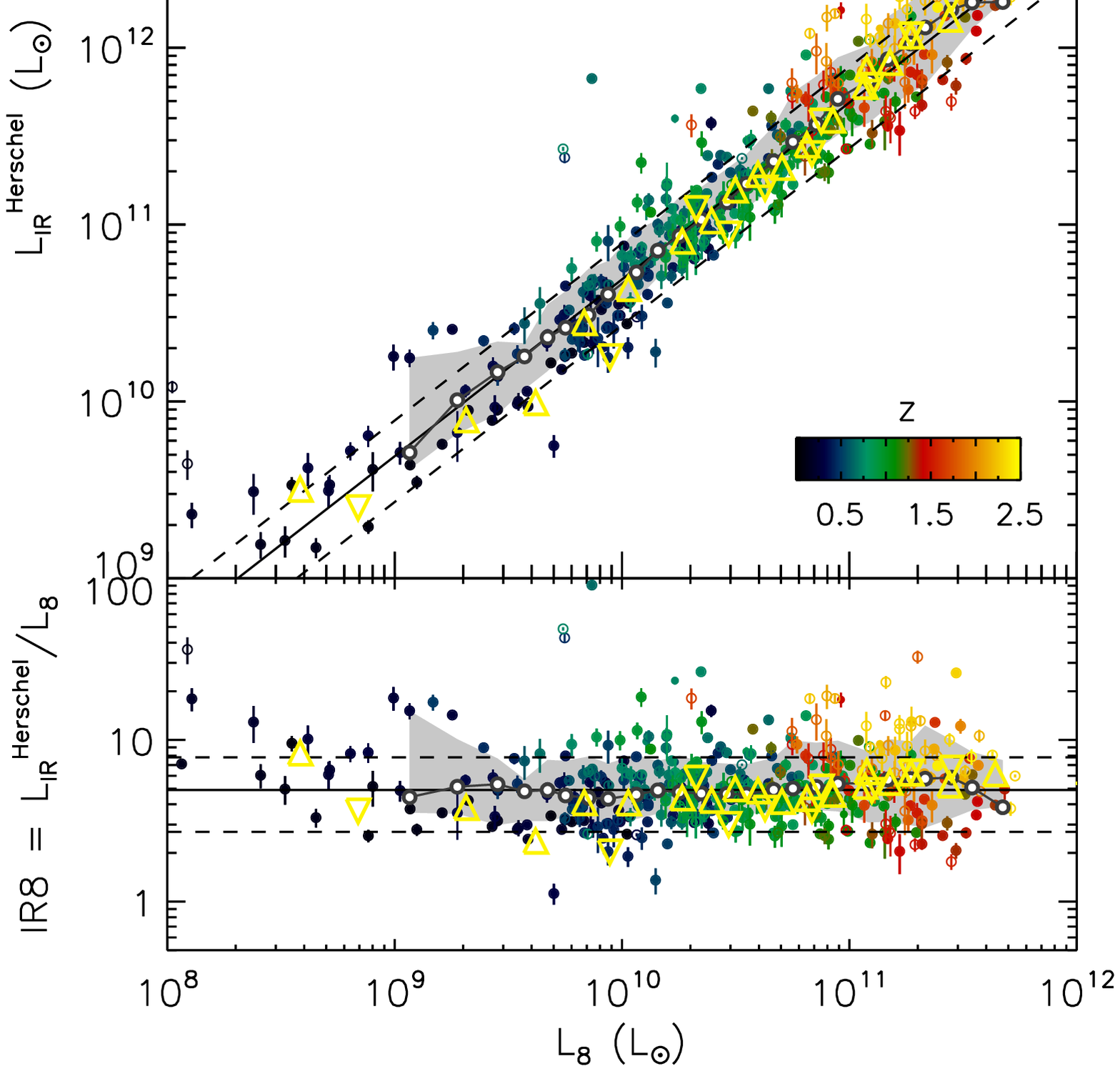}
      \caption{
\textbf{\textit{Left:}} Comparison of $L_{\rm IR}^{\rm tot}$ (8--1000\,$\mu$m) as directly measured from {\it Herschel} ($L_{\rm IR}^{\rm {\it Herschel}}$) with the value extrapolated from 24\,$\mu$m ($L_{\rm IR}^{\rm 24,CE01}$) using the CE01 technique. Only ``clean'' galaxies are represented (as defined in Sect.~\ref{SEC:GH}). Galaxies with spectroscopic and photometric redshifts (from both GOODS--north and south) are marked with filled and open symbols respectively. Colors range from black ($z$$\sim$0) to orange ($z$$\sim$2.5), passing through green ($z$$\sim$1) and red ($z$$\sim$2). The wavelength range sampled by the MIPS-24\,$\mu$m passband is shown in orange at the top of the figure where it is compared to the redshifted SED of M82. The dashed line in the left-hand side panels is the one-to-one correlation. The sliding median and 16th and 84th percentiles of the distribution are shown with white dots connected with a solid line and grey zone respectively. The {\bf bottom panel} shows the ratio of the actual over extrapolated total IR luminosity.
\textbf{\textit{Right:}} Comparison of $L_{\rm IR}^{\rm {\it Herschel}}$ with $L_8$ (rest-frame 8\,$\mu$m broadband) for ``clean'' galaxies. The observed bandpasses used to estimate $L_8$ are illustrated in the top of the figure and compared to the redshifted SED of M82. The sliding scale of the median and 68\,\% dispersion around it is shown with a grey zone which is fitted by the solid and dashed lines: $IR8$=4.9 [-2.2,+2.9]. Stacked measurements combined with detections weighted by number of objects per luminosity bin are represented by large yellow open triangles (GOODS--south: upside down, GOODS--north: upward). The {\bf bottom panel} shows the $IR8$ (=$L_{\rm IR}^{\rm {\it Herschel}}$/$L_8$) ratio which is found to remain constant with luminosity and redshift.
}
         \label{FIG:IR8}
   \end{figure*}

\subsection{The mid-infrared excess problem}
\label{SEC:MIRexcess}
Before the launch of {\it Herschel}, the derivation of $L_{\rm IR}^{\rm tot}$, hence also of the SFR, of distant galaxies had to rely on extrapolations from either mid-IR or sub-mm photometry. While there are many reasons why extrapolations from the mid-IR could be wrong (evolution in metallicity, geometry of star formation regions, evolution of the relative contributions of broad emission lines and continuum), it was instead found that they work relatively well up to $z$$\sim$1.5. Using shallower {\it Herschel} data than the present study, Elbaz et al. (2010) compared $L_{\rm IR}^{\rm tot}$, estimated from {\it Herschel} PACS and SPIRE, to $L_{\rm IR}^{24}$ -- the total IR luminosity extrapolated from the observed {\it Spitzer} mid-IR 24\,$\mu$m flux density -- and found that they agreed within a dispersion of only 0.15 dex. The CE01 technique used to extrapolate $L_{\rm IR}^{24}$ attributes a single IR SED per total IR luminosity. Hence a given 24\,$\mu$m flux density is attributed the $L_{\rm IR}^{\rm tot}$ of the SED that would yield the same flux 24\,$\mu$m flux density at that redshift.

Stacking {\it Spitzer} MIPS-70\,$\mu$m measurements at prior positions defined by 24\,$\mu$m sources in specific redshift intervals, Magnelli et al. (2009) found that the rest-frame 24\,$\mu$m/(1+$z$) and 70\,$\mu$m/(1+$z$) luminosities were perfectly consistent with those derived using the CE01 technique for galaxies at $z$$\leq$1.3.  Although the 70\,$\mu$m passband probes the mid-IR regime for redshifts $z$$\gtrsim$0.8, it presents the advantage of sampling the continuum IR emission of distant galaxies without being affected by the potentially uncertain contribution of PAHs, contrary to that at 24\,$\mu$m. At $z$$\geq$1.5 however, extrapolations from 24\,$\mu$m measurements using local SED templates were found to systematically overestimate the 70\,$\mu$m measurements (Magnelli et al. 2011).  This mid-IR excess, first identified by comparing $L_{\rm IR}^{24}$ with radio, MIPS-70\,$\mu$m and 160\,$\mu$m stacking (Daddi et al. 2007a, Papovich et al. 2007, Magnelli et al. 2011) has recently been confirmed with {\it Herschel} by Nordon et al. (2010) on a small sample of $z$$\sim$2 galaxies detected with PACS  and by stacking PACS images on 24\,$\mu$m priors (Elbaz et al. 2010, Nordon et al. 2010). 

Here, thanks to the unique depth of the GOODS--{\it Herschel} images, we are able to compare $L_{\rm IR}^{24}$ to $L_{\rm IR}^{\rm tot}$ for a much larger number of galaxies than in Elbaz et al. (2010) and, more importantly, for direct detections at $z$$>$1.5. In the left-hand part of Fig.~\ref{FIG:IR8}, we show that the mid-IR excess problem is not artificially produced by imperfections that could result from the indirect stacking measurements, but instead takes place for individually detected galaxies at $z$$>$1.5 and at high 24\,$\mu$m flux densities, corresponding to $L_{\rm IR}^{24}$$>$10$^{12}$ L$_{\odot}$. Although known AGN were not included in the sample, unknown AGN may still remain. Indeed it has been proposed that the mid-IR excess problem could be due to the presence of unidentified AGN affected by strong extinction, possibly Compton thick (Daddi et al. 2007b, see also Papovich et al. 2007). At these high redshifts, the re-processed radiation of a buried AGN may dominate the mid-IR light measured in the 24\,$\mu$m passband, while the far-IR emission probed by {\it Herschel} would be dominated by dust-reprocessed stellar light. Indeed, studies of local dusty AGN have demonstrated that their contribution to the IR emission of a galaxy drops rapidly above 20\,$\mu$m in the rest-frame (Netzer et al. 2007). However, this explanation for the mid-IR excess problem was recently called into question by mid-IR spectroscopy of $z$$\sim$2 galaxies obtained using the {\it Spitzer} IRS spectrograph showing the presence of strong PAH emission lines where one would expect hot dust continuum emission to dominate if this regime were dominated by a buried AGN (Murphy et al. 2009, Fadda et al. 2010) and by deeper Chandra observations (Alexander et al. 2011).

\subsection{Resolving the mid-IR excess problem: universality of $IR8$}
\label{SEC:univIR8}
We have seen that extrapolations of $L_{\rm IR}^{\rm tot}$ from 24\,$\mu$m measurements using the CE01 technique fail at $z$$>$1.5. We also find that using the same technique with another set of template SEDs, such as the DH02 ones, fails in a similar way.

We wish to test the main hypothesis on which the CE01 technique relies, namely, that IR SEDs do not evolve with redshift. If that was the case, then a single SED could be used to derive the $L_{\rm IR}^{\rm tot}$ of any galaxy whatever the rest-frame wavelength probed, as long as it falls in the dust reprocessed stellar light wavelength range. Indeed, local galaxies are observed to follow tight correlations between their mid-IR luminosities at 6.75, 12, 15, 25\,$\mu$m and $L_{\rm IR}^{\rm tot}$ (see CE01, Elbaz et al. 2002) as well as with their SFR as derived from the Pa$\alpha$ line (Calzetti et al. 2007) for the {\it Spitzer} passbands at 8 and 24\,$\mu$m. This technique fails at $z$$>$1.5, which has until now been interpreted as evidence that distant IR SEDs are different from local ones. However, in order to properly test the redshift evolution of the IR SEDs, it is necessary to compare measurements in the same wavelength range for galaxies at all redshifts. For that purpose, we now compute the same rest-frame mid-IR luminosity, $L_8$ (=$\nu$$L_{\nu}$[8\,$\mu$m]), defined as the luminosity that would be measured in the IRAC--8\,$\mu$m passband in the rest-frame. We choose this particular wavelength range because it can be computed from $z$$\sim$0 to 2.5 with minimum extrapolations using the IRAC-8\,$\mu$m filter for nearby galaxies ($z$$<$0.5), the IRS-16\,$\mu$m peak-up array for intermediate redshifts around $z$$\sim$1 (0.5$\leq$z$<$1.5) and the MIPS-24\,$\mu$m passband at $z$$\sim$2 (1.5$\leq$$z$$\leq$2.5). Even in these conditions, small k-corrections need to be applied in order to calculate $L_8$ for the same rest-frame passband. This was done using the mid-IR SED of M82 for all galaxies. We verified that using other SEDs, such as the CE01 or DH02 templates, would alter $L_8$ by factors that are small when compared with the dispersion of the observed $L_{\rm IR}^{\rm tot}$ -- $L_8$ relation.  The results are shown in the right-hand part of Fig.~\ref{FIG:IR8}. Surprisingly, when plotting galaxies at all redshifts and luminosities in the same wavelength range, we no longer see a discrepancy between galaxies above and below $z$$\sim$1.5. The sliding median of the $IR8$ ratio, defined as $IR8$=$L_{\rm IR}^{\rm tot}/L_8$, -- illustrated by white points connected with a solid grey line in the right-hand part of Fig.~\ref{FIG:IR8} -- remains flat and equal to $IR8$=4.9 [-2.2,+2.9] (solid and dashed lines in Fig.~\ref{FIG:IR8}-right) from $L_8$=10$^9$ to 5$\times$10$^{11}$ L$_{\odot}$ or equivalently from $L_{\rm IR}^{\rm tot}$=5$\times$10$^{9}$ to 3$\times$10$^{12}$ L$_{\odot}$. The 68\,\% dispersion around the median is only $\pm$0.2 dex.  
   \begin{figure}
   \centering
   \includegraphics*[width=9cm]{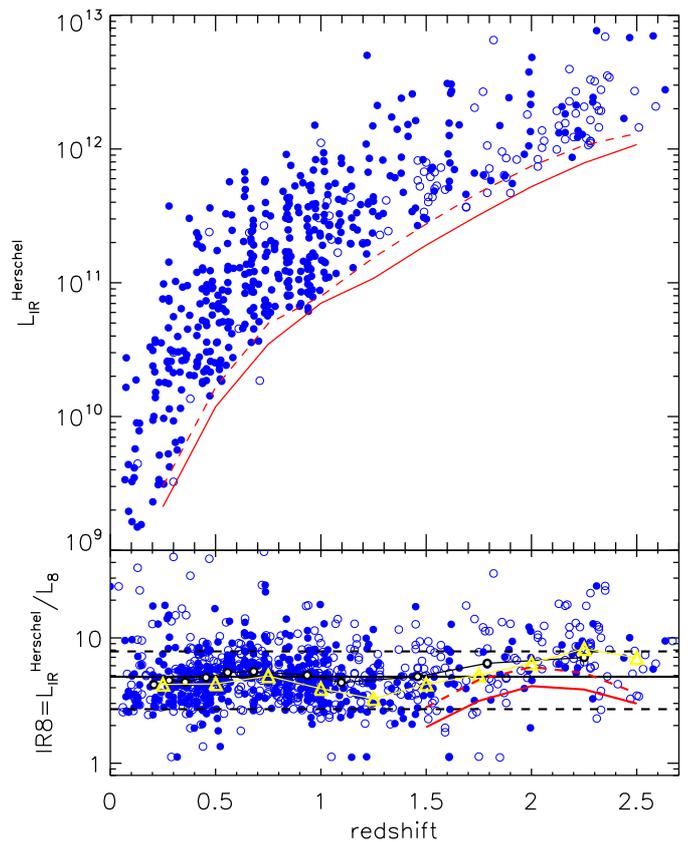}
      \caption{\textbf{\textit{Upper panel:}} Distribution of total IR luminosities ($L_{\rm IR}^{\rm {\it Herschel}}$) of the GOODS--{\it Herschel} galaxies classified as ``clean'' as a function of redshift. The solid and dashed red lines are the detection limits of the GOODS--S and GOODS--N images respectively. Spectroscopic and photometric redshifts are shown with filled and open dots respectively.
      \textbf{\textit{Bottom panel:}} $IR8$ ratio (=$L_{\rm IR}^{\rm {\it Herschel}}$/$L_8$) as a function of redshift. The solid and dashed horizontal black lines are the median and 16th and 84th percentiles of the distribution (Eq.~\ref{EQ:IR8}), i.e., $IR8$=$4.9$ $[-2.2,+2.9]$. The solid and dashed red lines show the detection limits of the GOODS--S and GOODS--N images above $z$=1.5.  The sliding median of the sources detected by {\it Herschel} is shown with black open circles connected with a solid line. A weighted combination of detections with stacked measurements (as in Fig.~\ref{FIG:IR8} and as described in Sect.~\ref{SEC:univIR8}) is shown with open yellow triangles (both fields combined).
      }
         \label{FIG:IR8z}
   \end{figure}
   \begin{figure}
   \centering
   \includegraphics*[width=9cm]{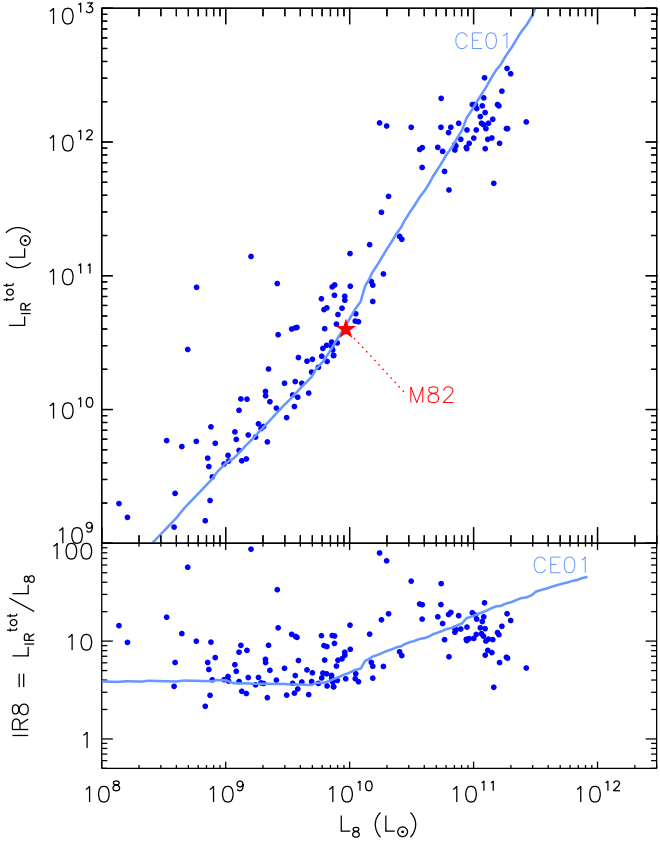}
      \caption{$L_{\rm IR}^{\rm tot}$[IRAS] versus $L_8$ for local galaxies including only the {\it ISO} sample of galaxies used to build the CE01 library of template SEDs and converted from 6.75\,$\mu$m to 8\,$\mu$m using the SED of M82. The light blue line shows the position of the CE01 SED templates, built to follow two power laws in the $L_{\rm IR}^{\rm tot}$ -- $L_8$ relation.}
         \label{FIG:IR8_iso}
   \end{figure}

In order to test possible selection effects on the galaxies used to determine the $IR8$ ratio, we combined {\it Herschel} detections with stacked measurements on 24\,$\mu$m prior positions. This was done by defining intervals of luminosity in $L_8$, e.g., from the 16\,$\mu$m band for sources around $z$$\sim$1 or 24\,$\mu$m for sources around $z$$\sim$2. In a given $L_8$ interval, we determined the median of the $L_{\rm IR}^{\rm tot}$ obtained for detections on one hand (white dots connected with a solid line in the right-hand part of Fig.~\ref{FIG:IR8}) and on the other hand measured average PACS 100\,$\mu$m and  160\,$\mu$m flux densities for the sources with no {\it Herschel} detection by stacking sub-images of 60\arcsec\,on a side at their 24\,$\mu$m prior positions. These sub-images were extracted from the residual images to avoid contamination by detections. The average stacked PACS-100\,$\mu$m and 160\,$\mu$m flux densities were converted into total IR luminosities using the CE01 library of template SEDs, selected based on luminosity at the median redshift of the galaxies in that $L_8$ luminosity interval. We found no systematic difference when deriving $L_{\mathrm IR}^{\mathrm tot}$ from the PACS 100~$\mu$m or 160~$\mu$m data when using the CE01 templates for the extrapolation (see also Elbaz et al. 2010) . Both PACS bands gave consistent values for $L_{\rm IR}^{\rm tot}$. The two values obtained for $L_{\rm IR}^{\rm tot}$ from detected and stacked undetected sources were then combined according to a weight depending on the number of sources in each group within this $L_8$ interval and on the signal-to-noise ratio of these measurements (quadratically), in order to avoid giving the same weight to both measurements if they have the same number of sources but very different S/N ratios. The resulting $L_{\rm IR}^{\rm tot}$ -- $L_8$ relation is shown with yellow open triangles separately for each GOODS field. Since the 100\,$\mu$m and 160\,$\mu$m gave similar results, we only present in the right-hand part of Fig.~\ref{FIG:IR8} the result obtained from the 100\,$\mu$m band. Again, the typical $IR8$ ratio appears to be flat, independent of both luminosity and redshift. The range of luminosities probed by GOODS--{\it Herschel} varies as a function of redshift as shown in the upper panel of Fig.~\ref{FIG:IR8z}, where we represent the distribution of total IR luminosities measured with {\it Herschel} as a function of redshift for the galaxies classified as ``clean'' (Sect.~\ref{SEC:data}). This is due to the combination of limited volume at low redshifts -- limiting the ability to detect rare luminous objects -- and depth at high redshifts -- limiting the ability to detect distant low luminosity objects. In the bottom panel of Fig.~\ref{FIG:IR8z}, we show the redshift evolution of the $IR8$ ratio. It is flat up to $z$$\sim$2 and then, due to the shallower detection limit of {\it Herschel} compared to {\it Spitzer}--24\,$\mu$m, it is slightly larger than the typical value, since only galaxies with high $L_{\rm IR}^{\rm tot}$/$L_8$ can be detected by {\it Herschel}.

Hence, we do not see a mid-IR excess when comparing systematically $L_{\rm IR}^{\rm tot}$ to 8\,$\mu$m rest-frame data. In particular, if AGN were playing a more important role at $z$$>$1.5 than at lower redshifts, we would expect to see a change in $IR8$ at this redshift cut-off contrary to what is actually observed. The cause for the mid-IR discrepancy is therefore not specific to galaxies at $z > 1.5$, but is instead due to the templates used to represent $z \sim 0$ ULIRGs. Locally, galaxies with $L_{\rm IR}^{\rm tot}$ $>$ 10$^{12}$ L$_{\odot}$ are very rare, most probably because galaxies today are relatively gas-poor compared to those at high redshift.  Moreover, they have infrared SEDs that are not typical of star-forming galaxies in general, including those of most distant ULIRGs. The majority of high-redshift galaxies, even ultraluminous ones, share the same IR properties as do local, normal, star-forming galaxies with lower total luminosities.  Galaxies with SEDs like those of local ULIRGs do exist at high redshift, but they do not dominate high redshift ULIRGs by number as they do in the present day.

\subsection{Origin of the ``mid-IR excess'' discrepancy}
\label{SEC:localIR8}
   \begin{figure*}[ht!]
   \centering
   \includegraphics*[width=8.5cm]{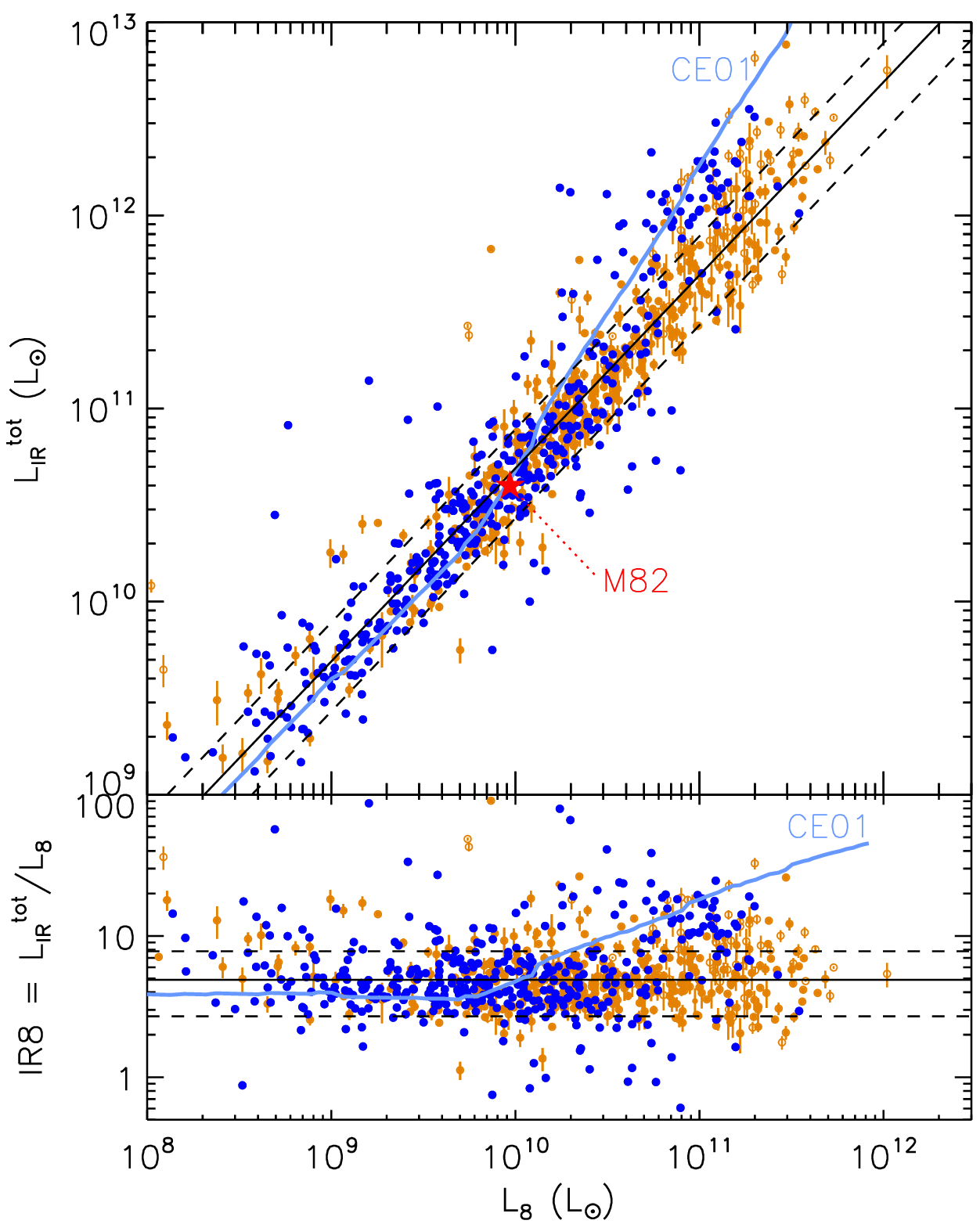}
   \includegraphics*[width=8.5cm]{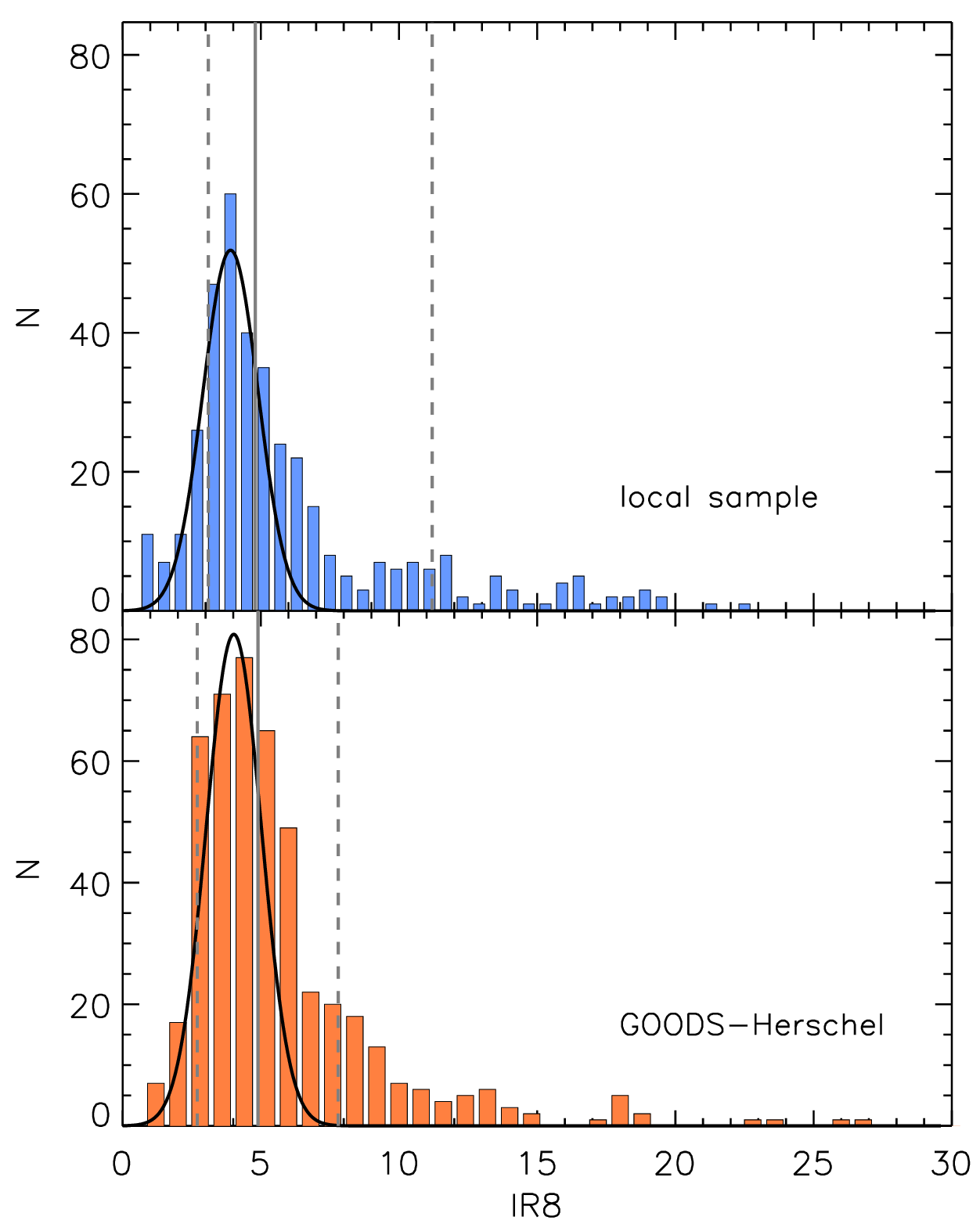}
      \caption{\textbf{\textit{Left:}}$L_{\rm IR}^{\rm tot}$[IRAS] versus $L_8$ for local {\it ISO} and {\it AKARI} galaxies (filled blue dots). The GOODS--{\it Herschel} galaxies are shown in the background with lighter orange symbols (same points as in Fig.~\ref{FIG:IR8}) together with their median (black solid line) and 68\,\% dispersion (black dashed lines). The light blue line shows the locus traced by the CE01 SED library.
      \textbf{\textit{Right:}} Histogram of the $IR8$ ratios for the local galaxy sample (blue, upper panel) and GOODS--{\it Herschel} sample (orange, bottom panel). The solid curves show Gaussians fit to the distributions.  The vertical grey lines indicate the  median (solid) and 68\,\% dispersion (dashed) for the full samples (as in Eqs.~\ref{EQ:IR8}, \ref{EQ:IR8_local}). 
}
         \label{FIG:IR8_local}
   \end{figure*}
Fig.~\ref{FIG:IR8_iso} shows the original $L_{\rm IR}^{\rm tot}$--$L_8$ data that were used to build the CE01 library of template SEDs. The solid line in the figure shows the relation traced by the SED templates. Originally, the mid-IR luminosity was computed from the ISOCAM--LW2 filter at 6.75\,$\mu$m, $L_{6.7}$, which we convert here to $L_8$ using the SED of M82. The conversion was validated by a sub-sample of galaxies for which we have measurements with both ISOCAM--LW2 and IRAC--8\,$\mu$m. While the trend followed by the CE01 templates is consistent with the GOODS--{\it Herschel} galaxies below $L_8\sim$10$^{10}$ L$_{\odot}$, there is a break above this luminosity threshold that was required to fit the local ULIRGs in this diagram.

In Fig.~\ref{FIG:IR8_local}-left, we supplement the original local {\it ISO} sample with the 287 {\it AKARI} galaxies introduced in Sect.~\ref{SEC:AKARIlocal}. 
With this larger sample, we see galaxies extending the low luminosity trend beyond the threshold of $L_8\sim$10$^{10}$ L$_{\odot}$, with a flat $IR8$ ratio. This trend is similar to the one found for the GOODS--{\it Herschel} galaxies (background larger orange symbols as in Fig.~\ref{FIG:IR8}) and the extended local sample is well contained within the 16th and 84th percentiles around the median of the GOODS--{\it Herschel} sample (solid and dashed lines in Fig.~\ref{FIG:IR8}).

The median of both samples are very similar (see Eqs.~\ref{EQ:IR8},\ref{EQ:IR8_local}),
\begin{equation}
IR8^{\rm local} = 4.8~~~[-1.7,+6.4] 
\label{EQ:IR8_local}
\end{equation}
\begin{equation}
IR8^{\rm GOODS-{\it Herschel}} = 4.9~~~[-2.2,+2.9] 
\label{EQ:IR8}
\end{equation}
Note, however, the large upper limit of the 68\,\% dispersion in Eq.~\ref{EQ:IR8_local}, which is mainly due to the elevated $IR8$ values of the local ULIRGs, as seen in the left-hand panel of Fig.~\ref{FIG:IR8_local}. The medians of both samples are shifted to higher values because of the asymmetric tails of galaxies with large values of $IR8$, as shown in the right-hand part of Fig.~\ref{FIG:IR8_local} where we compare the $IR8$ distribution for the local {\it ISO}$+${\it AKARI} galaxies (upper panel) with that of the GOODS--{\it Herschel} galaxies (lower panel). Both distributions present the same properties: they can be fitted by a Gaussian and a tail of high--$IR8$ values. The central values and widths $\sigma$ of the Gaussian distributions are very similar for both samples (Eqs.~\ref{EQ:IR8local},\ref{EQ:IR8gh}), 
\begin{equation}
IR8^{\rm local} ({\rm center~Gaussian})= 3.9~~~[\sigma=1.25] 
\label{EQ:IR8local}
\end{equation}
\begin{equation}
IR8^{\rm GOODS-{\it Herschel}} ({\rm center~Gaussian})= 4.0~~~[\sigma=1.6]~,
\label{EQ:IR8gh}
\end{equation}
again reinforcing the interpretation that the distant galaxies behave very similarly to local galaxies.  If the IR SED of galaxies were different at low and high redshift, then one would not expect them to have the same distributions in $IR8$.

Hence we do not find evidence for different IR SEDs in distant galaxies. Instead, we find that local and distant galaxies are both distributed in two quite well-defined regimes: a Gaussian distribution containing nearly 80\,\% of the galaxies, which share a universal $IR8$ ratio of $\sim$4, and a sub-population of $\sim$20\,\% of galaxies with larger $IR8$ values. The exact proportion of this sub-population is not absolutely determined from this analysis, since it depends on the flux limit used to define the local reference sample, while the distant sample mixes together galaxies spanning a large range of redshifts and luminosities. Nevertheless, the objects in the high-$IR8$ tail remain a minority at both low and high redshift compared with those in the Gaussian distribution. 

In the following, we call the dominant population ``main sequence'' galaxies, since they follow a Universal trend in $L_{\rm IR}^{\rm tot}$--$L_8$ valid at all redshifts and luminosities. We also justify this choice in the next sections by showing that this population also follows a main sequence in SFR -- $M_*$, while galaxies with an excess $IR8$ ratio systematically exhibit an excess sSFR (=SFR/$M_*$). In the local sample, ULIRGs are clearly members of the second population whereas $z$$\sim$2 ULIRGs mostly belong to the Gaussian distribution, hence are main sequence galaxies. It is therefore the weight of both populations that has changed with time and that is at the origin of the mid-IR excess problem. The CE01 SED library, illustrated by a blue line in Figs.~\ref{FIG:IR8_iso} and \ref{FIG:IR8_local}, reaches values of $IR8$ that are more than five times larger than the typical value for main sequence galaxies. This leads to an overestimate of $L_{\rm IR}^{\rm tot}$ when the SED templates for local ULIRGs are used to extrapolate from 24\,$\mu$m photometry for main sequence galaxies at $z$$\sim$2.  Note, however, that it is not necessary to call for a new physics for the IR SED of these galaxies that would justify, e.g., stronger PAH equivalent widths, since most of the distant LIRGs and ULIRGs belong to the same main sequence as local normal star-forming galaxies. It is well-known that local (U)LIRGs are experiencing a starburst phase, with compact star formation regions, triggered in most cases by major mergers (see e.g., Armus et al. 1987, Sanders et al. 1988, Murphy et al. 1996, Veilleux, Kim \& Sanders 2002 for ULIRGs and Ishida 2004 for LIRGs). This leads us to the investigation of the role of compactness presented in the next section. Indeed, if local ULIRGs are known to form stars in compact regions and are found to be atypical in terms of $IR8$, then it would be logical to expect that distant ULIRGs instead are less compact, perhaps as a result of their higher gas fractions. Note also that galaxies with an excess $IR8$ ratio are found at all luminosities and redshifts and are not only a characteristic of ULIRGs. 

\section{$IR8$ as a tracer of star formation compactness and ``starburstiness'' in local galaxies}
\label{SEC:compactness}
The size and compactness of the star formation regions in galaxies is a key parameter that can affect the IR SED of galaxies. Chanial et al. (2007) showed that the dust temperature (T$_{\rm dust}$) estimated from the {\it IRAS} 60 over 100\,$\mu$m flux ratio, R(60/100), is very sensitive to the spatial scale over which most of the IR light is produced. It is known that there is a rough correlation of R(60/100), hence T$_{\rm dust}$, with $L_{\rm IR}^{\rm tot}$ (Soifer et al. 1987): locally, the most luminous galaxies are warmer. This relation has recently been established with {\it AKARI} and {\it Herschel} in the local and distant Universe (Hwang et al. 2010a). Locally, where galaxies can be spatially resolved in the far-IR or radio, Chanial et al. (2007) showed that the dispersion in the $L_{\rm IR}$ -- T$_{\rm dust}$ relation was significantly reduced by replacing $L_{\rm IR}$ by the IR surface brightness, $\Sigma_{\rm IR}$. We extend this analysis to the relation between this star formation compactness indicator, $\Sigma_{\rm IR}$, and $IR8$, the far-IR over mid-IR luminosity ratio. In the present study, the term ``compactness'' is used to refer to the overall size of the starburst and not to the local clumpiness of the various star formation regions, which we cannot measure in most cases. 

An extension of the Chanial et al.\ analysis to the brighter IR luminosity range of (U)LIRGS has become possible thanks to the work of D\'iaz-Santos et al. (2010). They used \textit{Spitzer}/IRS data to derive the fraction of extended emission of the mid-IR continuum of the GOALS galaxy sample (Sect.~\ref{SEC:SPITZERlocal}) at 13.2$\,\mu$m.

\subsection{Determination of the projected star formation density}
\label{SEC:proj_density}
\subsubsection{Radio/Far-IR projected surface brightness}
\label{SEC:RADIOsizes}
Due to the limited angular resolution of far-IR data, we first estimate the sizes of star formation regions from radio imaging by cross-matching the local galaxy sample with existing radio continuum surveys and then convert them into far-IR sizes using a correlation determined from a small sample of galaxies resolved in both wavelength domains as in Chanial et al. (2007). 

\begin{figure}
   \centering
	\includegraphics*[width=9cm]{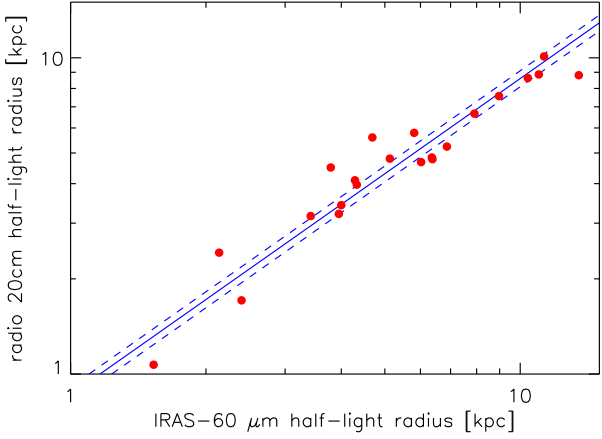}
      \caption{IRAS-60\,$\mu$m versus VLA 20 cm radio continuum half-light radius correlation.}
         \label{FIG:rir_radio}
   \end{figure}

The IRAS-60\,$\mu$m and VLA radio continuum (RC, 20 cm) azimuthally averaged surface brightness profiles of a sample of 22 nearby spiral galaxies was fitted by a combination of exponential and Gaussian functions by Mayya \& Rengarayan (1997). The angular resolutions of the 60\,$\mu$m and 20~cm maps used in that study was about 1\arcmin, so we deconvolved their synthesized profiles by a 1\arcmin\,beam and derived the intrinsic half-light radii r$_{\rm IR}$ (at 60\,$\mu$m) and r$_{\rm RC}$. The half-light radii estimates at both wavelengths are strongly correlated (Fig.~\ref{FIG:rir_radio});  a logarithmic bisector fit to the data is given in Eq.~\ref{EQ:radio_size}: 
\begin{equation}
r_{\rm IR} = (0.86 \pm 0.05)~~ r_{\rm RC}
\label{EQ:radio_size}
\end{equation}
Hence in the following, we estimate the far-IR sizes of the star formation regions of our local galaxy sample from their radio continuum half-light radius using Eq.~\ref{EQ:radio_size}. The existence of such correlation is not surprising, since the radio and far-IR emission of star-forming galaxies are known to present a tight correlation (Yun et al. 2001, de Jong et al. 1985, Helou, Soifer \& Rowan-Robinson 1985): the radio emission is predominantly produced by the synchrotron radiation of supernova remnants and the bulk of the far-IR emission is due to UV light from young and massive stars reprocessed by interstellar dust. Hence, we consider this size estimate to be a good proxy for the global size of the star formation regions of galaxies. This is obviously an approximation, since this does not account for the clumpiness or granularity of the region, but this is the best that we can do with existing datasets.

Our local galaxy sample was cross-matched with the NRAO VLA Sky Survey (NVSS, Condon et al. 1998) and the Faint Images of the Radio Sky at Twenty-cm (FIRST, Becker, White \& Helfand 1995), both obtained with the VLA at 20 cm. A total of 11, 47 and 58 galaxies have radio sizes in our {\it ISO}, {\it AKARI} and {\it Spitzer} local galaxy samples (see Table~\ref{TAB:local}). 

We computed the IR surface brightness using Eq.~\ref{EQ:surfIR}, 
\begin{equation}
\Sigma_{\rm IR} = \frac{L_{\rm IR}/2}{\pi r_{IR}^{2}}~,
\label{EQ:surfIR}
\end{equation}
where the IR luminosity is divided by 2 since r$_{\rm IR}$ is the far-IR (60\,$\mu$m) half-light radius, which is derived from the 20~cm radio measurements using Eq.~\ref{EQ:radio_size}.

\subsubsection{Mid-IR compactness}
\label{SEC:MIRsizes}
Using the low spectral resolution staring mode of the \textit{Spitzer}/IRS, D\'iaz-Santos et al. (2010) measured the spatial extent of the mid-IR continuum emission at 13.2$\,\mu$m for 211 local (U)LIRGs of the GOALS sample (see section 2.2.3). The 13.2$\,\mu$m emission probes the warm dust (very small grains, VSGs) heated by the UV continuum of young and massive stars, and hence traces regions of dust-obscured star formation. Instead of measuring the half-light radius of the sources at this wavelength, D\'iaz-Santos et al. (2010) calculated their fraction of extended emission, or FEE, which they defined as the fraction of light in a galaxy that does not arise from its spatially unresolved central component. Conversely, the compactness of a source can be defined as the percentage of light that is unresolved, that is, 100$\times$(1$-$FEE). The angular resolution of \textit{Spitzer}/IRS at 13.2$\,\mu$m is $\sim\,3.6\,\arcsec$ which, at the median distance of the sample used in this work, 91\,Mpc, results in a spatial resolution of 1.7\,kpc.

In the following, we consider galaxies as ``compact'' if their 13.2$\,\mu$m compactness is greater than 60\%. With this definition, we find that 55\% (117/211) of the GOALS galaxies are compact. Interestingly, while it is true that the fraction of galaxies showing compact star formation (i.e., compact hot dust emission) increases with increasing $L_{\rm IR}^{\rm tot}$ (hence also with SFR), the compact population is not systematically associated with the most luminous sources. On the contrary, galaxies with compact star formation can be found at all luminosities (see Figure~4 of D\'iaz-Santos et al. 2010).

   \begin{figure}
   \centering
   \includegraphics[width=9cm]{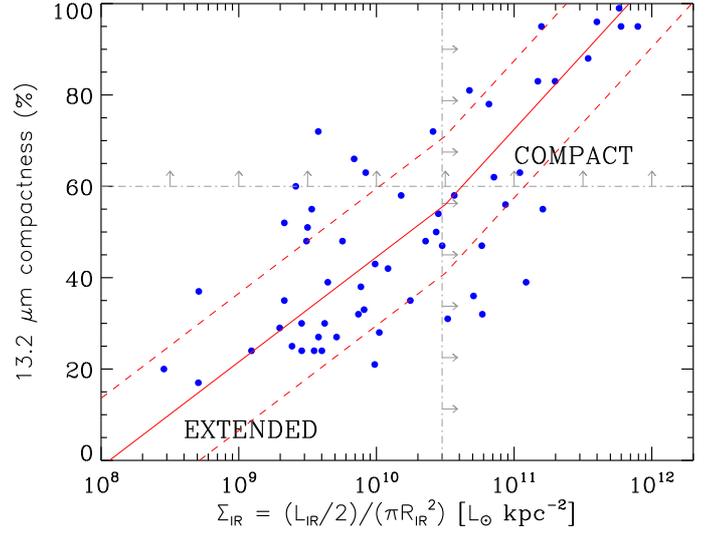}
      \caption{Comparison of the two compactness indicators: $\Sigma_{\rm IR}$ (=$L_{\rm IR}^{\rm tot}$/(2$\times$$\pi$$R_{\rm IR}^2$)), the IR surface brightness, and 13.2\,$\mu$m compactness (percentage of unresolved {\it Spitzer}/IRS light at 13.2\,$\mu$m, D\'iaz-Santos et al. 2010).}
         \label{FIG:compactness}
   \end{figure}
   \begin{figure}
   \centering
   \includegraphics*[width=9cm]{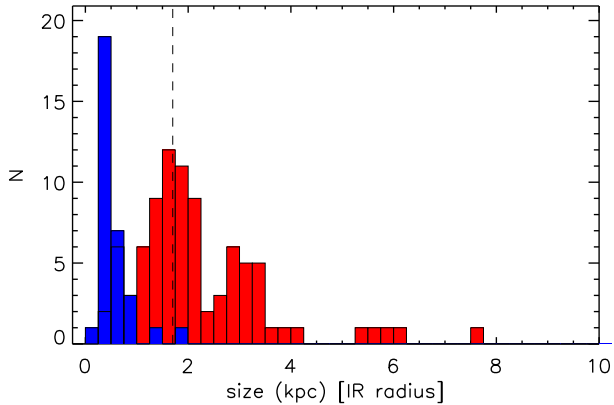}
      \caption{Distribution of IR sizes (half-light radius) of 119 local galaxies as derived from their 1.4 GHz radio continuum using Eq.~\ref{EQ:radio_size}. Galaxies with an IR surface brightness greater than $\Sigma_{\rm IR}$=3$\times$10$^{10}$ L$_{\odot}$kpc$^{-2}$, i.e., compact galaxies, are in blue, while extended galaxies are in red. The vertical dashed line indicates the typical resolution of $\sim$1.7 kpc of the mid-IR compactness at 13.2\,$\mu$m estimated by D\'iaz-Santos et al. (2010).}
         \label{FIG:IRsize}
   \end{figure}

\subsubsection{Identification of the galaxies with compact star formation}
\label{SEC:compactSF}
In order to check whether both star formation compactness indicators are consistent, we used the 58 galaxies from the GOALS sample for which we can determine both $\Sigma_{\rm IR}$ from the radio sizes (Sect.~\ref{SEC:RADIOsizes}) and a 13.2\,$\mu$m compactness (Sect.~\ref{SEC:MIRsizes}). The comparison of both compactness indicators shows a correlation with a dispersion of $\sim$0.45 dex (Fig.~\ref{FIG:compactness}). The critical threshold of 60\,\% in the 13.2\,$\mu$m compactness above which we classify galaxies as compact corresponds to $\Sigma_{\rm IR}$$\sim$3$\times$10$^{10}$ L$_{\odot}$ kpc$^{-2}$. Hence, we hereafter classify as compact the galaxies for which $\Sigma_{\rm IR}$$\ge$3$\times$10$^{10}$ L$_{\odot}$ kpc$^{-2}$. This threshold is more than two orders of magnitude lower than typical upper limits for star formation on small (kpc) scales (see Soifer et al. 2001). We note that if it were not averaged on large scales, the local star formation surface density could be much higher in many of these sources.

In Fig.~\ref{FIG:IRsize}, we present the distribution of far-IR sizes of extended (red) and compact (blue) galaxies, estimated from radio 20 cm imaging using Eq.~\ref{EQ:radio_size}. The median far-IR sizes of compact and extended galaxies are 0.5 kpc and 1.8 kpc respectively. The typical spatial resolution reached at 13.2\,$\mu$m, i.e., 1.7 kpc, is close to the typical size of extended galaxies and is significantly larger than the median size for compact galaxies. This contributes to the relatively high dispersion seen in Fig.~\ref{FIG:compactness}. With a linear resolution of $\sim$0.2 kpc at the average distance of the GOALS sample, the radio estimator is therefore a finer discriminant of compact galaxies when good quality radio data exist. 

In the following, we use both compactness indicators, i.e., radio and 13.2\,$\mu$m, with no distinction to define the projected IR surface brightness, $\Sigma_{\rm IR}$ (=$L_{\rm IR}^{\rm tot}$/(2$\times$$\pi$$R_{\rm IR}^2$)). For galaxies with a measured radio size,  $R_{\rm IR}$ is computed using Eq.~\ref{EQ:radio_size}, while for galaxies with a 13.2\,$\mu$m compactness estimate but no radio size we use the relation presented in Fig.~\ref{FIG:compactness}.

   \begin{figure}
   \centering
   \includegraphics*[width=8.5cm]{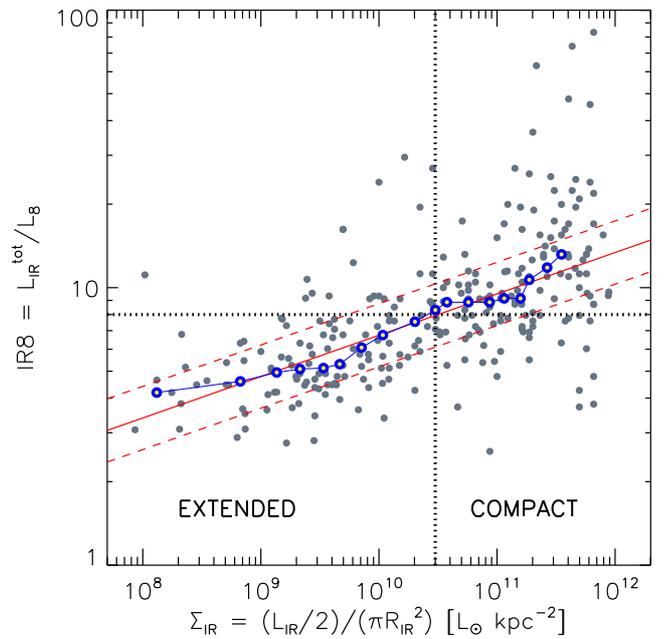}
      \caption{Dependence of $IR8$ (=$L_{\rm IR}^{\rm tot}$/$L$[8\,$\mu$m]) with $\Sigma_{\rm IR}$ (=$L_{\rm IR}^{\rm tot}$/(2$\times$$\pi$$R_{\rm IR}^2$)), the IR surface brightness. Galaxies to the right of the vertical dotted line are considered to be compact star-forming galaxies ($\Sigma_{\rm IR}$$\ge$3$\times$10$^{10}$ L$_{\odot}$ kpc$^{-2}$). Galaxies above the horizontal dotted line exhibit $IR8$ ratio that is two times larger than that of main sequence galaxies, which follow the Gaussian $IR8$ distribution shown in Fig.~\ref{FIG:IR8_local}). The sliding median is shown with open blue dots. It is fitted by the solid line in red and its 16th and 84th percentiles are fitted with the dashed red lines (see Eq.~\ref{EQ:IR8_sigma}).
      }
         \label{FIG:radio_compact}
   \end{figure}
   \begin{figure}
   \centering	
      \includegraphics*[width=9cm]{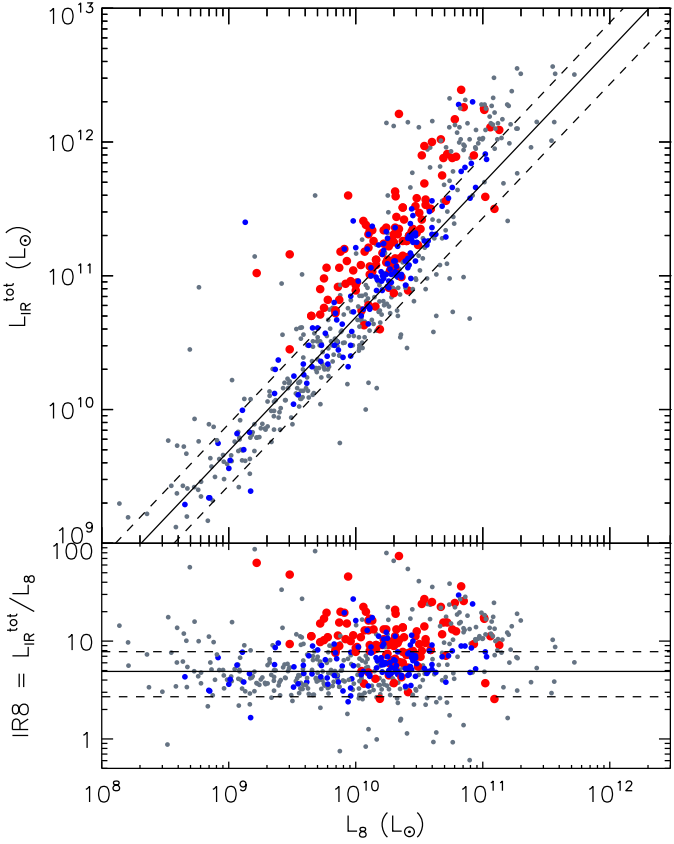}
      \caption{$L_{\rm IR}^{\rm tot}$ (top) and $IR8$ ratio (bottom) versus $L$[8\,$\mu$m] for the local galaxy sample (ISO, {\it AKARI}, {\it Spitzer}-GOALS). Galaxies with compact mid-IR (more than 60\,\% of the light emitted at 13.2\,$\mu$m within the resolution element of 3.6\arcsec) and radio ($\Sigma_{\rm IR}$$\geq$3$\times$10$^{10}$ L$_{\odot}$) light distributions are marked with large filled red dots. Here the solid and dashed lines present the median and 68\,\% dispersion around the Gaussian distribution as defined in Eq~\ref{EQ:IR8local}.
      }
         \label{FIG:setMSlocal}
   \end{figure}

\subsection{$IR8$, a star formation compactness indicator}
\label{SEC:IR8compact}

The $IR8$ ratio is compared to the IR surface brightness, $\Sigma_{\rm IR}$, in Fig.~\ref{FIG:radio_compact}. The number of galaxies presented in this figure is larger than in Fig.~\ref{FIG:compactness} because we include sources with no radio size estimate as well. We find that $IR8$ is correlated with $\Sigma_{\rm IR}$ for local galaxies following Eq.~\ref{EQ:IR8_sigma}, 
\begin{equation}
IR8 = 0.22~[-0.05,+0.06] \times \Sigma_{\rm IR}^{0.15}~,
\label{EQ:IR8_sigma}
\end{equation}
where $\Sigma_{\rm IR}$ is in L$_{\odot}$ kpc$^{-2}$. Hence $IR8$ is a good proxy for the projected IR surface brightness of local galaxies. Galaxies with strong $IR8$ ratios are also those which harbor the highest star-formation compactness.

We showed in Fig.~\ref{FIG:compactness} that galaxies having more than 60\,\% of their 13.2\,$\mu$m emission unresolved by {\it Spitzer}--IRS, defined as compact star-forming galaxies by D\'iaz-Santos et al. (2010), presented an IR surface brightness of $\Sigma_{\rm IR}$$\geq$3$\times$10$^{10}$ L$_{\odot}$kpc$^{-2}$. This threshold is illustrated in Fig.~\ref{FIG:radio_compact} by a vertical dotted line. It crosses the best-fitting relation of Eq.~\ref{EQ:IR8_sigma} at $IR8$=8, i.e., twice the central value of the Gaussian distribution of main sequence galaxies (Fig.~\ref{FIG:IR8_local} and Eq.~\ref{EQ:IR8local}). As a result, compact star-forming galaxies, with $\Sigma_{\rm IR}$$\geq$3$\times$10$^{10}$ L$_{\odot}$kpc$^{-2}$, systematically present an excess in $IR8$, whereas nearly all galaxies with extended star formation exhibit a 'normal' $IR8$, i.e., within the Gaussian distribution of Fig.~\ref{FIG:IR8_local}. This is illustrated in Fig.~\ref{FIG:setMSlocal}, reproducing the $IR8$--$L8$ diagram for local galaxies of Fig.~\ref{FIG:IR8_local}, this time including the GOALS sample. The sub-sample of galaxies with measured IR surface brightnesses are represented with large symbols, with blue and red marking galaxies with extended and compact star-formation respectively, i.e.,  $\Sigma_{\rm IR}$ lower and greater than 3$\times$10$^{10}$ L$_{\odot}$kpc$^{-2}$. Galaxies with compact star formation systematically lie above the typical range of $IR8$ values. The trend can be extended to the local ULIRGs with no size measurement, since they are known to experience compact starbursts driven by major mergers (Armus et al. 1987, Sanders et al. 1988, Murphy et al. 1996, Veilleux, Kim \& Sanders 2002).

   \begin{figure}
   \centering
   \includegraphics*[width=8.5cm]{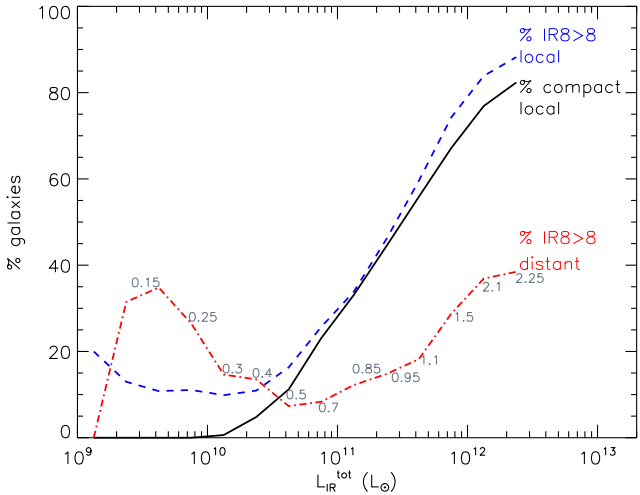}
      \caption{Fractions of compact local galaxies (black line), local galaxies with $IR8$$>$8 (blue dashed line) and GOODS--{\it Herschel} galaxies with $IR8$$>$8 (red dash-dotted line) as a function of total IR luminosity. The median redshift of the GOODS--{\it Herschel} galaxies varies with $L_{\rm IR}^{\rm tot}$, and is indicated by labels in grey along the red dash-dotted line.
      }
         \label{FIG:pcIR8}
   \end{figure}

Very interestingly, the proportion of galaxies with compact star formation rises with $L_{\rm IR}^{\rm tot}$ following a path very similar to the proportion of galaxies with $IR8$$>$8 (Fig.~\ref{FIG:pcIR8}), the 68\,\% upper limit of the GOODS--{\it Herschel} galaxies (Eq.~\ref{EQ:IR8}). Hence, $IR8$ can be considered as a good proxy of the star formation compactness of local galaxies. This can be very useful for galaxies with no radio size measurement. 

In comparison, the fraction of excess $IR8$ sources within the GOODS--{\it Herschel} sample remains low (around 20\,\%, see Fig.~\ref{FIG:IR8}-right and Fig.~\ref{FIG:IR8_local}) and never reaches such high proportions as seen in local ULIRGs. Due to the {\it Herschel} detection limit, however, only ULIRGs are individually detected at $z$$>$2. The $IR8$ parameter is found to be biased towards high values in these galaxies which are responsible for the increase in the compactness fraction in the {\it Herschel} sample at the highest redshifts from 20 to 40\,\%.

Globally, this analysis suggests that compact sources have been a minor fraction of star-forming galaxies at all epochs, but locally, due to the low gas content of galaxies, compact sources make the dominant population of ULIRGs. Extending the analysis of local galaxies to the distant ones, this also suggests that compact star formation takes place at all luminosities but does not dominate the majority of distant ULIRGs.
Conversely, knowing the compactness and mid-IR luminosity of a galaxy, one may optimize the determination of its total IR luminosity from mid-IR observations alone. This is discussed in Sect.~\ref{SEC:sed}. 

Finally, we have assumed in this section that the compactness measured either from the radio or from the mid-IR continuum is associated with star formation. We discuss the role of AGN in Sect.~\ref{SEC:AGN}, but we can already note that when an AGN contributes to the IR emission of a galaxy, it does so mainly at wavelengths shorter than 20\,$\mu$m (Netzer et al. 2007, Mullaney et al. 2011a). If AGN were contributing to the infrared emission, they would tend to boost $L_8$ relative to $L_{\rm IR}^{\rm tot}$, therefore reducing $IR8$.  Instead, we see that an increasing compactness corresponds to an increase in $IR8$ as well.  This reinforces the idea that we are dealing here with star formation compactness and not an effect produced by the presence of an active nucleus. In the next section, we show that compact star-forming galaxies are generally experiencing a starburst phase.
   \begin{figure*}
   \centering
\includegraphics*[width=9.cm]{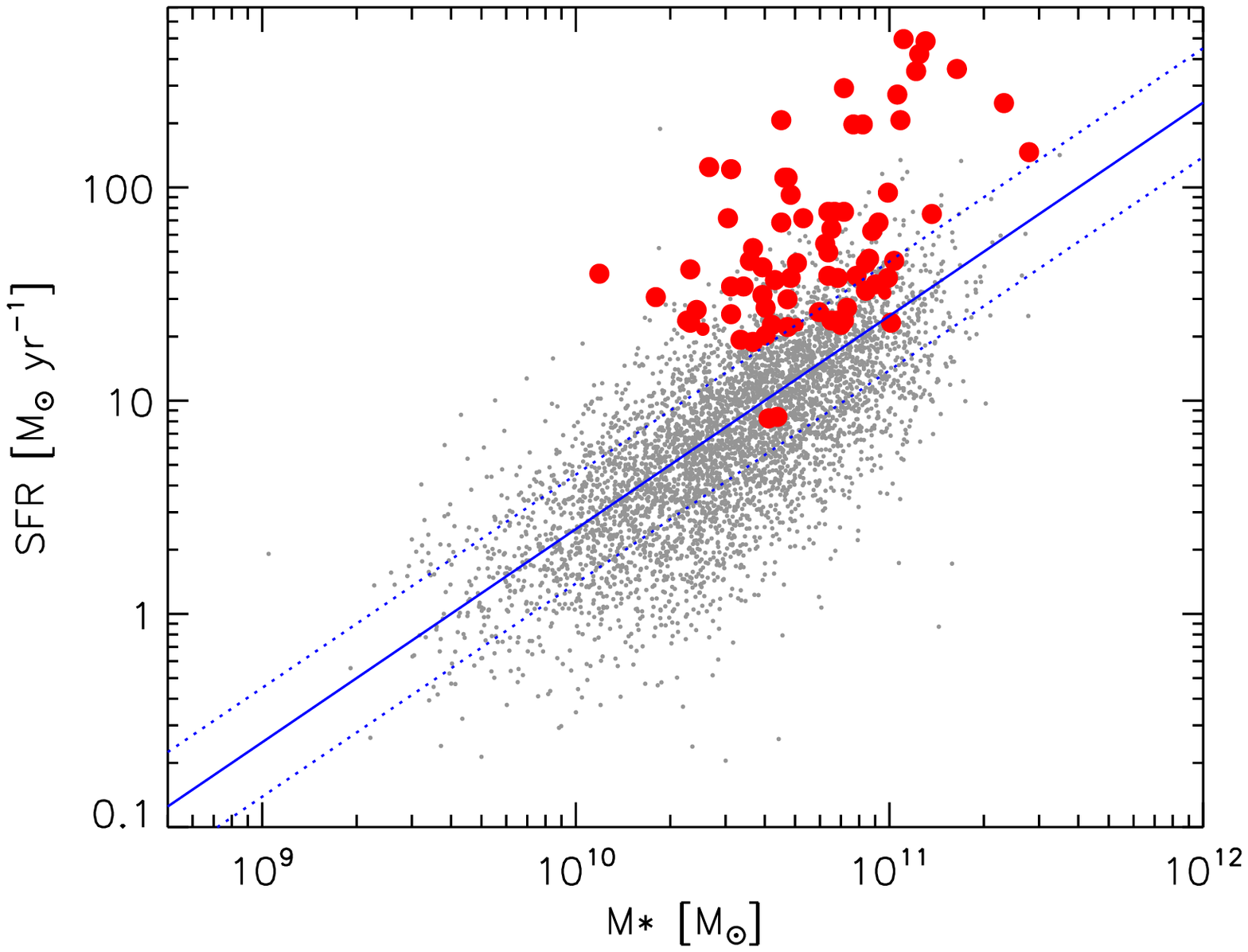}
\includegraphics*[width=9cm]{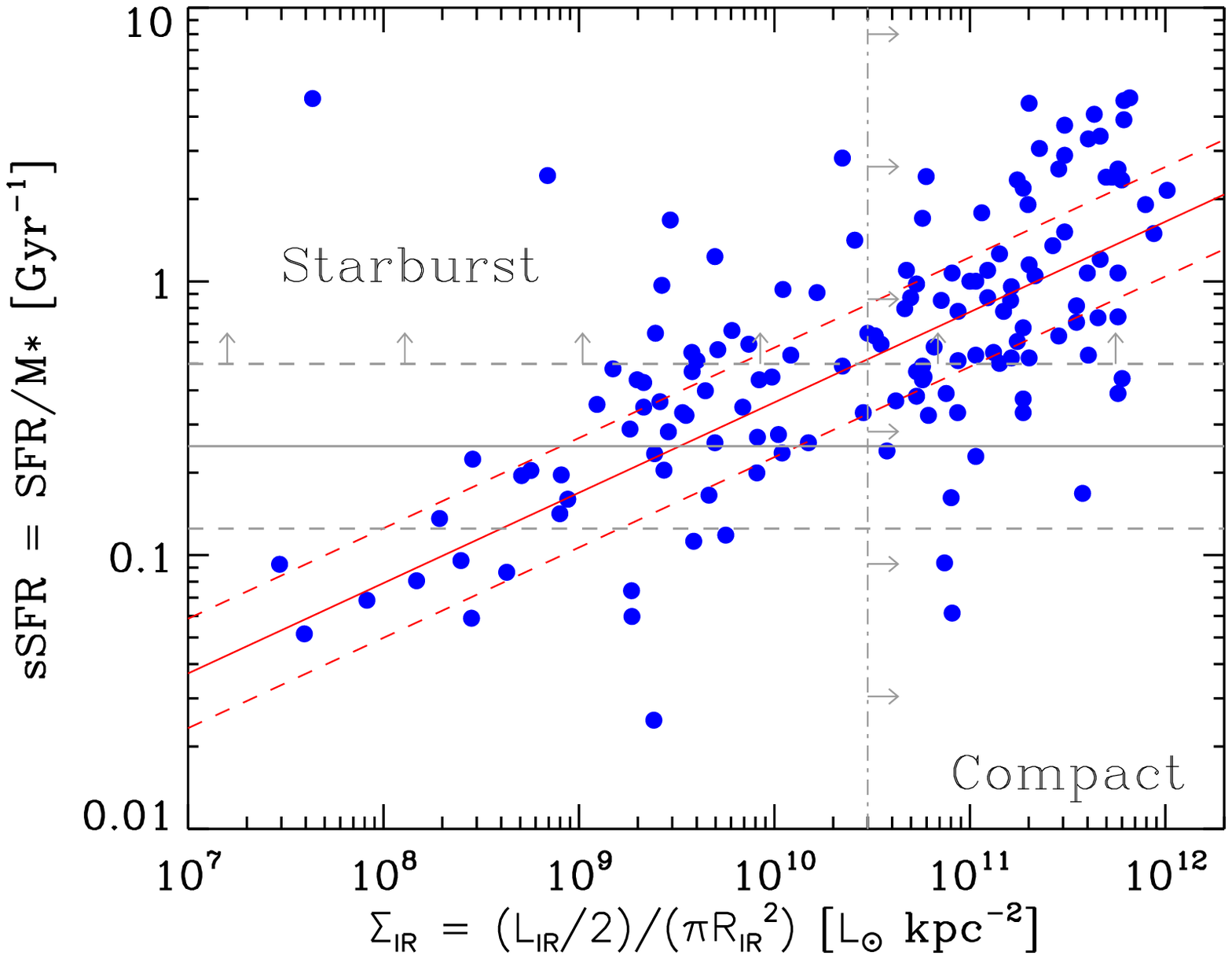}
      \caption{\textbf{\textit{Left:}} SFR -- $M_*$ correlation at $z$$\sim$0. Galaxies classified as compact are marked with large filled red dots. Solid line: fit to the main sequence SFR--$M_*$ relation: SFR$\propto$M$_{\star}/$[4$\times$10$^9$ M$_{\odot}]$. Dotted lines: 16th and 84th percentiles of the distribution around the sliding median (0.26 dex).
      \textbf{\textit{Right:}}Relation of the sSFR and IR surface brightness of galaxies for which a radio size was estimated. The vertical dashed line illustrates the threshold above which galaxies have been classified as compact. The solid and dashed red lines are a fit to the sliding median of the relation (Eq.~\ref{EQ:sSFR_Sir}) and its 68\,\% dispersion.}
         \label{FIG:sfrAKARI}
   \end{figure*}
\subsection{$IR8$, a starburst indicator}
\label{SEC:IR8starburst}
In the previous section, we have seen that high $IR8$ values were systematically found in galaxies with compact star formation regions. We now show that these galaxies are experiencing a starburst phase. 
In the following, a star-forming galaxy is considered to be experiencing a starburst phase if its ``current SFR'' is twice or more stronger than its ``averaged past SFR'' ($<$SFR$>$), i.e., if its birthrate parameter $b$=SFR/$<$SFR$>$ (Kennicutt 1983) is greater than 2. Here $<$SFR$>$=$M_*$/$t_{\rm gal}$, where $t_{\rm gal}$ is the age of the galaxy. Alternatively, a star-forming galaxy may be defined as a starburst if the time it would take to produce its current stellar mass, hence its stellar mass doubling timescale, $\tau$, defined in Eq.~\ref{EQ:t2}, 
\begin{equation}
\tau~[{\rm Gyr}]=M_{\star} ~[{\rm M}_{\odot}]~/~SFR~[{\rm M}_{\odot} ~{\rm Gyr}^{-1}] = 1 / sSFR~[{\rm Gyr}^{-1}]~,
\label{EQ:t2}
\end{equation}
is small when compared to its age. Both definitions are equivalent if one assumes that galaxies at a given epoch have similar ages.

In recent years, a tight correlation between SFR and $M_*$ has been discovered which defines a typical specific SFR, (sSFR = SFR/$M_*$), for ``normal star-forming galaxies'' as opposed to ``starburst galaxies''. This relation evolves with redshift but a tight correlation between SFR and $M_*$ is observed at all redshifts from $z$$\sim$0 to 7 (Brinchmann et al. 2004, Noeske et al. 2007, Elbaz et al. 2007, Daddi et al. 2007a, 2009, Pannella et al. 2009, Magdis et al. 2010a, Gonzalez et al. 2011). Hence, we use the sSFR definition of a starburst since it can be applied at all lookback times. In the present section, we consider only local star-forming galaxies.
The SFR -- $M_*$ relation for local {\it AKARI} galaxies is shown in the left-hand part of Fig.~\ref{FIG:sfrAKARI}. The best fit to this relation is a one-to-one correlation (0.26 dex--rms), hence a constant sSFR$\sim$0.25 Gyr$^{-1}$ or $\tau$$\sim$4 Gyr  (Eq.~\ref{EQ:t2}). Local galaxies with compact star formation (large red dots), as defined in the previous section, are systematically found to have higher sSFR, i.e., $\tau$$<$ 1 Gyr, than that of normal star-forming galaxies. 

Since there is a continuous distribution of galaxies ranging from the normal mode of star formation, with $\tau$$\sim$4 Gyr, to extreme starbursts, that can double their stellar masses in $\tau$$\sim$50 Myr, we quantify the intensity of a starburst by the parameter $R_{\rm SB}$, which measures the excess in sSFR of a star-forming galaxy (which we label its ``starburstiness''), as defined in Eq.~\ref{EQ:starburstiness}: 

\begin{equation}
R_{\rm SB} = sSFR / sSFR_{\rm MS} = \tau_{\rm MS} / \tau~~~~[>~2~~{\rm for~starbursts}]~,
\label{EQ:starburstiness}
\end{equation}
where the subscript MS indicates the typical value for main sequence galaxies at the redshift of the galaxy in question.  A starburst is defined to be a galaxy with $R_{\rm SB}$$\geq$2.  75\,\% of the galaxies with compact star formation ($\Sigma_{\rm IR}$$\geq$3$\times$10$^{10}$ L$_{\odot}$ kpc$^{-2}$) have $R_{\rm SB}$$>$2, hence are also in a starburst mode, and 93\,\% of them have $R_{\rm SB}$$>$1.  Conversely, 79\,\% of the starburst galaxies are ``compact''. Globally, starburst galaxies with sSFR $>$ 2 $\times$ $<$sSFR$>$ have a median $\Sigma_{\rm IR}$$\sim$1.6$\times$10$^{11}$ L$_{\odot}$ kpc$^{-2}$, hence more than 5 times higher than the critical IR surface brightness above which galaxies are compact. The size of their star-forming regions is typically 2.3 times smaller than that of galaxies with sSFR$_{\rm MS}$. 
   \begin{figure}
   \centering
   \includegraphics*[width=8.5cm]{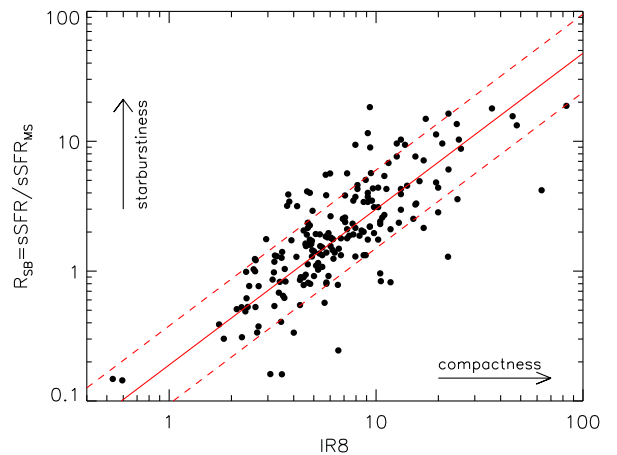}
      \caption{$R_{\rm SB}$=sSFR/sSFR$_{MS}$ versus $IR8$ (=$L_{\rm IR}^{\rm tot}$/$L_8$) for $z$$\sim$0 galaxies ({\it AKARI} and GOALS samples). The red line is the best fit (plain) and its 0.3 dex dispersion (dashed). 
      }
         \label{FIG:sSFR_lirl8}
   \end{figure}

The sSFR and $\Sigma_{\rm IR}$ are correlated with a 0.2 dex dispersion (Fig.~\ref{FIG:sfrAKARI}-right) following Eq.~\ref{EQ:sSFR_Sir}, where sSFR is in Gyr$^{-1}$ and $\Sigma_{\rm IR}$ in L$_{\odot}$kpc$^{-2}$:
\begin{equation}
sSFR = 1.81~[-0.66,+1.05]\times10^{-4}~\times~\Sigma_{\rm IR}^{0.33}
\label{EQ:sSFR_Sir}
\end{equation}
Both parameters measure specific quantities related to the SFR: the sSFR is measured per unit stellar mass, while $\Sigma_{\rm IR}$ is related to the SFR (derived from $L_{\rm IR}^{\rm tot}$) per unit area. Note, however, that it was not obvious {\em a priori} that these quantities should be correlated, since the stellar mass of most galaxies is dominated by the old stellar population, whereas the IR (or radio) size used to derive $\Sigma_{\rm IR}$ measures the spatial distribution of young and massive stars. 

Because the starburstiness and the $IR8$ ratio are both enhanced in compact star-forming galaxies, they are also correlated as shown by Fig.~\ref{FIG:sSFR_lirl8}. The fit to this correlation is given in Eq.~\ref{EQ:R_IR8}:
\begin{equation}
R_{\rm SB} = (IR8/4)^{1.2}
\label{EQ:R_IR8}
\end{equation}
The dispersion in this relation is 0.3 dex. Hence we find that it is mostly compact starbursting galaxies that present atypically strong $IR8$ bolometric correction factors, although there is not a sharp separation of both regimes, but instead a continuum of values.
\section{$IR8$ as a tracer of star formation compactness and ``starburstiness'' in distant galaxies}
\label{SEC:MSSB}
In the previous section, we have defined two modes of star formation:
\begin{itemize}
\item a normal mode that we called the infrared main sequence, in which galaxies present a universal $IR8$ bolometric correction factor and a moderate star formation compactness, $\Sigma_{\rm IR}$, and
\item a starburst mode, identified by an excess SFR per unit stellar mass, hence sSFR, as compared to the typical sSFR of most local galaxies. 
\end{itemize}
Galaxies with an enhanced $IR8$ ratio were systematically found to be forming their stars in the starburst mode and to show a strong star formation compactness ($\Sigma_{\rm IR}$$>$3$\times$10$^{10}$ L$_{\odot}$kpc$^{-2}$). In order to separate these two modes of star formation in distant galaxies as well, we first need to define the typical sSFR of star-forming galaxies in a given redshift domain. This definition has become possible since the recent discovery that star-forming galaxies follow a tight correlation between their SFR and $M_*$ with a typical dispersion of 0.3 dex over a large range of redshifts: $z$$\sim$0 (Brinchmann et al. 2004), $z$$\sim$1 (Noeske et al. 2007, Elbaz et al. 2007), $z$$\sim$2 (Daddi et al. 2007a, Pannella et al. 2009), $z$$\sim$3 (Magdis et al. 2010a), $z$$\sim$4 (Daddi et al. 2009, Lee et al. 2011) and even up to $z$$\sim$7 (Gonzalez et al. 2011).

\subsection{Evolution of the specific SFR with cosmic time and definition of main sequence versus starburst galaxies}
   \begin{figure}[t!]
   \centering
   \includegraphics*[width=9.2cm]{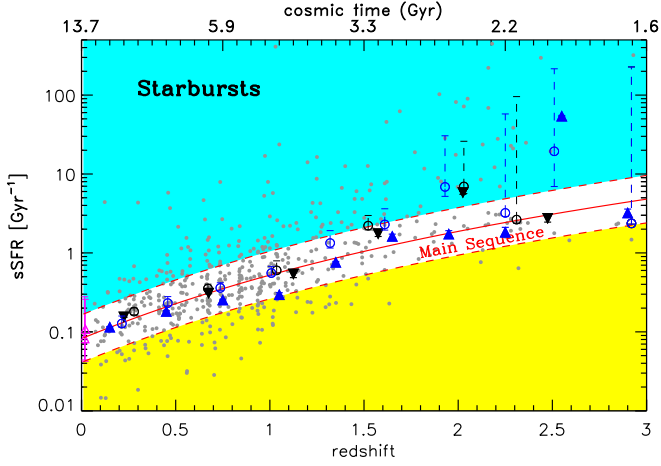}
      \caption{Redshift evolution of the median specific SFR (sSFR=SFR/$M_*$) of star-forming galaxies.   Values for individual GOODS--{\it Herschel} galaxies are shown as grey points.   Median sSFR values in redshift bins are shown with open circles (blue for GOODS--N and black for GOODS--S).  Values combining individual detections and stacking measurements for undetected sources are shown with filled triangles (blue upward for GOODS--N and black downward for GOODS--S). The red solid line is the fit shown in Eq.~\ref{EQ:sSFRz}, and the dashed lines are a factor 2 above and below this fit. Starbursts are defined as galaxies with a sSFR$>$2$\times$sSFR$_{\rm MS}$ (blue zone). The yellow zone shows the galaxies with significantly lower sSFR values. 
}
         \label{FIG:sSFRz}
   \end{figure}
In the following, we assume that the slope of the SFR -- $M_*$ relation is equal to 1 at all redshifts, hence that the specific SFR, sSFR (=SFR/$M_*$), is independent of stellar mass at fixed redshift. A small departure from this value would not strongly affect our conclusions and the same logic may be applied for a different slope. At $z$$\sim$0, our local reference sample is well fitted by a constant sSFR (see Fig.~\ref{FIG:sfrAKARI}-left), although the best-fitting slope is 0.77 (Elbaz et al. 2007). At $z$$\sim$1$\pm$0.3, Elbaz et al. (2007) find a slope of 0.9 but we checked that the dispersion of the data allows a nearly equally good fit with a slope of 1. At $z$$\sim$2, Pannella et al. (2009) find a slope of 0.95 consistent with the value obtained by Daddi et al. (2007a) in the same redshift range. Lyman-break galaxies at $z$$\sim$3 (Magdis et al. 2010a) and $z$$\sim$4 (Daddi et al. 2009) are also consistent with a slope of unity. From a different perspective, Peng et al. (2010a) argue that a slope of unity is required to keep an invariant Schechter function for the stellar mass function of star-forming galaxies from $z$$\sim$0 to 1 as observed from COSMOS data, while non-zero values would result in a change of the faint-end slope of the mass function that would be inconsistent with the observations.

However, the slope of the SFR -- $M_*$ relation is sensitive to the technique used to select the sample of star-forming galaxies. Karim et al. (2011) find two different slopes depending on the selection of their sample: a slope lower than 1 for a mildly star-forming sample, and a slope of unity when selecting more actively star-forming galaxies (see their Fig.13). Using shallower {\it Herschel} data than the present observations, Rodighiero et al. (2010) found a slope lower than unity. 

Assuming that the slope of the SFR -- $M_*$ relation remains equal to 1 at all redshifts, a main sequence mode of star formation can be defined by the median sSFR in a given redshift interval, sSFR$_{\rm MS}(z)$. The starburstiness, described in Eq.~\ref{EQ:starburstiness}, measures the offset relative to this typical sSFR. Since at any redshift -- at least in the redshift range of interest here, i.e., $z$$<$3 -- most galaxies belong to the main sequence in SFR -- $M_*$, we assume that the median sSFR measured within a given redshift interval is a good proxy to the sSFR$_{\rm MS}(z)$ defining the MS. Galaxies detected with {\it Herschel} follow the trend shown with open circles in Fig.~\ref{FIG:sSFRz} (blue for GOODS--N and black for GOODS--S). We have performed the analysis independently for both GOODS fields in order to check the impact of cosmic variance on our result. To correct for incompleteness, we performed stacking measurements as for Fig.~\ref{FIG:IR8} but in redshift intervals. The stacking was done on the PACS-100\,$\mu$m images using the 24\,$\mu$m sources as a list of prior positions. The resulting values (blue upward triangles for GOODS--N and black downward triangles for GOODS--S) were computed by weighting detections and stacking measurements by the number of sources used in both samples per redshift interval. The SFR was derived from $L_{\rm IR}^{\rm tot}$ extrapolated from the PACS-100\,$\mu$m band photometry using the CE01 technique. The CE01 method works well for 100\,$\mu$m measurements up to $z$$\sim$3 as noted already in Elbaz et al. (2010), and we confirm this agreement with the extended sample of detected sources in the present analysis (Sect.~\ref{SEC:IRtot}). The trends found for both fields are in good agreement. The stacking $+$ detection measurements for GOODS--N are slightly lower than those obtained for GOODS--S which may result from a combination of cosmic variance and the fact that the GOODS--S image is deeper. 

The redshift evolution of sSFR$_{\rm MS}(z)$ (Fig.~\ref{FIG:sSFRz}), accounting for both detections and stacked measurements, is well fitted by Eq.~\ref{EQ:sSFRz}, 
\begin{equation}
sSFR_{\rm MS}~[{\rm Gyr}^{-1}]=26 \times t_{\rm cosmic}^{-2.2}~,
\label{EQ:sSFRz}
\end{equation}
where $t_{\rm cosmic}$ is the cosmic time elapsed since the Big Bang in Gyr. 
A starburst can be defined by its sSFR following Eq.~\ref{EQ:sSFRz_SB}, 
\begin{equation}
sSFR_{\rm SB}~[{\rm Gyr}^{-1}]>52 \times t_{\rm cosmic}^{-2.2}~.
\label{EQ:sSFRz_SB}
\end{equation}
The intensity of such starbursts, or ``starburstiness'', is then defined by the excess sSFR:  $R_{\rm SB}$=sSFR$_{\rm SB}$/sSFR$_{\rm MS}$. Due to the evolution observed with cosmic time, a galaxy with a sSFR twice as large as the local MS value would be considered a starburst today, but a galaxy with the same sSFR at $z$$\sim$1 would be part of the main sequence.

   \begin{figure}[h!]
   \centering
   \includegraphics*[width=9cm]{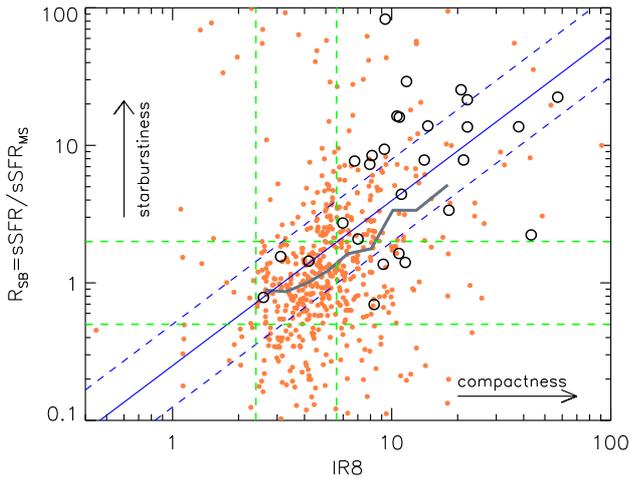}
      \caption{$R_{\rm SB}$=sSFR/sSFR$_{MS}$ versus $IR8$ (=$L_{\rm IR}^{\rm tot}$/$L_8$) for the distant GOODS--{\it Herschel} galaxies. The green lines show the range of values occupied by main sequence galaxies (68\,\% dispersion) in sSFR (horizontal lines; $R_{\rm SB}$=1$\pm$1) and $IR8$ (vertical lines; $IR8_{MS}$=4$\pm$1.6, see Eq.~\ref{EQ:IR8gh}). The thick grey line show the sliding median for the GOODS--{\it Herschel} galaxies. The diagonal blue lines are the best fit (solid) and 68\,\% dispersion (dashed) for local galaxies as in Fig.~\ref{FIG:sSFR_lirl8}. Large open dots show the position of sub-mm galaxies from Menendez-Delmestre et al. (2009) and Pope et al. (2008a).
}
         \label{FIG:sSFR_lirl8_distant}
   \end{figure}
We have seen that for local galaxies, the starburstiness and $IR8$ are correlated (see Fig.~\ref{FIG:sSFR_lirl8}). The same exercise for distant GOODS--{\it Herschel} galaxies, mixing galaxies of all luminosities and redshifts, is shown in Fig.~\ref{FIG:sSFR_lirl8_distant}. Distant galaxies exhibit a non negligible dispersion, but their sliding median, shown by a thick grey line in Fig.~\ref{FIG:sSFR_lirl8_distant}, is coincident with the best fit relation for local galaxies (solid and dashed blue lines). 

We find that 80\,\% of the galaxies which belong to the SFR -- $M_*$ main sequence -- with 0.5$\leq$$R_{\rm SB}$$\leq$2 -- also belong to the main sequence in $IR8$ -- $IR8_{\rm MS}$=4$\pm$1.6 (Eq.~\ref{EQ:IR8gh}). Hence we confirm that the two definitions of ``main sequence galaxies'' are similar and that on average they represent the same galaxy population. We note also that even though there is a tail toward stronger starburstiness and compactness, i.e., increased $R_{\rm SB}$ and $IR8$, this regime of parameter space is only sparsely populated in the GOODS--{\it Herschel} sample, which suggests that analogs to the local compact starbursts predominantly produced by major mergers remain a minority among the distant galaxy population. Finally, we see that sub-mm galaxies (large open circles in Fig.~\ref{FIG:sSFR_lirl8_distant}) also follow the same trend. 

\subsection{Star formation compactness of distant galaxies}
\label{SEC:distant_compactness}
We have shown that local galaxies with high $\Sigma_{\rm IR}$ values also exhibit high $IR8$ ratios. We do not have IR or radio size estimates for the distant galaxy population, but we can use the high resolution {\it HST}--ACS images to study the spatial distribution of the rest-frame UV light in the populations of MS and SB galaxies. It has been suggested that distant (U)LIRGs at 1.5$<$$z$$<$2.5 (Daddi et al. 2007a) and at $z$$\sim$3 (Magdis et al. 2010b) are not optically thick since the SFR derived from the UV after correcting for extinction using the Calzetti et al. (2000) law is consistent with the SFR derived from radio stacking measurements at these redshifts (see also Nordon et al. 2010). 

   \begin{figure}[ht!]
   \centering
   \includegraphics*[width=7.5cm]{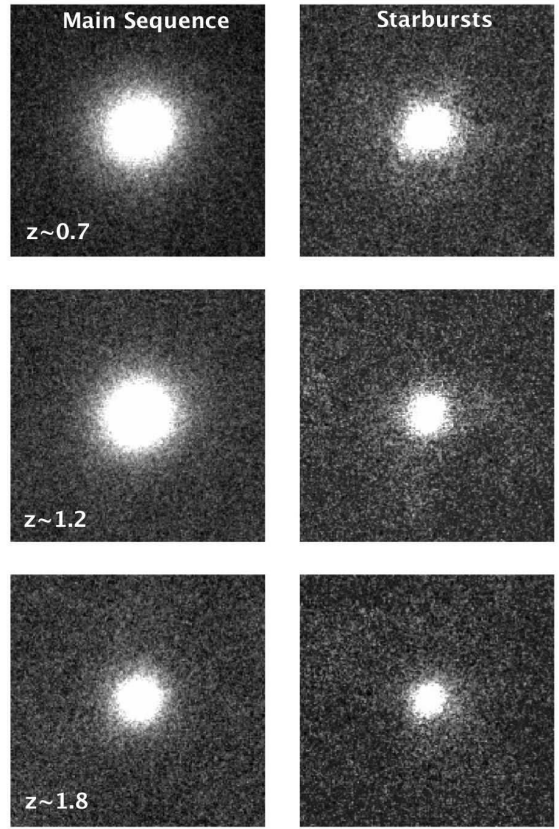}
      \caption{Stacked images (5\arcsec\ on a side) centered on main sequence (\textit{left column}) and starburst (\textit{right column}) galaxy positions. Typical MS galaxies are selected to have $R_{\rm SB}$$=$1$\pm$0.1 and SB galaxies $R_{\rm SB}$$\geq$3. Each image results from the stacking of {\it HST}--ACS images in the B (4350\,\AA), V (6060\,\AA) and I (7750\,\AA) bands corresponding to the rest-frame UV at $\sim$2700\,\AA\ at $z$=0.7 (first line), $z$=1.2 (second line) and $z$=1.8 (third line) respectively.
}
         \label{FIG:stacksUV}
   \end{figure}
\begin{table}[ht!]
\caption{UV -- 2700\,\AA\ half-light radii of distant main sequence and starburst galaxies.}
\begin{center}
\begin{tabular}{cccc}
\hline
\hline
Redshift             & Main Sequence    &  \multicolumn{2}{c}{Starburst}  \\ 
                         &                           &  $R_{\rm SB}$$>$2 & $R_{\rm SB}$$>$3 \\
\hline
0.7                      & 5.2 kpc              &  3.9 kpc & 2.5 kpc\\
1.2                      & 4.4 kpc              &  3.3 kpc & 2.5 kpc\\ 
1.8                      & 3.0 kpc              &  2.5 kpc & 2.0 kpc\\ 
\hline
\end{tabular}
\end{center}
\label{TAB:uvsizes}
\end{table}

We use {\it HST}--ACS images in the $B$ (4350\,\AA), $V$ (6060\,\AA) and $I$ (7750\,\AA) bands to sample the same rest-frame UV wavelength of $\sim$2700\,\AA\ at $z$=0.7, 1.2 and 1.8 respectively. MS galaxies are selected to have $R_{\rm SB}$$=$1$\pm$0.1 (Eq.~\ref{EQ:starburstiness}) whereas SB galaxies are defined as galaxies with $R_{\rm SB}$$\geq$2. We also tested a stricter definition for starbursts, $R_{\rm SB}$$\geq$3 to avoid contamination from MS galaxies (Table~\ref{TAB:uvsizes}). The result of the stacking of {\it HST}--ACS sub-images is shown in Fig.~\ref{FIG:stacksUV} for MS (left column) and SB galaxies with $R_{\rm SB}$$\geq$3 (right column). It is clear that the sizes of the starbursts are more compact than those of the main sequence galaxies. The half-light radius of each stacked image was measured with GALFIT (Peng et al. 2010b) and is listed in Table~\ref{TAB:uvsizes}. 

These sizes are consistent with those obtained by Ferguson et al. (2004). SB galaxies typically exhibit half-light radii that are two times smaller than those of MS galaxies, implying projected star formation densities that are 4 times larger. We verified that this was not due to a mass selection effect by matching the stellar masses in both samples and obtained similar results, although with larger uncertainties. These sizes are larger than the radio-derived IR half-light radii of the local sample of MS (1.8 kpc) and SB (0.5 kpc) galaxies. However, since the distant galaxy sample has a different mass and luminosity selection than that of the local reference sample, we cannot directly compare their sizes. However, the difference in the {\em relative} sizes among the high-redshift galaxies confirms that star formation in distant starbursts is more concentrated than that in distant main sequence galaxies. This, again, is strong evidence for a greater concentration of star formation in galaxies with higher specific SFRs. Since we have seen that sSFR and $IR8$ are correlated (Fig.~\ref{FIG:sSFR_lirl8_distant}), this implies that in distant galaxies, like in local ones, galaxies with strong $IR8$ ratios are likely to be compact starbursts.

This result is consistent with the work of Rujopakarn et al. (2011), who measured IR luminosity surface densities for distant (U)LIRGs similar to those found in local normal star-forming galaxies. However, we find that this is the case for most but not all high redshift (U)LIRGs. Compact starbursts do exist in the distant Universe, even among (U)LIRGs, but they are not the dominant population.
\begin{figure*}[htbp]
  \centering
\includegraphics*[width=9cm]{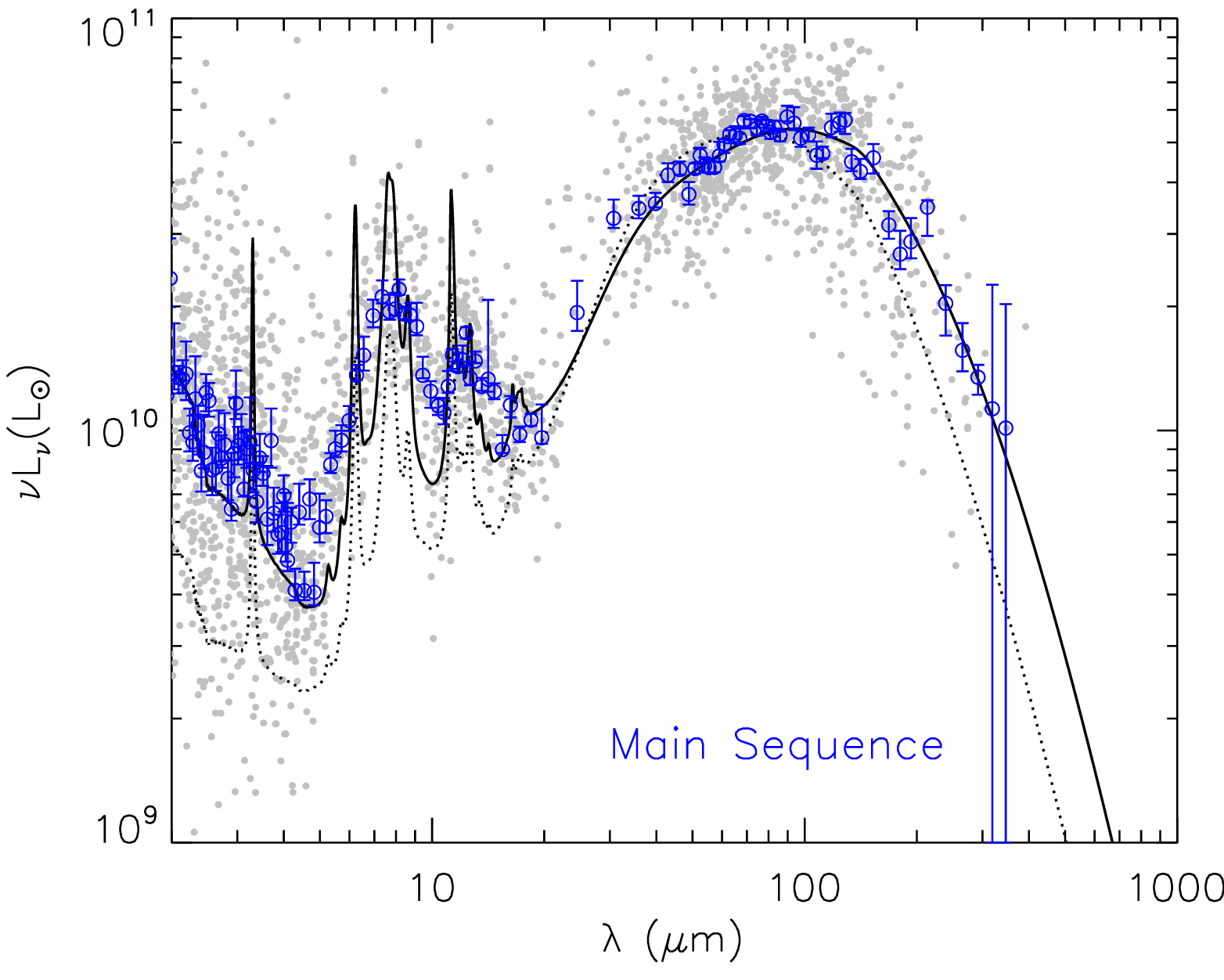}
\includegraphics*[width=9cm]{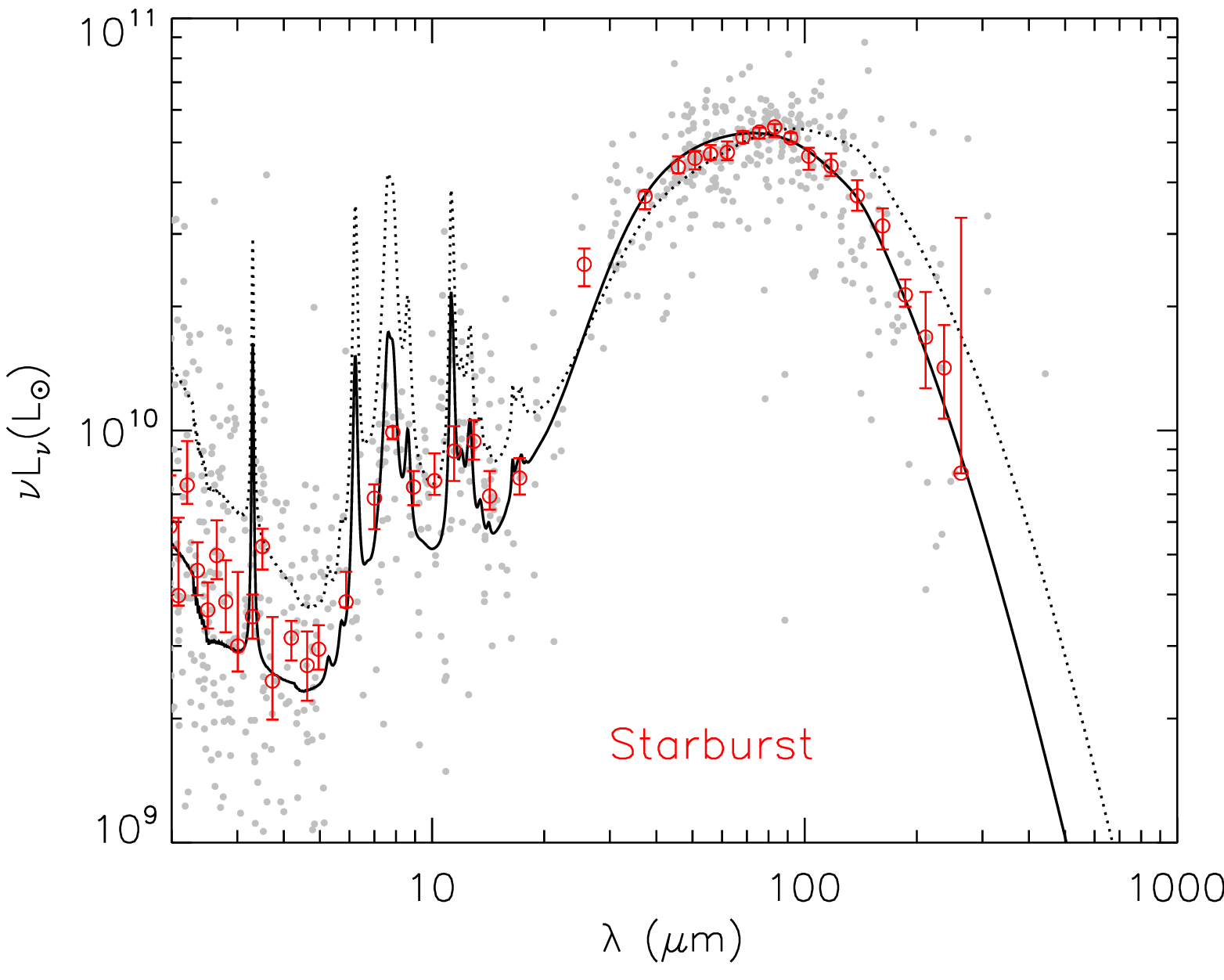}
  \caption{Composite spectral energy distribution of the typical main sequence galaxy (\textit{\textbf{left}}; $IR8$=4$\pm$2, see Eq.~\ref{EQ:IR8gh}) and starburst (\textit{\textbf{right}}; $IR8$$>$8, i.e., above 2$\sigma$). Light grey dots: individual GOODS--{\it Herschel} galaxies normalized to $L_{\rm IR}^{\rm tot}=10^{11}\;\rm L_\odot$. The large filled symbols with error bars are the median and associated uncertainty of the MS (left figure, blue dots) and SB (right figure, red dots) galaxies computed in intervals of wavelengths defined to contain a fixed number of 25$\pm$5 galaxies. The uncertainty on the median values is derived from the 16th and 84th percentiles around the median divided by the square root of the number of galaxies. The model fit to each SED is shown with a solid black line while the opposing SED (MS or SB) is shown with a dotted black line for comparison. 
           }
  \label{FIG:SEDs}
\end{figure*}

\section{Toward a universal IR SED for Main Sequence and Starburst galaxies}
\label{SEC:sed}

\subsection{Medium resolution IR SED for main sequence ($IR8$$\sim4$) and starburst ($IR8$$>8$) galaxies}
At $z$$<$2.5 -- where we can estimate the rest-frame $L_8$ from {\it Spitzer} IRAC, IRS and MIPS photometry as well as reliable $L_{\rm IR}$ from {\it Herschel} measurements at rest-frame $\lambda$$>$30\,$\mu$m -- the $IR8$ (=$L_{\rm IR}$/$L_8$) ratio follows a Gaussian distribution centered on $IR8$$\sim$4 (Eq.~\ref{EQ:IR8gh}, Fig.~\ref{FIG:IR8_local}), with a tail skewed toward higher values for compact starbursts. This defines two populations of star-forming galaxies or, more precisely, two modes of star formation: the MS and SB modes. Galaxies in the MS mode form the Gaussian part of the $IR8$ distribution and present typical sSFR values (i.e., $R_{\rm SB}$$\sim$1) while SB exhibit stronger $IR8$ values (see Fig.~\ref{FIG:IR8_local}) and a stronger ``starburstiness'' ($R_{\rm SB}$$>$2).

$IR8$ is universal among MS galaxies of all luminosities and redshifts. This suggests that these galaxies share a common IR SED. In the local Universe, the rest-frame $L_{12}$, $L_{25}$, $L_{60}$, $L_{100}$ from {\it IRAS} and $L_{15}$ from ISOCAM were also found to be nearly directly proportional to $L_{\rm IR}^{\rm tot}$ (see CE01 and Elbaz et al. 2002), hence reinforcing this idea. To produce the typical IR SED of MS and SB galaxies, we use $k$-correction as a spectroscopic tool. We separate MS and SB galaxies by their $IR8$ ratios: $IR8$=4$\pm$2 for MS galaxies (as in Eq.~\ref{EQ:IR8gh}) and $IR8$$>$8 (hence $>$2$\sigma$ away from the MS) for SB galaxies. We then normalize the individual IR SEDs by a factor 10$^{11}$/$L_{\rm IR}^{\rm tot}$ so that all galaxies are normalized to the same reference luminosity of $L_{\rm IR}^{\rm tot}$=10$^{11}$ L$_{\odot}$. The result is shown with light grey dots in the left-hand part of Fig.~\ref{FIG:SEDs} for MS galaxies and in the right-hand part of Fig.~\ref{FIG:SEDs} for SB galaxies. A sliding median was computed in wavelength intervals which always encompass 25$\pm$5 galaxies (blue points for MS in Fig.~\ref{FIG:SEDs}-left and red points for SB in Fig.~\ref{FIG:SEDs}-right). As a result, the typical MS and SB IR SEDs have an effective resolution of $\lambda$/$\Delta \lambda$=25 and 10 respectively, nearly homogeneously distributed in wavelength from 3 to 350\,$\mu$m. 

The typical MS IR SED in the left-hand part of Fig.~\ref{FIG:SEDs} has a broad far-IR bump centered around 90\,$\mu$m, suggesting a wide range of dust temperatures around an effective value of $\sim$30 K, and strong PAH features in emission. Instead, the typical IR SED for SB galaxies (Fig.~\ref{FIG:SEDs}-right) presents a narrower far-IR bump peaking around $\lambda$$\sim$70--80\,$\mu$m, corresponding to an effective dust temperature of $\sim$40 K, and weak PAH emission lines. We note however, that these prototypical IR SEDs result from the combination of 267 and 111 galaxies for the MS and SB modes, respectively.  They therefore should be considered as average SEDs, acknowledging that there is a continuous transition from one to the other with increasing $IR8$ or star-formation compactness. In the next Section, we provide a model fit to these SEDs to better describe their properties. 

\subsection{SED decomposition of main sequence and starburst galaxies}
\label{SEC:decomp}
\begin{figure}
\centering
\includegraphics*[width=9cm]{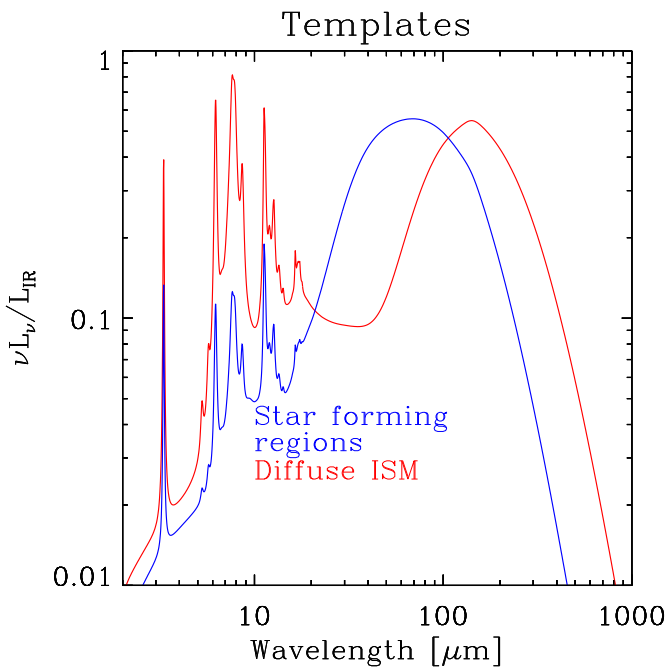}
  \caption{{\sl Components used in the fit of Fig.~\ref{FIG:SEDs}}.
           These two components have been constrained by the simultaneous fit 
           of the two SEDs (Fig.~\ref{FIG:SEDs}).
           The main sequence and starburst SEDs are the linear composition of
           these two components.
           The luminosity ratio of these components controls IR8.
           }
  \label{FIG:SED_decomposition}
\end{figure}

In order to interpret the physical nature of the MS and SB SEDs derived in the previous section, we adopt a simple phenomenological approach.
We decompose the two classes of SEDs with the linear combination of two templates, shown in Fig.~\ref{FIG:SED_decomposition}: (1) a ``{\it star-forming region}'' component including 
H$\,${\sc ii} regions and the surrounding photo-dissociation region (labeled SF), and (2) a ``{\it diffuse ISM}'' (interstellar medium) component accounting for the quiescent regions (labeled ISM).
The luminosity ratio of the two components controls the {\it IR8} parameter. This SED decomposition is not unique and the two components used here are not rigorously associated with physical regions of the galaxies.

The SED of each sub-component is given by the model of Galliano et al. (2011, in prep.; also presented by Galametz et al. 2009).
This model adopts the Galactic dust properties of Zubko, Dwek \& Arendt (2004).
To account for the diversity of physical conditions within a galaxy, we combine the emission of grains exposed to different starlight intensities, $U$ (normalized to the solar neighborhood value of $2.2\times10^{-5}\;\rm W\,m^{-2}$). We assume, following Dale et al. (2001), that the mass fraction of dust exposed to a given starlight intensity follows a power-law (index $\alpha$): $dM_{\rm dust}/dU\propto U^{-\alpha}$. The two cutoffs are $U_{min}$ and $U_{min}+\Delta U$. We fit the two SEDs simultaneously, varying only the luminosity ratio of the two components. We add a stellar continuum to fit the short wavelengths (see Galametz et al. 2009 for a description).
This component is a minor correction. In summary, the free parameters for the fit are: 
\begin{itemize}
  \item the starlight intensity distribution parameters ($\alpha$, $U_{min}$ and $\Delta U$) of each sub-component;
  \item the PAH mass fraction and charge of each sub-component;
  \item the luminosity ratio of the two components for the main sequence and for the starburst;
  \item the contribution of the stellar continuum (negligible here)
\end{itemize}

The fits are shown with solid black lines in Fig.~\ref{FIG:SEDs} while the derived templates that we used for the decomposition are shown in blue and red lines in Fig.~\ref{FIG:SED_decomposition}. The most relevant parameters are summarized in Table~\ref{tab:decomp}. The ``{\it diffuse ISM}'' component has colder dust and a larger PAH mass fraction than the ``{\it star-forming region}'' SED. 

The main differences between galaxies in the MS and SB modes are:
\begin{itemize}
  \item the effective T$_{\rm dust}$ of galaxies in the SB mode is warmer than that of MS galaxies, i.e., $\sim$40 K versus $\sim$31 K;
  \item the contribution of diffuse ISM emission to the SB SED is negligible (8\,\%), consistent with the strong compactness seen both for local starbursts (in radio and mid-IR imaging, see Sect.~\ref{SEC:compactness}) and for high redshift analogs (in the rest-frame UV, see Sect.~\ref{SEC:distant_compactness});
  \item the MS SED requires a wider distribution of dust temperatures, typically ranging from 15 to 50 K;
  \item the stronger contribution of PAH lines to the broadband mid-IR emission in the MS SED is the main cause for the difference in $IR8$ ratios between the two populations.
\end{itemize}
Galaxies are distributed continuously between the MS and various degrees of SB strength, hence this decomposition technique can be used in the future to produce SEDs suitable for ranges of $IR8$ or sSFR values, in the form of a new library of template SEDs. We note, however, that this decomposition of the typical MS and SB SEDs is not unique. For example, the SB SED is very similar to the CE01 template for a local galaxy with $L_{\rm IR} = 6 \times 10^{11}$~L$_{\odot}$ galaxy in the local Universe, which turns out to be close to the observed median luminosity of the starbursts.  Instead, the MS SED is closer to the CE01 SED for a 4$\times$10$^{9}$ L$_{\odot}$ galaxy in the local Universe.

We note also that a direct fit of the Rayleigh-Jeans portions of both SEDs would favor an effective emissivity index of $\beta$=1.5 for the MS and $\beta$=2 for the SB. However, this is a degenerate problem. Indeed, the effective emissivity index $\beta$ is not necessarily equal to the intrinsic $\beta$ of the grains. A temperature distribution of grains having an intrinsic $\beta = 2$ would flatten the sub-mm SED and can give an effective $\beta$ of $\simeq1.5$, as it is the case for our star-forming region. Finally, it is also not possible to disentangle some potential contribution from an AGN, particularly for the SB SED. Indeed, AGN are known to be ubiquitous in LIRGs (Iwasawa et al. 2011) and ULIRGs (Nardini et al. 2010), and they may contribute in part to the mid-IR continuum, mostly in SB SEDs, since those are both more compact and exhibit lower PAH equivalent widths than they do MS galaxies. However, even if AGN may contribute to some fraction of the light in these galaxies, they cannot dominate both in the mid and far-IR regimes since we find evidence that PAHs dominate around 8\,$\mu$m in both MS and SB galaxy types, even if they are stronger in the MS SED.  The high $IR8$ values measured for SBs also suggest that star formation dominates the IR emission in these galaxies. In Sect.~\ref{SEC:AGN} we present a technique to search for hidden AGN activity in the GOODS--{\it Herschel} galaxies.

\begin{table}[t!]
\caption{Main parameters relative to the SED decomposition.}
\begin{tabular}{lrrr}
    \hline\hline
      \multicolumn{4}{c}{\sc Dust Conditions} \\
    \hline
      & \it ISM & \it Star Forming & [units] \\
    \hline
      $\langle U\rangle$ & 1.8 & 757
        & $[2.2\times10^{-5}\;\rm W\,m^{-2}]$ \\
      $T_{\rm eff}$ & 19 & 53 & [K] \\ 
      $f_{\rm PAH}$ & 1.9 & 0.3 & $[4.6\,\%]$ \\
      {\it IR8} & 15 & 3 & \ldots \\
    \hline
      \multicolumn{4}{c}{\sc SED properties} \\
    \hline
      & Main Sequence & Starburst & \\
      $\phi$ & 62 & 92 & [$\%$] \\
      {\it IR8} & 5 & 11 & \ldots \\
      $T_{\rm eff}^{\rm peak}$ & 31 & 40 & [K] \\
\hline
\end{tabular}
\textbf{Notes.} $\langle U\rangle$ is the luminosity averaged starlight intensity;
           $T_{\rm eff}\simeq \langle U\rangle^{1/6}\times 17.5$~K is the 
           corresponding effective temperature of the grains;
           $f_{\rm PAH}$ is the PAH-to-total-dust mass fraction;
           $\phi=L_{\rm SF}/(L_{\rm ISM}+L_{\rm SF})$ is the luminosity fraction
           of the star-forming component. $T_{\rm eff}^{\rm peak}$ is the effective dust 
           temperature corresponding to the peak of the far-IR bump using Wien's law.
  \label{tab:decomp}
\end{table}

\begin{figure}
\centering
\includegraphics[width=9.25cm]{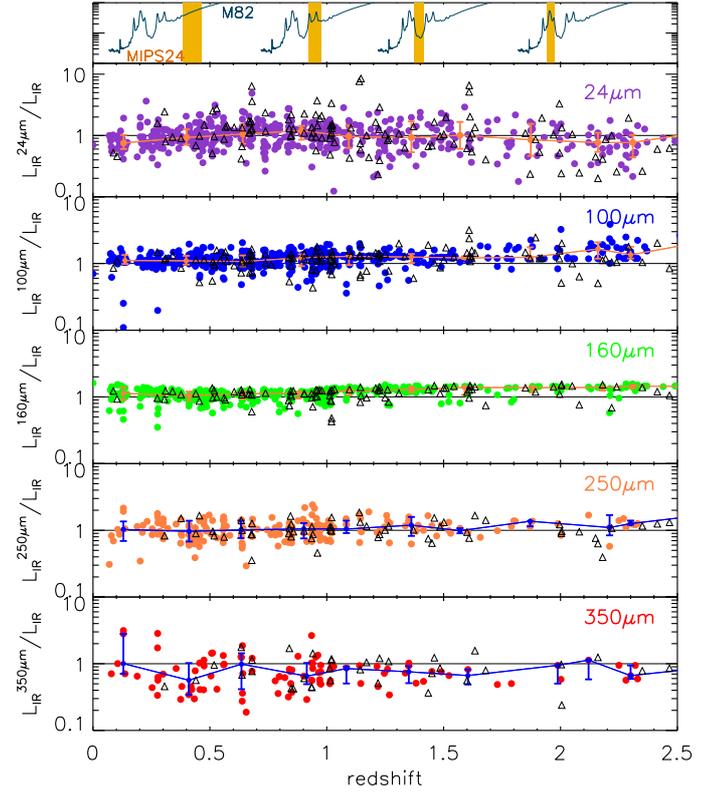}
\caption{Ratio of the extrapolated ($L_{\rm IR}^{\lambda}$) over {\it Herschel} total IR luminosity as a function of redshift, for all clean GOODS--{\it Herschel} galaxies. $L_{\rm IR}^{\lambda}$ is computed by normalizing the main sequence SED to the broadband photometric measurement at $\lambda$. The 5 passbands used for the extrapolation are, from the top to bottom, {\it Spitzer} MIPS--24\,$\mu$m, {\it Herschel} PACS--100\,$\mu$m \& 160\,$\mu$m, {\it Herschel} SPIRE--250\,$\mu$m \& 350\,$\mu$m. $L_{\rm IR}$=$L_{\rm IR}^{\rm Herschel}$ is measured using the full set of {\it Herschel} measurements at rest-frame wavelengths $\lambda$$>$30\,$\mu$m to normalize the main sequence SED and by integrating over 8--1000\,$\mu$m.  Black triangles: AGN. The solid lines with error bars are the sliding median and the 16th and 84th percentiles around it. Upper panel: SED of M82 and MIPS 24\,$\mu$m filter at $z$=0.25, 0.9, 1.4, 2.}
\label{FIG:IRtot}
\end{figure}

\subsection{Derivation of total IR luminosities from monochromatic measurements}
\label{SEC:IRtot}
Now that we have defined a typical IR SED for main sequence galaxies, this SED may be used to extrapolate the total IR luminosity of galaxies for which only one measurement exists. Ideally, one would need to know the value of $IR8$ or equivalently the starburstiness, $R_{\rm SB}$, of a galaxy, to know whether to use the MS or SB SED. But this would require to already know the actual SFR of a galaxy, which is what we are looking for. An alternative technique would consist in using the star formation compactness of a galaxy, or $\Sigma_{\rm IR}$, to determine if it is in the main sequence or starburst mode. 

Assuming that all galaxies share the same MS SED, in Fig.~\ref{FIG:IRtot} we compare the total IR luminosity that can be extrapolated from a single passband using the main sequence IR SED ($L_{\rm IR}^{\lambda}$) with the value ($L_{\rm IR}^{\rm tot}$) measured from an SED fit to galaxies with ``clean'' {\it Herschel} detections in several bandpasses (see Sect.~\ref{SEC:clean}). Both luminosities agree with an average uncertainty of $\sim$35\,\% when $L_{\rm IR}^{\lambda}$ is extrapolated from 24\,$\mu$m and $\sim$20\,\% when $L_{\rm IR}^{\lambda}$ is derived from one of the 100 to 350\,$\mu$m wavelengths. This is remarkable since we only used a single IR SED to extrapolate $L_{\rm IR}^{\lambda}$ for all galaxies. The MS SED does a better job than the CE01 technique (see Fig.3 in Elbaz et al. 2010), which overestimates $L_{\rm IR}^{\rm tot}$ from 24\,$\mu$m measurements at $z$$>$1.5 as well as from SPIRE 250 and 350\,$\mu$m measurements at $z$$<$1.3. Note however that individual galaxies do present a wide range of IR SEDs with different dust temperatures. 

Finally, we note that these extrapolations work nearly equally well for X-ray AGN (black open triangles in Fig.~\ref{FIG:IRtot}) on average, although the dispersion is slightly larger for these galaxies. This suggests that star formation dominates the IR emission in the hosts of typical AGN in deep-field X-ray surveys. We discuss the properties of AGN in detail in Sect.~\ref{SEC:AGN}.

\subsection{Interpretation of the connection between compactness, starburstiness and $IR8$}
At fixed redshift, a normal galaxy forms stars at a rate proportional to its gas mass divided by the free-fall time, as expressed in Eq.~\ref{EQ:SFRinterp},
\begin{equation}
\begin{array}{l}
SFR \propto M_{\rm gas} / \tau_{\rm free-fall} \propto M_{\rm gas} ~~ \rho_{\rm gas + stars}^{0.5} \\
{\rm since}~\tau_{\rm free-fall} \propto 1 / \sqrt{G \rho_{\rm gas + stars}}~,
\end{array}
\label{EQ:SFRinterp}
\end{equation}
If one assumes that the free-fall time is dominated by the gas density, then it follows that the right term of the equation is proportional to $\rho_{\rm gas}^{1.5}$, as in the Schmidt-Kennicutt relation and close to the value of 1.4 found by Kennicutt (1998b) for projected gas and SFR densities. However the role of stars (in the free-fall time) may not be negligible in some conditions and may partly explain why including them in the relation may reduce the observed dispersion, as proposed by Shi et al. (2011).

If, instead, we consider separately the roles played by the density and gas mass in Eq.~\ref{EQ:SFRinterp}, we may interpret that SB galaxies form stars more efficiently as a result of a greater $\rho_{\rm gas + stars}$, hence shorter $\tau_{\rm free-fall}$, possibly due to a merger. In the case of more distant galaxies, a galaxy with a similar stellar mass will naturally possess a higher gas fraction, and hence gas mass for its stellar mass, which will result in a greater sSFR. In this framework, where the difference of sSFR with redshift comes from a greater gas mass content in the past, but with similar gas densities, it is natural that MS galaxies exhibit similar $IR8$ values and share a common prototypical IR SED. In the case of a SB, where the density is increased (e.g., by a merger), the $IR8$ ratio is very sensitive to the geometry of the young stellar population (see, e.g., Galliano, Dwek \& Chanial 2008, in particular their Fig.6). Increasing the compactness of the young stellar population, hence $\Sigma_{\rm IR}$, in the case of SB would increase the radiation field and push the photo-dissociation region farther away where molecules such as PAHs can survive and emit their light. The equivalent width of PAHs would then be reduced and the contribution of continuum emission increased, resulting in greater $IR8$ values. 

   \begin{figure*}[ht!]
   \centering
\includegraphics[width=8.5cm]{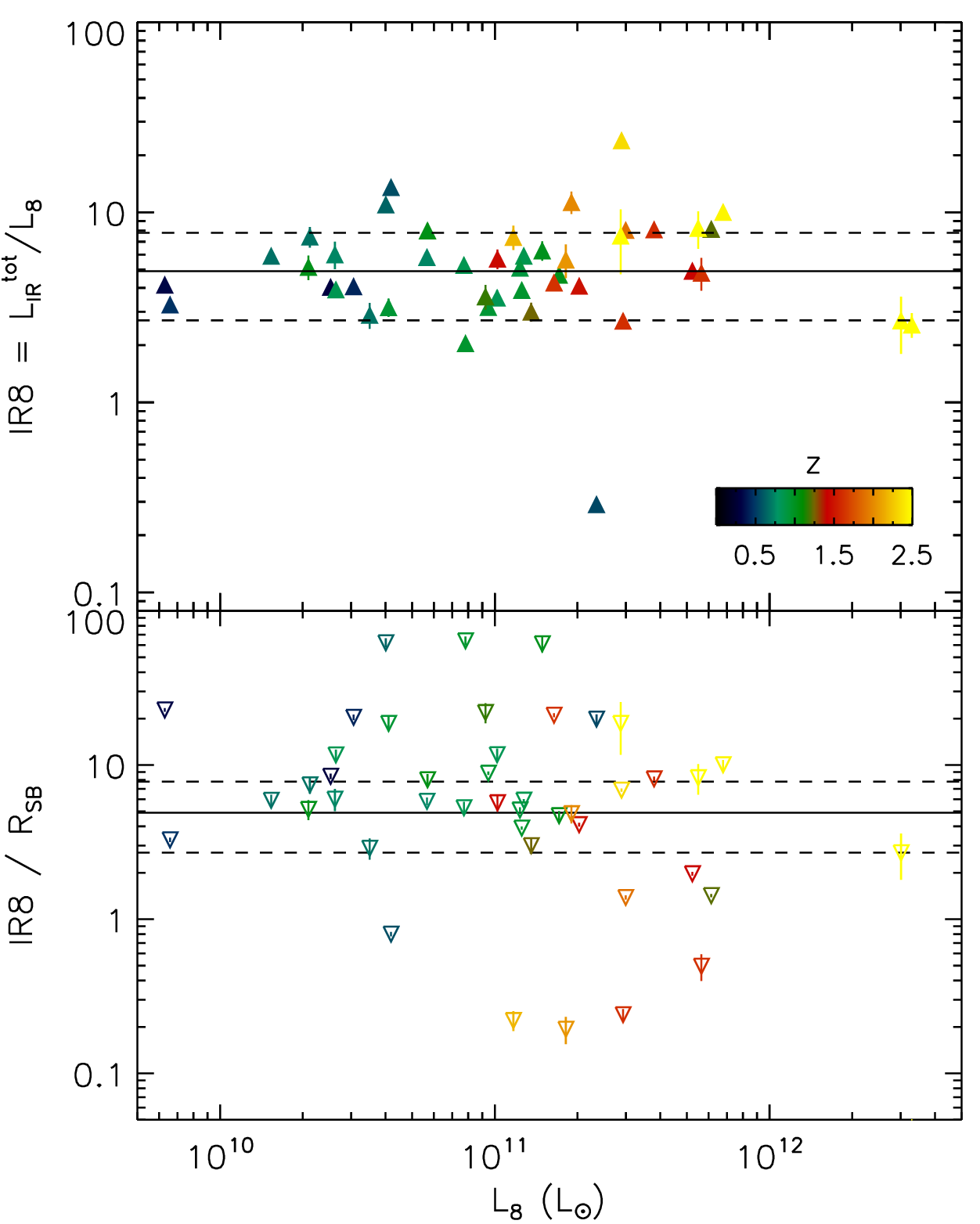}
\includegraphics[width=8.5cm]{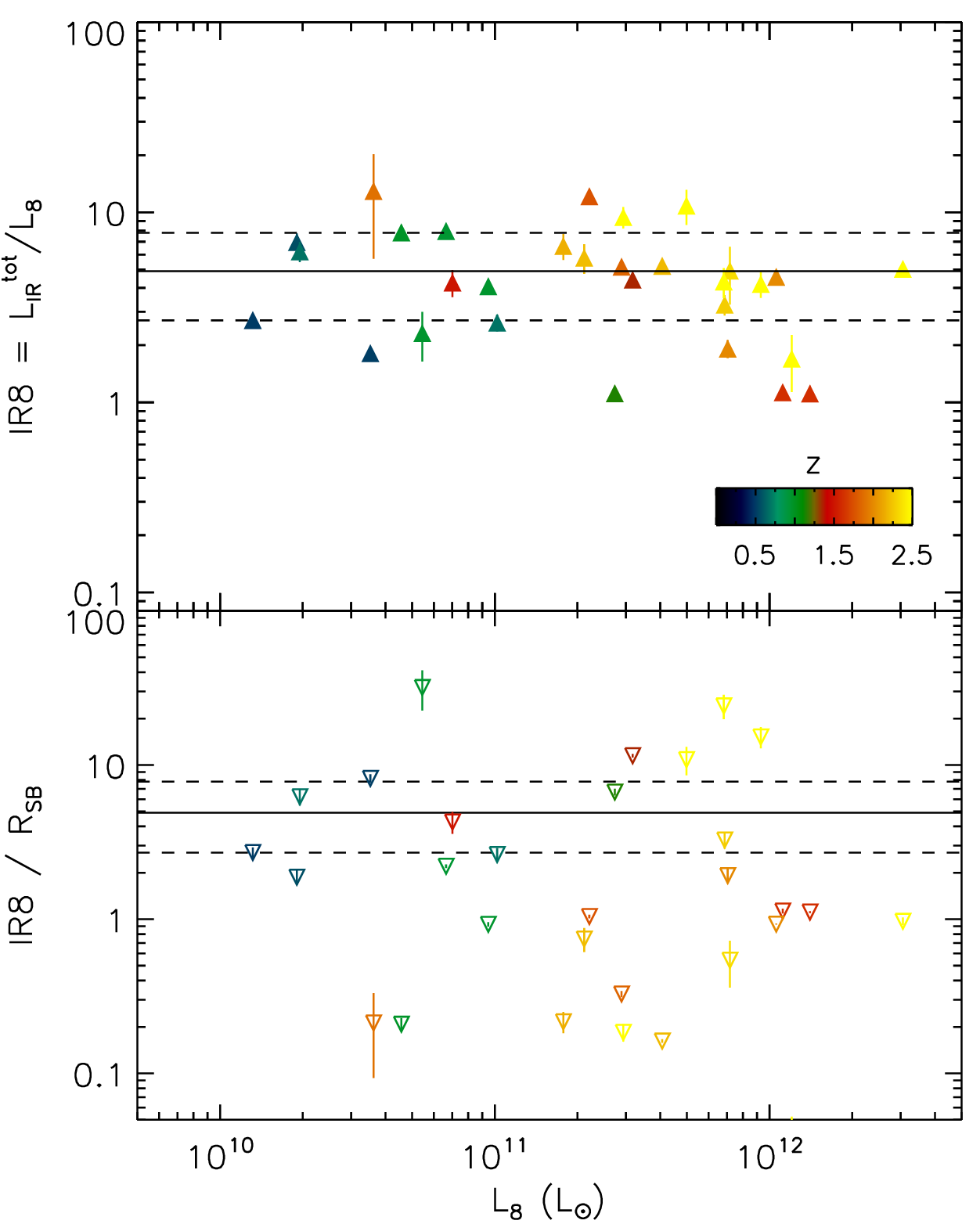}
\caption{\textbf{\textit{Left:}} $IR8$=$L_{\rm IR}^{\rm tot}$/$L_8$ ratio as a function of $L_8$ for X-ray/optical AGN excluding power-law AGN. \textit{Upper panel:} As-measured $IR8$ ratios of the X-ray/optical AGN. They fall on the same sequence as star-forming galaxies. \textit{Bottom panel:} The same galaxies after de-boosting their $IR8$ ratios for starburstiness. The galaxies present a slightly different configuration, but basically remain centered on the trend of star-forming galaxies, suggesting that in these AGN the IR emission is dominated by star formation.
      \textbf{\textit{Right:}} $IR8$ ratio as a function of $L_8$ for power-law AGN. \textit{Upper panel:} As-measured $IR8$ ratios of the power-law AGN. They fall on the same sequence as star-forming galaxies, like the X-ray/optical AGN at left. \textit{Bottom panel:} The same galaxies after de-boosting their $IR8$ ratios for starburstiness. A significant fraction of the galaxies now fall systematically below the main sequence $IR8$ ratio, showing that their 8\,$\mu$m emission is boosted by the hot dust heated by the AGN. 
      }
         \label{FIG:AGN}
   \end{figure*}

\section{Unveiling dusty AGN within starburst galaxies}
\label{SEC:AGN}
\subsection{The $IR8$ bolometric correction factor for X-ray and power-law AGN} 
It is known that AGN can heat the dust that surrounds them to temperatures of several hundreds of Kelvin and produce a spectrum that can dominate the mid-IR emission of a galaxy, but which typically falls off steeply at wavelengths longer than 20\,$\mu$m (Netzer et al. 2007; see also Mullaney et al. 2011a).  AGN can therefore make quite a significant contribution to the emission around 8\,$\mu$m.  In this study, we have identified AGN using all available criteria and have excluded them from most aspects of the analysis up to this point.  Only highly obscured and unidentified AGN may still be present in the sample that we used to carry out the $IR8$ analysis. 

Before trying to define techniques to identify those hidden AGN, we should examine the properties of known and well-recognized AGN in the GOODS--{\it Herschel} data.  We divide the known AGN into two populations (defined in Sect.~\ref{SEC:GH}): the X-ray/optical AGN and the infrared power-law AGN. The reason for this separation is that infrared power-law sources already show evidence that part of their mid-IR emission is powered by an AGN, since they have been identified as galaxies with a rising mid-infrared continuum from 3.6 to 8\,$\mu$m. We used the criteria given in Eq.~\ref{EQ:PLagn}, taken from Le Floc'h et al. (in prep.; technique similar to that from Ivison et al. 2004 and Pope et al. 2008b for sub-mm galaxies):
\begin{equation}
\begin{array}{l}
S_{\nu}[4.5\,\mu m] > S_{\nu}[3.6\,\mu m] \\
log_{10}\left(\frac{S_{\nu}[24\,\mu m]}{S_{\nu}[8\,\mu m]}\right) < 3.64 \times log_{10}\left(\frac{S_{\nu}[8\,\mu m]}{S_{\nu}[4.5\,\mu m]} \right) + 0.15
\end{array}
\label{EQ:PLagn}
\end{equation}
X-ray/optical AGN that also meet the infrared power-law criteria are counted as power-law sources in the following discussion. 

Surprisingly, AGN of both types exhibit $IR8$ ratios that are centered on the median of star-forming galaxies (upper panels of Fig.~\ref{FIG:AGN}), i.e., $IR8$$\sim$4.9 (Eq.~\ref{EQ:IR8}). This behavior may be understood for the non-power-law X-ray/optical AGN, since there is no evidence that high amounts of radiation from the AGN heats the surrounding dust in these galaxies. Hence for those galaxies, it is the star formation that is most probably responsible for both the 8\,$\mu$m and far-IR emission.  

The situation is different for power-law AGN, however. In this case, we know, by definition, that a hot dust continuum is present that exceeds the stellar continuum emission at wavelengths shorter than 8\,$\mu$m. Still, the $IR8$ ratios observed for these galaxies remains similar to that found for star-forming galaxies (upper panel of Fig.~\ref{FIG:AGN}). This is consistent with Fig.~\ref{FIG:IRtot}, where we showed that extrapolations of the total IR luminosity from any single photometric measurement between the observed 24 to 350\,$\mu$m passbands were nearly as accurate for AGN as for star-forming galaxies with no X-ray or optical AGN signatures.

However, we have seen in Sect.~\ref{SEC:MSSB} that compact starbursts have larger $IR8$ ratios than do normal star-forming galaxies. Hence it is possible that two mechanisms act in opposite ways: some contribution from the hot dust heated by an AGN to the mid-IR light may be counterbalanced by the presence of a starburst that increases the far-IR over mid-IR ratio. To test this possibility, we correct for the effect of starbursts on $IR8$ in the next section.
\subsection{Correction of the effect of starbursts on $IR8$} 
\label{SEC:SBcorrection}
We showed that a starburst induces an enhancement of the far-IR emission at fixed 8\,$\mu$m luminosity, whereas an AGN may induce an increase of the 8\,$\mu$m luminosity at fixed far-IR luminosity. The enhancement of $IR8$ in the presence of a starburst is proportional to its intensity, as measured by the starburstiness, $R_{\rm SB}$ (see Fig.~\ref{FIG:sSFR_lirl8} and Fig.~\ref{FIG:sSFR_lirl8_distant}). Hence it can be corrected by normalizing $IR8$ by $R_{\rm SB}$, i.e., replacing $IR8$ by $IR8$/$R_{\rm SB}$.

The de-boosted $IR8$ ratios galaxies with known AGN are shown in the lower panels of Fig.~\ref{FIG:AGN} (open triangles). Interestingly, we find that the two AGN populations behave differently. The $IR8$ ratios of X-ray/optical AGN (from which we have excluded power-law AGN) remain centered on the region of star-forming galaxies. The fraction of AGN falling below the lower limit of main sequence star-forming galaxies increases from 11\,\% to 22\,\% but the same happens above the upper limit, which illustrates that this is just a result of the enhanced dispersion produced when correcting by $R_{\rm SB}$. This suggests that the IR emission of X-ray AGN is predominantly powered by star formation (as also confirmed by Mullaney et al. 2001b). 

The case of the power-law AGN is very different. The fraction of galaxies falling below the lower limit in $IR8$ of MS star-forming galaxies rises from 33\,\% (already three times higher than for the X-ray AGN) to 70\,\% after dividing by  $R_{\rm SB}$. Hence the majority of the power-law AGN show evidence for an 8\,$\mu$m excess, but this excess mid-infrared emission was disguised by the presence of a concurrent starburst.

Two important conclusions can be derived from this observation:
\begin{enumerate}
\item most of the IR emission from non-power-law X-ray/optical AGN appears to be powered by dust heated by stars. Hence their IR luminosities may be used to derive SFRs;
\item the bulk of power-law AGN host both an obscured AGN and a compact starburst. 
\end{enumerate}

The second point suggests that there is a physical link between both activities, the obscured AGN and the starburst, since they take place at the same time. It makes sense that infrared power-law AGN are associated with both compact starbursts and obscured active nuclei since compactness is required both to explain the excess sSFR of these galaxies as well as their dust obscuration. We note that the power-law criterion has not been demonstrated to be a perfect tracer of dusty AGNs, hence some of the galaxies selected by this criterion may just be purely star-forming galaxies.

\subsection{Searching for unknown obscured AGN}
   \begin{figure}
   \centering
\includegraphics[width=9cm]{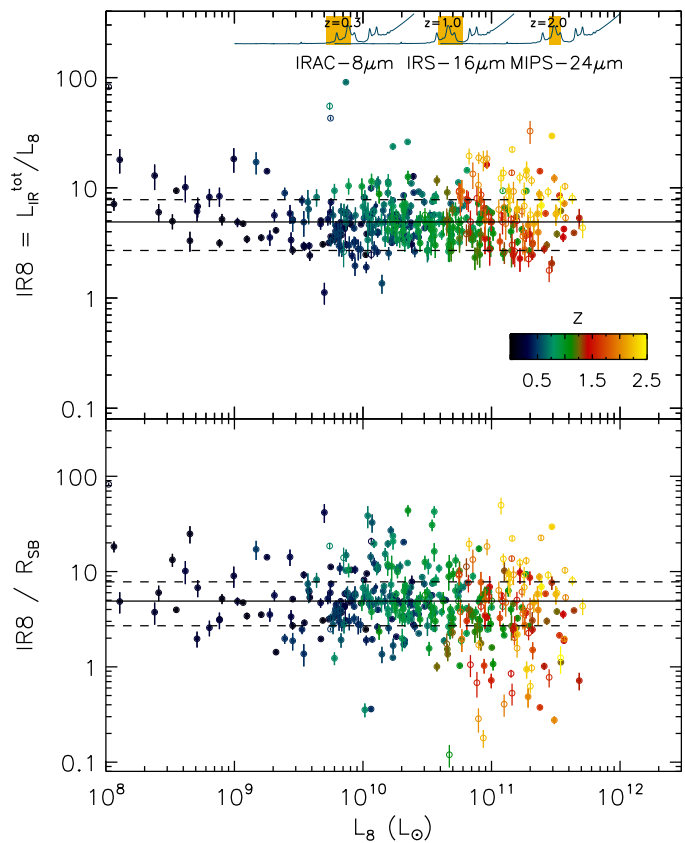}
      \caption{$IR8$=$L_{\rm IR}^{\rm tot}$/$L_{8 \mu m}$ ratio as a function of $L_{8 \mu m}$ for star-forming galaxies, excluding recognized X-ray/optical AGN and infrared power-law AGN. The plain and dashed horizontal lines show the center and width of the Gaussian distribution of main sequence galaxies (Fig.~\ref{FIG:IR8_local} and Eq.~\ref{EQ:IR8gh}). \textbf{\textit{Upper panel:}} position of the star-forming galaxies as measured. Most fall on the same sequence as local star-forming galaxies. \textbf{\textit{Bottom panel:}} position of the same galaxies after correcting their $IR8$ ratio for starburstiness (unless 0.5$\leq$$R_{\rm SB}$$\leq$2, where $R_{\rm SB}$ is consistent with unity within the error bars). 
      }
         \label{FIG:deboostdistgals}
   \end{figure}

We have seen in Sect.~\ref{SEC:SBcorrection} that the presence of a starburst could hide the signature of a dusty AGN on $IR8$, and that this could be corrected by normalizing $IR8$ by $R_{\rm SB}$. In Sect.~\ref{SEC:SBcorrection}, we applied this correction only to known AGNs. We now consider the possibility that galaxies with neither X-ray nor optical evidence of an AGN may harbor a dust-obscured AGN whose signature is masked by the co-existence of a starburst.

In the upper panel of Fig.~\ref{FIG:deboostdistgals} we present the $IR8$ ratios of GOODS--{\it Herschel} galaxies as in the bottom panel of the right-hand part of Fig.~\ref{FIG:IR8}. The central value and width of the Gaussian distribution of main sequence galaxies (Fig.~\ref{FIG:IR8_local} and Eq.~\ref{EQ:IR8gh}) is shown with plain and dashed lines respectively. Only 2\,\% of the galaxies fall below the lower limit of the IR main sequence. After correcting $IR8$ by $R_{\rm SB}$, as in Sect.~\ref{SEC:SBcorrection}, we find that the fraction of galaxies falling below the $IR8$=2.4 lower limit of the main sequence is increased by a factor of 8, reaching 17\,\%, whereas the number of sources above the main sequence was reduced by 15\,\%. The effect is stronger in (U)LIRGs, i.e., above $L_8$$\sim$3$\times$10$^{10}$ L$_{\odot}$, where the fraction of galaxies below $IR8$=2.4 reaches 25\,\%. These galaxies present a starburstiness coefficient that would normally put them in the high-$IR8$ tail of the distribution. Instead, they exhibit normal values of $IR8$, and fall down to low values of $IR8$/$R_{\rm SB}$. This suggests that part of their 8\,$\mu$m rest-frame radiation is powered by a dust-obscured AGN that was not identified from X-ray or optical signatures. These candidate obscured AGN behave similarly to power-law AGN, but they were not identified as such because of the presence of a starburst. They are very good candidates for the missing Compton Thick AGN needed to explain the peak emission of the cosmic X-ray background around 30 keV (Gilli, Comastri \& Hasinger 2007). 

Finally, we applied this technique to a sample of sub-millimeter galaxies (SMGs) from Menendez-Delmestre et al. (2009) and Pope et al. (2008a). The $IR8$ and $R_{\rm SB}$ values of these galaxies are shown in Fig.~\ref{FIG:sSFR_lirl8_distant}. While most objects follow the trend defined for local galaxies (albeit with a wide dispersion), 11 out of a total of 28 SMGs exhibit a starburstiness ($R_{\rm SB}$) higher than expected for their $IR8$. Hence if we correct $IR8$ for starburstiness in these galaxies, we find evidence for the presence of hidden AGN activity. The most extreme cases are the SMGs known as ``C1'', ``GN39a'' and ``GN39b'' from Pope et al. (2008a). C1, or SMM~J123600+621047, is a $z$$\sim$2.002 SMG which has the strongest mid-IR continuum and weakest PAH emission lines in a sample of SMGs with {\it Spitzer} IRS spectroscopy analyzed by Pope et al. (2008a).  Those authors interpret this as evidence that 80\,\% of the mid-IR emission from this object arises from an AGN. It is undetected in the 2 Ms CDF-N data (Alexander et al. 2003), and its X-ray to 6\,$\mu$m luminosity ratio indicates that it hosts a Compton-thick AGN (Alexander et al. 2008). GN39a and GN39b are both classified as obscured AGN based on their strong X-ray hardness ratios, although their mid-IR spectra only show 10-40\,\% AGN contribution. Hence the technique appears to be efficient even in the case of extreme systems such as distant SMGs.

\section{Discussion and conclusion}
The mode in which galaxies form their stars seems to follow some fairly simple scaling laws.
The Schmidt-Kennicutt law, which connects the surface densities of gas and star formation in the local Universe (Kennicutt 1998b), has been recently extended to the study of distant galaxies. Two star formation modes
have thus been identified:  
so-called ``normal'' star formation,
and an accelerated mode, where the star formation efficiency (SFE) is increased, probably due to the merger of two galaxies (Daddi et al. 2010, Genzel et al. 2010). The different SFEs of these two modes are difficult to recognize because of the observational challenges associated with measuring the mass and density of molecular gas at high redshift. Indeed, the CO luminosity to H$_2$ conversion factor is poorly known and is based on $^{12}$CO $J$=1--0 emission locally, while observations of distant galaxies rely on higher-$J$ transitions (see, e.g., Ivison et al. 2011). Similarly, star-forming galaxies follow another scaling law: the SFR -- $M_*$ relation,
which measures the characteristic time to double the stellar mass of a galaxy. At each cosmic epoch, one can identify a typical sSFR for star-forming galaxies.  This distinguishes a main sequence of ``normal'' star-forming galaxies from a minority population of starburst galaxies with elevated sSFR.

Our analysis of the deep surveys carried out in the open time key program GOODS--{\it Herschel} allowed us to establish a third scaling law for star-forming galaxies relating the total IR luminosity of galaxies, $L_{\rm IR}$, hence their SFR, to the broadband 8\,$\mu$m luminosity, $L_8$. We showed that the 8\,$\mu$m bolometric correction factor, $IR8$$\equiv$$L_{\rm IR}$/$L_8$, exhibits a Gaussian distribution containing the vast majority of star-forming galaxies both locally and up to $z$$\sim$2.5, centered on $IR8$$\sim$4. This defines an IR main sequence for star-forming galaxies. Outliers from this main sequence produce a tail skewed toward higher values of $IR8$. We find that this sub-population ($<$20\,\%) is due to galaxies experiencing compact star-formation in a starburst mode.

The projected star-formation densities of present-day galaxies were estimated from their IR surface brightnesses, $\Sigma_{\rm IR}$, as measured from the size of their radio and/or 13.2\,$\mu$m continuum emission. For distant galaxies, we stacked rest-frame UV--2700\,\AA\ images from {\it HST}--ACS in the $B$, $V$ and $I$ filters for galaxies located at $z$$\sim$ 0.7, 1.2 and 1.8 respectively. We find that at all times the projected star-formation density of galaxies in the high-$IR8$ tail is more compact ($\Sigma_{\rm IR}$$>$3$\times$10$^{10}$ L$_{\odot}$kpc$^{-2}$ at $z$$\sim$0) than in galaxies belonging to the IR main sequence, which includes distant (U)LIRGs as well.

Using the more accurate SFRs derived from {\it Herschel} data for galaxies at 0$<$$z$$<$3, we established the evolution of the typical specific SFR (sSFR=SFR/$M_*$) for star-forming galaxies. This allowed us to separate main sequence and starbursting galaxies thanks to a new parameter, labeled ``starburstiness'' (R$_{\rm SB}$), which measures the excess sSFR with respect to the SFR -- $M_*$ main sequence, i.e., R$_{\rm SB}$=sSFR/sSFR$_{\rm MS}$($z$). We find that galaxies belonging to this main sequence (R$_{\rm SB}$=1$\pm$2) also belong to the one defined by the Gaussian distribution of $IR8$, and that the compact, star-forming galaxies that make up the high-$IR8$ tail fall systematically above the SFR -- $M_*$ relation, with strong starburstiness (R$_{\rm SB}$$>$2--3).  Indeed, we find that $IR8$ is strongly correlated with R$_{\rm SB}$ in general.  Hence $IR8$ appears to be a good proxy for identifying compact starbursts, most probably triggered by merger events. In the present-day Universe, most (U)LIRGs are found to be experiencing compact star-formation during a starburst phase, which is not the case for most distant (U)LIRGs.  Most probably, the very high SFRs of local (U)LIRGs can only be achieved during mergers, whereas distant galaxies are more gas-rich and can sustain these large SFRs in other ways.  As a result of this difference, previous studies that have used local (U)LIRG SED templates, with their large, starbursting $IR8$ ratios, to extrapolate from MIPS 24\,$\mu$m photometry of galaxies at $z > 1.5$ have overestimated their total infrared luminosities and SFRs, resulting in the so-called ``mid-IR excess'' issue.

Using $k$-correction as a spectrophotometric tool for converting broadband photometric measurements at various redshifts into a medium resolution IR SED, we were able to determine the prototypical IR SED of MS and SB galaxies with a resolution of $\lambda$/$\Delta \lambda$=25 and 10 respectively. 
The SED of MS galaxies presents strong PAH emission line features, a broad far-IR bump resulting from a combination of emission from dust at different temperatures ranging typically from 15 to 50 K, and an effective dust temperature of 31 K, as derived from the peak wavelength of the IR SED. Galaxies that inhabit the SB regime instead exhibit weak PAH equivalent widths and a sharper far-IR bump with an effective dust temperature of $\sim$40 K. Although PAHs are stronger in MS than in SB galaxies, they are found to be present in the SEDs of both, implying that the IR emission in both populations is primarily powered by star formation and not AGN activity.

Finally, we present evidence that the mid-to-far IR emission of X-ray active galactic nuclei is dominantly produced by star formation, and that power-law AGNs systematically occur in compact, dusty starbursts. After correcting for the excess $IR8$ due to star formation -- estimated from the starburstiness R$_{\rm SB}$ which we showed correlates with $IR8$ -- we identify candidate members of for a sub-population of extremely obscured AGN that have not been identified as such by any other method.

Future studies will be dedicated to understanding the origin of the increase of $IR8$ with compactness and starburstiness, by separating the relative contributions of PAH lines and continuum emission, relating the starburstiness with the local environment of galaxies as well as with their dust temperature. ALMA and eMERLIN will soon provide powerful tools to measure the spatial distribution of star formation in distant galaxies at high angular resolution, making it possible not only to understand the compactness but also the clumpiness of the star-forming regions.

Spectro-imaging with integral field instruments like VLT/SINFONI will also be essential for measuring kinematic signatures to assess whether galaxy interactions play an important role in this process (see, e.g., F\"orster Schreiber et al. 2009, Shapiro et al. 2009). Most star formation takes place among main sequence galaxies, which suggests that internal, secular processes dominate over the role of mergers.  This also accounts more naturally for the long duty cycle of their star-forming phase (Noeske et al. 2007, Daddi et al. 2007a, 2010). Finally, we note that this technique of separating MS and SB galaxies based on their IR star formation compactness will be extremely useful to extrapolate accurate total IR luminosities and SFRs of distant galaxies when using the mid-IR camera MIRI on board the {\it James Webb Space Telescope}.

\begin{acknowledgements}
We wish to thank R.Gobat for generating the three color images of the GOODS fields and our referee Kai Noeske for his constructive comments that helped improving the paper. D.Elbaz and H.S.Hwang thank the Centre National d'Etudes Spatiales (CNES) for their support. D.Elbaz wishes to thank the French National Agency for Research (ANR) for their support (ANR-09-BLAN-0224). VC would like to acknowledge partial support from the EU ToK grant 39965 and FP7-REGPOT 206469. Support for this work was also provided by NASA through an award issued by JPL/Caltech. PACS has been developed by a consortium of institutes led by MPE (Germany) and including UVIE (Austria); KU Leuven, CSL, IMEC (Belgium); CEA, LAM (France); MPIA (Germany); INAFIFSI/OAA/OAP/OAT, LENS, SISSA (Italy); IAC (Spain). This development has been supported by the funding agencies BMVIT (Austria), ESA-PRODEX (Belgium), CEA/CNES (France), DLR (Germany), ASI/INAF (Italy), and CICYT/MCYT (Spain). SPIRE has been developed by a consortium of institutes led by Cardiff University (UK) and including Univ. Lethbridge (Canada); NAOC (China); CEA, LAM (France); IFSI, Univ. Padua (Italy); IAC (Spain); Stockholm Observatory (Sweden); Imperial College London, RAL, UCL-MSSL, UKATC, Univ. Sussex (UK); and Caltech, JPL, NHSC, Univ. Colorado (USA). This development has been supported by national funding agencies: CSA (Canada); NAOC (China); CEA, CNES, CNRS (France); ASI (Italy); MCINN (Spain); Stockholm Observatory (Sweden); STFC (UK); and NASA (USA).
\end{acknowledgements}

\end{document}